\documentclass[12pt,letterpaper,a4paper]{article}
\usepackage[includeheadfoot,
marginratio={1:1,2:3},
width=500pt,
height=720pt,]{geometry}

\usepackage{amsmath}
\usepackage{amsfonts}
\usepackage{amssymb}
\usepackage{graphicx}
\usepackage{cancel}
\usepackage{empheq}
\usepackage{color}
\usepackage{hyperref}

\numberwithin{equation}{section}
\usepackage{float}
\restylefloat{table}
\restylefloat{figure}
\usepackage[utf8]{inputenc}
\usepackage{amsmath, amsthm, amssymb, amsfonts}
\usepackage[font=small, labelfont=bf]{caption}
\usepackage{amsmath, amsthm, amssymb, amsfonts}
\usepackage{multirow}
\usepackage{graphicx}
\usepackage{float}
\usepackage[numbers,sort&compress]{natbib}
\usepackage{enumerate}
\usepackage{amsmath}
\usepackage{amsfonts}
\usepackage{amssymb}
\usepackage{graphicx}
\usepackage{cancel}
\usepackage{empheq}
\usepackage{color}
\usepackage{hyperref}
\usepackage{float}
\restylefloat{table}
\restylefloat{figure}

\newcommand{\nc}{\newcommand}
\nc{\beq}{\begin{equation}}
\nc{\eeq}{\end{equation}}
\nc{\bea}{\begin{eqnarray}}
\nc{\eea}{\end{eqnarray}}

\def\ov{\overline}

\begin{document}
	{\hfill
		%
		arXiv:1909.10993}
	
	\vspace{1.0cm}
	\begin{center}
		{\Large
			Rigid nongeometric orientifolds and the swampland}
		\vspace{0.4cm}
	\end{center}
	
	\vspace{0.35cm}
	\begin{center}
		Pramod Shukla\footnote{Email: pramodmaths@gmail.com}
	\end{center}

	\vspace{0.1cm}
	\begin{center}
		{ICTP, Strada Costiera 11, Trieste 34151, Italy.}
	\end{center}
	
	\vspace{1cm}
	
\abstract{Nongeometric flux compactifications with frozen complex structure moduli have been recently studied for several phenomenological purposes. In this context, we analyze the possibility of realizing de-Sitter solutions in the context of ${\cal N} =1$ type II nongeometric flux compactifications using the ${\mathbb T}^6/({\mathbb Z}_3 \times {\mathbb Z}_3)$ toroidal orientifolds. For the type IIB case, we observe that the Bianchi identities are too strong to simultaneously allow both the NS-NS three-form flux ($H_3$) and the nongeometric ($Q$) flux to take non-zero values, which makes this model irrelevant for phenomenology due to the no-scale structure. For the type IIA case, we find that all the (nongeometric) flux solutions satisfying the Bianchi identities result in de-Sitter no-go scenarios except for one case in which the no-go condition can be evaded. However for this case also, in our (limited) numerical investigation we do not find any de-Sitter vacua using the integer fluxes satisfying all the Bianchi identities.}

	\clearpage
	
	\tableofcontents
	
\section{Introduction}
\label{sec_intro}
In the context of moduli stabilization, the four-dimensional effective potentials arising from type II flux compactifications have attracted a lot attention since more than a decade \cite{Kachru:2003aw, Taylor:1999ii, Blumenhagen:2003vr, Grimm:2004ua, Grimm:2004uq, Denef:2005mm, Grana:2005jc, Balasubramanian:2005zx, Blumenhagen:2006ci, Douglas:2006es, Blumenhagen:2007sm}. In particular, the study of nongeometric flux compactifications and their four-dimensional scalar potentials have led to a continuous progress in various phenomenological aspects such as towards moduli stabilization, in constructing de-Sitter vacua and also in realizing the minimal aspects of inflationary cosmology \cite{Aldazabal:2006up, Ihl:2006pp, Ihl:2007ah, Font:2008vd, Guarino:2008ik, Aldazabal:2008zza, deCarlos:2009qm,  Danielsson:2012by, Blaback:2013ht, Damian:2013dwa, Damian:2013dq, Hassler:2014mla, Blumenhagen:2014gta, Blumenhagen:2015qda,  Blumenhagen:2015jva, Li:2015taa, Blumenhagen:2015kja, Blumenhagen:2015xpa, Blaback:2015zra}. In the context of Type II supergravity theories, such (non-)geometric fluxes arise from a successive application of the T-duality on the NS-NS three-form flux $H_3$ \cite{Hellerman:2002ax,Dabholkar:2002sy,Hull:2004in,Derendinger:2004jn, Derendinger:2005ph, Shelton:2005cf, Wecht:2007wu}, and the exciting part about studying these fluxes is the fact that they can generically induce tree level contributions to the scalar potential for all the moduli and hence can subsequently help in dynamically stabilizing them through the lowest order effects. Moreover, the common presence of the nongeometric fluxes in Double Field Theory (DFT), superstring flux-compactifications, and the gauged supergravities has helped in understanding a variety of interconnecting aspects in these three formulations along with opening new windows for exploring some phenomenological aspects as well \cite{Derendinger:2004jn, Derendinger:2005ph, Shelton:2005cf, Wecht:2007wu, Aldazabal:2006up, Dall'Agata:2009gv, Aldazabal:2011yz, Aldazabal:2011nj, Geissbuhler:2011mx, Grana:2012rr, Dibitetto:2012rk, Andriot:2013xca, Andriot:2014qla, Blair:2014zba, Andriot:2012an, Geissbuhler:2013uka, Guarino:2008ik, Blumenhagen:2013hva, Villadoro:2005cu, Robbins:2007yv, Lombardo:2016swq, Lombardo:2017yme}. Moreover, the nongeometric flux compactification scenarios also present some interesting utilisations of the symplectic geometries \cite{Ceresole:1995ca, D'Auria:2007ay} to formulate the effective scalar potentials; e.g. see \cite{Shukla:2015hpa, Gao:2017gxk, Shukla:2016hyy}, which generalize the work of \cite{Taylor:1999ii, Blumenhagen:2003vr} by including the nongeometric fluxes. The ten-dimensional origin of the four-dimensional nongeometric scalar potentials have been explored recently via an iterative series of works in the supergravity theories \cite{Villadoro:2005cu, Blumenhagen:2013hva, Gao:2015nra, Shukla:2015rua, Shukla:2015bca, Andriot:2012wx, Andriot:2012an,Andriot:2014qla, Blair:2014zba, Andriot:2013xca, Andriot:2011uh, D'Auria:2007ay} and through some robust realization of the Double Field Theory (DFT) reduction on Calabi Yau threefolds \cite{Blumenhagen:2015lta}. Moreover, a concrete connection among the type II effective potentials derived from DFT reductions and those of the symplectic approach has been established in \cite{Shukla:2015hpa, Gao:2017gxk}. A recent review on the developments in the nongeometric flux compactifications can be found in \cite{Plauschinn:2018wbo}. 

Although a consistent incorporation of the various nongeometric fluxes enriches the compactification backgrounds creating more/better possibilities for model building, it is still not fully known how many and which type of fluxes can be simultaneously turned-on. However, it has been observed through explicit constructions that it is quite constraining to consistently satisfy all the quadratic flux constraints coming from the various Bianchi identities and the tadpole cancellation conditions. In this regard, there have been two formulations for computing the Bianchi identities; the standard one mostly applicable to toroidal orientifolds involves fluxes with non-cohomology indices \cite{Shelton:2005cf, Ihl:2007ah, Aldazabal:2008zza}, while in the cohomology formulation fluxes are represented using the non-trivial cohomology indices \cite{Ihl:2007ah, Benmachiche:2006df, Grana:2006hr}. For obvious reasons the later one is utilized for simplifying the scalar potentials in the models developed in the beyond toroidal settings. However, a mismatch between the two sets of constraints arising from the Bianchi identities of these two formulations have been observed in \cite{Ihl:2007ah,Robbins:2007yv} and studied in some good detail in \cite{Shukla:2016xdy,Gao:2018ayp}. The additional identities in the cohomology formulation might be relevant in the recent interesting studies performed in \cite{Blumenhagen:2015qda, Blumenhagen:2015kja,Blumenhagen:2015jva, Blumenhagen:2015xpa, Li:2015taa, Betzler:2019kon}. 

In the context of type IIA flux compactifications, some significant initiatives for the moduli stabilization and the study of scalar potential have been taken, e.g. see \cite{Grimm:2004ua, Villadoro:2005cu, DeWolfe:2005uu, Camara:2005dc, Ihl:2006pp, Ihl:2007ah, Palti:2008mg, Dibitetto:2011qs, Blaback:2013fca, Escobar:2018rna, Marchesano:2019hfb}, and in the meantime, several no-go scenarios forbidding de-Sitter and inflationary realizations have been also found \cite{Maldacena:2000mw, Hertzberg:2007wc, Hertzberg:2007ke, Haque:2008jz, Flauger:2008ad, Caviezel:2008tf,  Covi:2008ea, deCarlos:2009fq, Caviezel:2009tu, Danielsson:2009ff, Danielsson:2010bc, Wrase:2010ew,  Shiu:2011zt, McOrist:2012yc, Dasgupta:2014pma, Gautason:2015tig, Junghans:2016uvg, Andriot:2016xvq, Andriot:2017jhf, Danielsson:2018ztv}.  In close connections with these type IIA no-go scenarios, the swampland conjecture of \cite{Ooguri:2006in} about obstructing the de-Sitter realization in a consistent theory of quantum gravity has been recently promoted as a bound involving the scalar potential ($V$) and its derivatives \cite{Obied:2018sgi},
\bea
\label{eq:old-swamp}
& & |\nabla V| \geq \frac{c}{M_p} \, V\,, 
\eea
where the constant $c > 0$ is an order one quantity. This bound in eqn. (\ref{eq:old-swamp}) does not only forbid the de-Sitter minima but also the de-Sitter maxima as well, and several counter examples were known \cite{Haque:2008jz, Danielsson:2009ff, Danielsson:2011au, Chen:2011ac, Danielsson:2012et} or have been proposed soon after the proposal was made \cite{Andriot:2018wzk, Andriot:2018ept, Garg:2018reu, Denef:2018etk, Conlon:2018eyr, Roupec:2018mbn, Murayama:2018lie, Choi:2018rze, Hamaguchi:2018vtv, Olguin-Tejo:2018pfq, Blanco-Pillado:2018xyn} reflecting the need of refining the conjecture. Subsequently a refined version of the conjecture has been proposed which states that at least one of the following two constraints should always hold \cite{Ooguri:2018wrx},
\bea
\label{eq:new-swamp}
& & |\nabla V| \geq \frac{c}{M_p}\,V\, , \qquad \qquad {\rm min} \Bigl[\nabla_i \nabla_j V\Bigr] \leq  - \frac{c'}{M_p^2}\,V \,,
\eea
where $c$ and $c' > 0$ are order one constants. Following from the first eqn. of (\ref{eq:new-swamp}), one can observe that the conjecture is trivial for non-positive potentials, as well as in the limit $M_p \to \infty$ when gravity decouples. Moreover, these two parameters $c$ and $c'$ are correlated with the well known inflationary parameters expressed in terms of the derivatives of the potential, namely the $\epsilon_V$ and $\eta_V$ parameters; e.g. see \cite{BlancoPillado:2006he, Hertzberg:2007ke}, where $\epsilon_V \geq \frac{1}{2}\, c^2$ and $|\eta_V| \geq c'$, and these parameters are needed to be sufficiently smaller than one, i.e. $\epsilon_V \ll 1$ and  $|\eta_V| \ll 1$, for having the slow-roll inflation. In the meanwhile, there have been a surge in related studies in connection with this conjecture \cite{Denef:2018etk, Conlon:2018eyr, Garg:2018reu, Kinney:2018nny, Roupec:2018mbn, Murayama:2018lie, Choi:2018rze, Hamaguchi:2018vtv, Olguin-Tejo:2018pfq, Blanco-Pillado:2018xyn, Achucarro:2018vey, Kehagias:2018uem, Kinney:2018kew, Andriot:2018wzk, Andriot:2018ept, Lin:2018kjm, Han:2018yrk, Raveri:2018ddi, Danielsson:2018qpa, Dasgupta:2018rtp, Andriolo:2018yrz, Dasgupta:2019gcd, Andriot:2019wrs}; see \cite{Palti:2019pca} also for a recent review. The swampland conjecture \cite{Ooguri:2006in} has been also found to be in close connections with the allowed inflaton field range in a trustworthy effective field theory description as it has been argued that massive tower of states can get excited after a certain limit to the inflaton excursions \cite{Blumenhagen:2017cxt, Blumenhagen:2018nts, Blumenhagen:2018hsh, Palti:2017elp, Conlon:2016aea, Hebecker:2017lxm, Klaewer:2016kiy, Baume:2016psm, Landete:2018kqf, Cicoli:2018tcq, Font:2019cxq, Grimm:2018cpv, Hebecker:2018fln, Banlaki:2018ayh, Junghans:2018gdb}. However, let us also note that in contrary to the (minimal) de-Sitter no-go scenarios, there have been tremendous amount of efforts leading to several proposals for realizing stable de-Sitter vacua in the context of string model building \cite{Kachru:2003aw, Burgess:2003ic, Achucarro:2006zf, Westphal:2006tn, Silverstein:2007ac, Rummel:2011cd, Cicoli:2012fh, Louis:2012nb, Cicoli:2013cha, Cicoli:2015ylx, Cicoli:2017shd, Akrami:2018ylq, Antoniadis:2018hqy}; see \cite{Heckman:2019dsj, Heckman:2018mxl} also for the $F$-theoretic initiatives taken in this regard. We also refer the interested readers to a recent review in \cite{Cicoli:2018kdo}. 

Toroidal orientifolds have been utilized as basic toolkits for several model building purposes. Being simple they facilitate a playground for performing explicit computations for nongeometric flux compactifications as well. In connection to that, nongeometric flux compactifications with rigid (CY) threefolds have recently witnessed a significant amount of interest in studying some phenomenological aspects \cite{Blumenhagen:2015kja, Blumenhagen:2015qda, Blumenhagen:2015jva, Shukla:2015rua, Damian:2018tlf}. In the absence of any active complex structure moduli, these setups are quite simple for explicit computations. In this article we consider one such rigid construction using  the ${\mathbb T}^6/({\mathbb Z}_3 \times {\mathbb Z}_3)$ orbifold with the aim to perform a deep investigation on finding de-Sitter vacua using nongeometric fluxes satisfying all the NS-NS Bianchi identities. For that purpose, firstly we will explore the mismatch of identities for the ${\cal N} = 2$ rigid compactifications using the ${\mathbb T}^6/({\mathbb Z}_3 \times {\mathbb Z}_3)$ orbifold, and subsequently for the ${\cal N} =1$ type IIA and type IIB orientifolds to be used later on. It has been a quite conventional anticipation that the inclusion of nongeometric fluxes can (always) provide a window for evading the classical de-Sitter no-go theorems, however using the $T$-dual dictionary of \cite{Shukla:2019wfo} and the mirror arguments, several no-go scenarios have been derived in the context of nongeometric type IIB compactifications in \cite{Shukla:2019dqd}. However, given the fact that mirror of a rigid CY is not a CY \cite{Hosono:1994av, Hori:2000kt, Sethi:1994ch}, a separate analysis for the type II scalar potentials obtained from the rigid compactifications is necessary to explore along the lines of de-Sitter swampland conjecture, and we aim to fill this gap in this article.

The article is organized as follows: In section \ref{sec_basics} we present a short review on the generic type II nongeometric flux compactifications with the necessary ingredients relevant for the upcoming sections. Section \ref{sec_BIs-Z3xZ3} provides an explicit computation of the ${\cal N}=2$ Bianchi identities appearing in the type II orbifold compactification using a ${\mathbb T}^6/({\mathbb Z}_3 \times {\mathbb Z}_3)$ threefold, where we also present the full set of flux constraints for the orientifolded type IIA and type IIB theories. Section \ref{sec_IIA} contains some detailed investigation of the de-Sitter no-go scenarios in type IIA nongeometric compactifications using the above toroidal model. Section \ref{sec_conclusions} presents summary and conclusions. In addition, we have appended two sections \ref{sec_appendix1} and \ref{sec_appendix2} towards the end of the article, in which section \ref{sec_appendix1} includes some details on the explicit computations for the allowed flux components, identities etc. for the ${\mathbb T}^6/({\mathbb Z}_3 \times {\mathbb Z}_3)$ orientifolds while the section \ref{sec_appendix2} presents a concise review on the type IIA scalar potential which we use in section \ref{sec_IIA} for studying the no-go's in rigid orientifolds in particular.

\section{Nongeometric flux compactifications}
\label{sec_basics}
In the generic ${\cal N} =2$ type II compactifications to four-dimensions, let us first fix the notations and conventions. We consider $\mu_A$ as the basis of harmonic $(1,1)$-forms and $\tilde\mu^A$ as the respective basis of the dual $(2,2)$-forms. Further, ${\cal A}_\Lambda$ and ${\cal B}^\Delta$ form the bases of $(2,1)$-forms. In addition, we denote the zero-form as {\bf 1} and the six-form as $\Phi_6$. Further, the triple intersection numbers and the normalization of the various forms are fixed as,as follows
\bea
\label{eq:intersectionBases}
& & \hskip-1cm \int_{X_3} \mu_A \wedge \tilde{\mu}^B \equiv \delta_A{}^B, \qquad \qquad \int_{X_3} {\cal A}_\Lambda \wedge {\cal B}^\Delta = \delta_\Lambda{}^\Delta\,,\\
& & \hskip-1cm \int_{X_3} \mu_A \wedge \mu_B \wedge \mu_C = \kappa_{ABC}, \qquad \qquad \int_{X_3} \Phi_6 = 1\,. \nonumber
\eea
To incorporate the effects of various NS-NS fluxes, we consider the twisted differential operator ${\cal D}$ defined as follows,
\bea
\label{eq:twistedD}
& & {\cal D} = d - H \wedge . - \omega \triangleleft . - Q \triangleright . - R \bullet . \, ,
\eea
where the origin of geometric flux ($\omega$) and the nongeometric fluxes ($Q$ and $R$) is motivated by a successive application of $T$-duality on the three-form $H$-flux of the type II supergravities. It turns out that a chain with geometric and nongeometric fluxes appears in the following manner under three $T$-duality transformations \cite{Shelton:2005cf},
\bea
\label{eq:Tdual}
& & H_{ijk} \longrightarrow \omega_{ij}{}^k  \longrightarrow Q_i{}^{jk}  \longrightarrow R^{ijk} \, .
\eea
Note that the action of various (non-)geometric fluxes in eqn. (\ref{eq:twistedD}) are introduced via operations $\triangleleft$, $\triangleright$ and $\bullet$ on a $p$-from which changes it into a $(p+1)$-form, a $(p-1)$-form and a $(p-3)$-form respectively. The explicit forms of the various flux-actions on a $p$-form $A_p$ can be given as follows \cite{Robbins:2007yv,Shelton:2006fd},
\bea
\label{eq:action0}
& & \hskip-0.75cm (\omega \triangleleft A)_{i_1i_2...i_{p+1}} = \frac{p(p+1)}{2} \, \omega_{[\underline{i_1 i_2}}{}^{j} A_{j|\underline{i_3.....i_{p+1}}]}, \\ 
& & \hskip-0.75cm (Q \triangleright A)_{i_1i_2...i_{p-1}} = \frac{p-1}{2} \, Q_{[\underline{i_1}}{}^{jk}{} A_{jk|\underline{i_2.....i_{p-1}}]}, \nonumber\\ 
& & \hskip-0.75cm (R \bullet A)_{i_1i_2...i_{p-3}} = \frac{p-3}{3!} \, R^{jkl} A_{jkl \, i_1.....i_{p-3}} \, ,\nonumber
\eea 
where the underlined indices inside the brackets $[..]$ are anti-symmetrized, and in addition we have assumed that the components of $\omega$ flux and the $Q$-flux with one free index are absent, i.e. $\omega_{ij}{}^i = 0$ and $Q_i{}^{ij}  = 0$. This is something called as ``tracelessness condition" \cite{Shelton:2006fd, Wecht:2007wu} and in the literature it is very common to impose this constraint, especially for the case of a Calabi Yau (CY) compactification as a CY does not have any homologically non-trivial one- as well as five-cycles, and hence for the CY orientifold cases it would be well justified to require that all flux components having effectively one (real six-dimensional) free-index to be trivial. 

\subsection{Cohomology formulation}
In order to study the nongeometric flux models beyond the toroidal settings, it is important to express the ingredients in what we call as ``cohomology formulation". In this formulation, the non-trivial flux actions on various harmonic forms can be given as \cite{Grana:2006hr},
\bea
\label{eq:fluxAction2}
& & \hskip-0.75cm H \wedge {\bf 1} = -\, H^\Lambda\, {\cal A}_\Lambda + H_\Lambda\, {\cal B}^\Lambda, \qquad \quad H \wedge {\cal A}_{\Lambda} = - \, H_{\Lambda} \, \Phi_6, \qquad H \wedge {\cal B}^{\Lambda} = - \, H^{\Lambda} \, \Phi_6, \\
& & \hskip-0.75cm \omega \triangleleft \mu_A = \, \omega_{A}{}^{\Lambda}\, {\cal A}_\Lambda - \omega_{A\Lambda}\, {\cal B}^\Lambda, \qquad  \quad \omega \triangleleft {\cal A}_{\Lambda} = - \, \omega_{A\,\Lambda} \, \tilde\mu^{A}, \qquad \omega \triangleleft {\cal B}^{\Lambda} = - \, \omega_A{}^{\Lambda} \, \tilde\mu^A, \nonumber\\
& & \hskip-0.75cm Q \triangleright \tilde\mu^A = - \, Q^{A\,\Lambda} \, {\cal A}_\Lambda + Q^{A}{}_{\Lambda} \, {\cal B}^\Lambda, \qquad Q \triangleright {\cal A}_{\Lambda} = - \, Q^{A}{}_{\Lambda} \, \mu_A, \quad \, \, \, Q \triangleright  {\cal B}^{\Lambda} = - \, Q^{A\, \Lambda} \, \mu_A, \nonumber\\
& & \hskip-0.75cm R \bullet \Phi_6 = R^{\Lambda} \, {\cal A}_\Lambda - R_{\Lambda} \, {\cal B}^\Lambda, \qquad \quad \quad R \bullet {\cal A}_{\Lambda} = - \, R_{\Lambda} \, {\bf 1}, \qquad \quad \, \, \, R \bullet {\cal B}^{\Lambda} = - \, R^{\Lambda} \, {\bf 1}\,. \nonumber
\eea
Some rough estimates on the counting of the (maximum) possible twist-invariant flux components in cohomology formulation are  presented in table \ref{tab_Neq2fluxcount}. Moreover, after the orientifolding is imposed, it is easy to convince that the number of flux components in the cohomology formulation can be simply given as $2\, (h^{1,1} + 1)\, (h^{2,1}+1)$, which is half of the number before orientifolding \cite{Ihl:2007ah, Gao:2018ayp}. However, all these estimates and the countings would be further constrained by the Bianchi identities and the tadpole cancellation conditions as we will discuss in the upcoming sections. 
\begin{table}[H]
 \centering
 \begin{tabular}{|c||c|c|}
\hline
 & & \\
Flux type & Flux type  & Max. number of \\
 (standard) & (cohomology) & flux components \\
\hline
 & & \\
$H_{ijk}$ \qquad  & \quad $H_{\Lambda}, \quad H^{\Lambda}$ \qquad & \, $2(h^{2,1}+1)$ \qquad\\
 &  & \\
$\omega_{ij}{}^k$  \qquad  & \quad $\omega_{A\Lambda}, \quad \omega_{A}{}^{\Lambda}$ \qquad & $2\,h^{1,1}\, (h^{2,1}+1)$ \qquad \\
 &  & \\
$Q_i{}^{jk}$    \qquad  & \quad $Q^A{}_\Lambda, \quad Q^{A \Lambda}$ \qquad & $2\,h^{1,1}\, (h^{2,1}+1)$ \qquad\\
 &  & \\
$R^{ijk}$      \qquad & \quad $R_\Lambda, \quad R^\Lambda$ \qquad & $2(h^{2,1}+1)$ \qquad \\
&  & \\
 \hline
&  & \\
& \qquad {\bf Total } \qquad & $4\, (h^{1,1} + 1)\, (h^{2,1}+1)$ \qquad  \\
&  & \\
 \hline
 \end{tabular}
\caption{Maximum possible number of flux components in cohomology formulation.}
 \label{tab_Neq2fluxcount}
\end{table}
\noindent
In addition, let us also mention the following moduli space metrics relevant for writing down the effective scalar potentials in the four-dimensional theory \cite{Grimm:2004uq},
\bea
\label{eq:Neq2Kmatrices1}
& & \int_{X_3} \mu_A \wedge \ast \mu_B = {\cal G}_{AB}, \qquad \int_{X_3} \, \tilde{\mu}^A \wedge \ast \tilde{\mu}^B = {\cal G}^{AB}, \\
& & \int_{X_3} \beta^I \wedge \ast \, \beta^J = \, {\rm Im}{\cal M}^{IJ} \,, \qquad  \int_{X_3} \alpha_I \wedge \ast \, \beta^J  = \, {\rm Re}{\cal M}_{IK} \, \, {\rm Im}{\cal M}^{KJ} \, , 
\nonumber\\
& & \int_{X_3} \alpha_I \wedge \ast \alpha_J = \, \left( {\rm Im}{\cal M}_{IJ} + {\rm Re}{\cal M}_{IK} \, \, {\rm Im}{\cal M}^{KL} \, \, {\rm Re}{\cal M}_{LJ} \right)\, , \nonumber
\eea
where
\bea
\label{eq:Neq2Kmatrices2}
& & K_{A \ov B} = \frac{\kappa_A \, \kappa_B - 4\, {\cal V} \, \kappa_{AB}}{16\, {\cal V}^2} \equiv  \frac{1}{4{\cal V}} \, {\cal G}_{AB}\,, \\
& &  K^{A \ov B} = 2 \, \, t^A \, t^B - 4\, {\cal V} \, \, \kappa^{AB}\,\equiv  {4{\cal V}} \, {\cal G}^{AB}\,, \nonumber\\
& & {\cal M}_{IJ} = \ov{\cal F}_{IJ} + 2\, i\, \frac{({\rm Im}\,{\cal F})_{IK} \, {\cal X}^K \, ({\rm Im}\,{\cal F})_{JL} \, {\cal X}^L }{{\cal X}^K({\rm Im}\,{\cal F})_{KL} \, {\cal X}^L} \,. \nonumber
\eea
Here $K$ is the K\"ahler potential depending the K\"ahler moduli $T^A = t^A + i\, b^A$ where $t^A$ denotes the volume of the two-cycle moduli and $b^A$ denotes the NS-NS $B_2$ axions. Further, ${\cal M}_{IJ}$ presents the moduli space metric depending on the complex structure moduli written in terms of the derivatives of a pre-potential ${\cal F}$, which is a homogeneous function of degree two in the complex coordinates ${\cal X}^I$. In addition, the shorthand notations such as $\kappa_A\, t^A = 6\, {\cal V} =  \kappa_{ABC} \, t^A \, t^B \, t^C, \, \kappa_{AB} = \kappa_{ABC}\, t^C, \, \kappa_{A} = \kappa_{ABC} \, t^B \, t^C = 2 \, \sigma_A$ as well as $\kappa^{AB}$ as the inverse of $\kappa_{AB}$, will be used whenever needed. 

\subsection{Bianchi identities}
For studying moduli stabilization and any subsequent phenomenology, a very crucial step to follow is to impose the constraints from various NS-NS Bianchi identities as well as RR tadpoles to get the {\it true} non-vanishing contribution to the effective four dimensional scalar potential. We have two formulations for representing the (NS-NS) Bianchi identities, and we emphasise here that both sets of Bianchi identities have their own advantages and limitations. In the `cohomology formulation' fluxes are expressed in terms of cohomology indices counted via respective Hodge numbers while in the `standard formulation', all the fluxes are written out using the real six-dimensional indices, e.g. $H_{ijk}$ etc.

Note that the various flux actions on the harmonic-forms given in eqn. (\ref{eq:fluxAction2}) can be rewritten in a more compact manner by introducing a couple of multi-forms and accordingly clubbing some flux components given as \cite{Grana:2006hr, Blumenhagen:2015lta},
\bea
\label{eq:defs}
& \mu_{\rm \bf A} = \left\{\Phi_6, \, \, \mu_A \right\}, & \qquad \qquad \tilde\mu^{\rm \bf A} = \left\{{\bf 1}, \, \, \tilde\mu^A \right\}\,,\\
& \omega_{0\Lambda} = R_\Lambda, \quad \omega_0{}^\Lambda = R^\Lambda, &\qquad \qquad Q^0{}_\Lambda = H_\Lambda, \quad Q^{0\Lambda} = H^\Lambda\,. \nonumber
\eea
Subsequently, the flux actions in eqn. (\ref{eq:fluxAction2}) can be expressed using the twisted differential ${\cal D}$ in the following simpler form,
\bea
\label{eq:fluxAction1}
& & \hskip-0.75cm {\cal D} \, {\cal A}_\Lambda = \, \, Q^{\rm \bf A}{}_\Lambda\, \mu_{\rm \bf A} + \omega_{{\rm \bf A} \, \Lambda}\, \tilde\mu^{\rm \bf A}, \quad \, \, \, \quad  {\cal D} \, {\cal B}^\Lambda = Q^{{\rm \bf A}\, \Lambda}\, \mu_{\rm \bf A} + \omega_{\rm \bf A}{}^{\Lambda}\, \tilde\mu^{\rm \bf A}, \\
& & \hskip-0.75cm {\cal D} \, \mu_{\rm \bf A} = - \, \omega_{\rm \bf A}{}^\Lambda\, {\cal A}_\Lambda + \, \omega_{{\rm \bf A}\Lambda}\, {\cal B}^\Lambda, \qquad {\cal D} \, \tilde\mu^{\rm \bf A} = Q^{{\rm \bf A}\Lambda}\, {\cal A}_\Lambda +\, Q^{\rm \bf A}{}_\Lambda\, {\cal B}^\Lambda\,.  \nonumber
\eea
Imposing the nilpotency of twisted differential operator ${\cal D}$ via ${\cal D}^2 A_p = 0$ where $A_p$ corresponds to the various harmonic forms, and using the flux actions in eqn. (\ref{eq:fluxAction1}) one finds the following six quadratic flux constraints \cite{Grana:2006hr},
\bea
\label{eq:Neq2BIsCohom}
& & \hskip-1cm \omega_{{\rm \bf A} \Lambda} \,\, \omega_{{\rm \bf B}}{}^{\Lambda} = \omega_{{\rm \bf B} \Lambda} \,\, \omega_{{\rm \bf A}}{}^{\Lambda}, \qquad Q^{\rm \bf A}{}_\Lambda \,\, Q^{{\rm \bf B}\Lambda} = Q^{\rm \bf B}{}_\Lambda \,\, Q^{{\rm \bf A}\Lambda}, \qquad \omega_{{\rm \bf A} \Lambda} \,\, Q^{{\rm \bf B}\Lambda} = \omega_{{\rm \bf A}}{}^{\Lambda} \, \, Q^{\rm \bf B}{}_\Lambda \,, \\
& & \hskip-1cm \omega_{{\rm \bf A} \Lambda} \,\,Q^{\rm \bf A}{}_\Sigma = \omega_{{\rm \bf A} \Sigma} \,\,Q^{\rm \bf A}{}_\Lambda, \qquad  \omega_{{\rm \bf A}}{}^{\Lambda} \, \, Q^{\rm \bf A}{}_\Sigma = \omega_{{\rm \bf A} \Sigma} \,\,  Q^{{\rm \bf A}\Lambda}, \qquad \omega_{{\rm \bf A}}{}^{\Lambda} \, \,Q^{{\rm \bf A}\Sigma} = \omega_{{\rm \bf A}}{}^{\Sigma} \, \,Q^{{\rm \bf A}\Lambda} \,. \nonumber
\eea
To be more specific in terms of the conventional $H, \, \omega, \, Q$, and $R$ flux, using the definitions in eqn. (\ref{eq:defs}), the compactly written six constraints in (\ref{eq:Neq2BIsCohom}) turn out to represent five classes of Bianchi identities which have a total 11 constraints as collected in table \ref{tab_Neq2BIsCohom}.
\begin{table}[h!]
  \centering
 \begin{tabular}{|c||c|c|}
\hline
& & \\
Class & Bianchi Identities of the & Maximum no. of \\
 & Cohomology formulation & identities \\
 \hline
 & & \\
{\bf (I)} & $H_\Lambda\, \omega_A{}^\Lambda = H^\Lambda\, \omega_{A\Lambda}$ & $h^{1,1}$ \\
& & \\
{\bf (II)} & $H_\Lambda\, Q^{A\Lambda} = H^\Lambda\, Q^A{}_\Lambda$ & $h^{1,1}$ \\
& $\omega_A{}^\Lambda \, \omega_{B\Lambda} = \omega_B{}^\Lambda \, \omega_{A\Lambda}$ & $\frac{1}{2} \, h^{1,1}\left(h^{1,1} -1\right)$ \\
& & \\
{\bf (III)} & $H^\Lambda\, R_\Lambda = H_\Lambda\, R^\Lambda$ & 1 \\
& $\omega_A{}^\Lambda \, Q^B{}_\Lambda = \omega_{A\Lambda}\, Q^{B\Lambda}$ & $\left(h^{1,1}\right)^2$ \\
& $R^\Lambda \, H^\Sigma + \omega_A{}^\Lambda\, Q^{A\Sigma} = H^\Lambda \, R^\Sigma + Q^{A\Lambda}\,\omega_A{}^\Sigma$ & $\frac{1}{2}\,h^{2,1}\left(h^{2,1} +1\right)$ \\
& $R_\Lambda \, H_\Sigma + \omega_{A\Lambda} \, Q^A{}_\Sigma = H_\Lambda \, R_\Sigma + Q^A{}_\Lambda\,\omega_{A\Sigma}$ & $\frac{1}{2}\,h^{2,1}\left(h^{2,1} + 1\right)$ \\
& $R_\Lambda \, H^\Sigma + \omega_{A\Lambda}\,Q^{A\Sigma} = H_\Lambda \, R^\Sigma + Q^A{}_\Lambda\,\omega_A{}^\Sigma$ & $\left(h^{2,1} + 1\right)^2$ \\
&& \\
{\bf (IV)} & $Q^A{}_\Lambda \, Q^{B\Lambda} = Q^B{}_\Lambda \, Q^{A\Lambda}$ & $\frac{1}{2} \, h^{1,1}\left(h^{1,1} -1\right)$ \\
& $R_\Lambda\, \omega_A{}^\Lambda = R^\Lambda\, \omega_{A\Lambda}$ & $h^{1,1}$ \\
& & \\
{\bf (V)} & $R_\Lambda\, Q^{A\Lambda} = R^\Lambda\, Q^A{}_\Lambda$ & $h^{1,1}$ \\
& & \\
\hline
\hline
& & \\
& {\bf Total} & $\left(h^{1,1}+1\right)\left(2\,h^{1,1}+1\right)$ \\
& & $+ \, \left(h^{2,1}+1\right)\left(2\,h^{2,1}+1\right)$\\
& & \\
 \hline
  \end{tabular}
  \caption{Bianchi identities of the cohomology formulation and their counting}
  \label{tab_Neq2BIsCohom}
 \end{table}

Similarly, imposing the nilpotency of twisted differential operator ${\cal D}$ via ${\cal D}^2 A_p = 0$ where $A_p = \frac{1}{p!} X_{i_1 ....i_p} dx^1 \wedge dx^2 ....\wedge dx^{p}$, one gets the five sets of Bianchi identities along with an `extra constraint' in what we call the `standard formulation'. These are given as follows \cite{Shelton:2005cf,Robbins:2007yv,Shelton:2006fd}:
\bea
\label{eq:Neq2BIsStandard}
& \hskip-2cm {\bf (I)} & \qquad H_{m[\underline{ij}} \, \omega_{\underline{kl}]}{}^{m}= 0,\\ 
& \hskip-2cm {\bf (II)} & \qquad \omega_{[\underline{ij}}{}^{m} \, {\omega}_{\underline{k}]m}{}^{l} \, = \, {Q}_{[\underline{i}}{}^{lm} \, {H}_{\underline{jk}]m},\nonumber\\ 
& \hskip-2cm {\bf (III)} & \qquad {H}_{ijm} \, {R}^{mkl} + {\omega}_{{ij}}{}^{m}{} \, {Q}_{m}{}^{{kl}} = 4\, {Q}_{[\underline{i}}{}^{m[\underline{k}} \, {\omega}_{\underline{j}]m}{}^{\underline{l}]},\nonumber\\ 
& \hskip-2cm {\bf (IV)} & \qquad {Q}_{m}{}^{[\underline{ij}}  \, {Q}_{l}{}^{\underline{k}]m} \, = \, {\omega}_{lm}{}^{[\underline{i}}\,\, {R}^{\underline{jk}]m},\nonumber\\ 
& \hskip-2cm {\bf (V)} & \qquad R^{m [\underline{ij}}\,  Q_{m}{}^{\underline{kl}]} \, =0,\nonumber\\
& \hskip-2cm {\bf  Constraint:} & \qquad \frac{1}{6}\, H_{ijk} \, R^{ijk} + \frac{1}{2} \, \omega_{ij}{}^k \, Q_k{}^{ij} = 0\,.\nonumber
\eea
Also let us mention that for our current interest, we consider the fluxes to be constant parameters, however for the non-constant fluxes and in the presence of sources, these Bianchi identities in eqn. (\ref{eq:Neq2BIsStandard}) are modified \cite{Geissbuhler:2013uka, Aldazabal:2013sca, Blumenhagen:2013hva, Andriot:2014uda}.

\section{Solutions of Bianchi identities for ${\mathbb T}^6/({\mathbb Z}_3 \times {\mathbb Z}_3)$ setups}
\label{sec_BIs-Z3xZ3}
In this section, we discuss the Bianchi identities and their solutions for the ${\cal N}=2$ theory obtained in nongeometric compactifications using ${\mathbb T}^6/({\mathbb Z}_3 \times {\mathbb Z}_3)$ orbifold, and subsequently we will present them for the type IIA and type IIB orientifolds as well.

\subsection{Capturing the ${\cal N} = 2$ missing Bianchi identities}
Using the cohomology formulation Bianchi identities given in table \ref{tab_Neq2BIsCohom}, we find the following set of constraints for this rigid toroidal orbifold,

\bea
& & H_0 \, \omega_A{}^0 = H^0 \, \omega_{A0}, \qquad H_0\, Q^{A0} = H^0\, Q^A{}_0, \qquad \omega_{A0}\, \omega_B{}^0 = \omega_{B0}\, \omega_A{}^0, \\
& & R_0 \, Q^{A0} = R^0 \, Q^A{}_0, \qquad R_0\, \omega_A{}^0 = R^0\, \omega_{A0}, \qquad \, \, Q^{A0}\, Q^B{}_0 = Q^{B0}\, Q^A{}_0\,, \nonumber\\
& & R_0\, H^0 = H_0\, R^0, \qquad  \quad \omega_{A0} \,Q^{A0} = \omega_A{}^0\, Q^A{}_0, \quad \, \, \omega_{A0} \,Q^{B0} = \omega_A{}^0\, Q^B{}_0, \qquad \forall A, B = 1, 2, 3. \nonumber
\eea
As expected the total number of cohomology formulation identities is 29, which follows from $(h^{1,1}+1)(2\,h^{1,1}+1) + (h^{2,1}+1)(2\,h^{2,1}+1)$ as $h^{1,1}=3$ and $h^{2,1} =0$ for the present orbifold construction. While we place the explicit computations in the appendix \ref{sec_appendix1}, the various flux conversion relations from the standard formulation to the cohomology formulation are given as follows,
\bea
\label{eq:Neq2flux}
& & H_0 = H_{135}, \qquad \qquad \quad H^0 = H_{246}\,,\\
& & \omega_{10} = -\, \omega_{46}{}^2, \qquad \qquad \omega_{20} = -\, \omega_{62}{}^4, \qquad \qquad \omega_{30} = -\, \omega_{24}{}^6, \nonumber\\
& & \omega_{1}{}^0 = -\, \omega_{46}{}^1, \qquad \qquad \omega_{2}{}^0 = -\, \omega_{62}{}^3, \qquad \qquad \omega_{3}{}^0 = -\, \omega_{24}{}^5, \nonumber\\
& & Q^1{}_0 = Q_2{}^{45}, \qquad \qquad \, \, \, Q^2{}_0 = Q_4{}^{61}, \qquad \qquad \, \, \, \, Q^3{}_0 = Q_6{}^{23}, \nonumber\\
& & Q^{10} = -\,Q_2{}^{46}, \qquad \quad \, \, \, \, Q^{20} = -\,Q_4{}^{62}, \qquad \quad \, \, \, \, \, Q^{30} = -\,Q_6{}^{24}, \nonumber\\
& & R_0 = R^{246}, \qquad \qquad \quad \, \, \, R^0 = - R^{135}\,. \nonumber
\eea
In order to capture the missing Bianchi identities, we consider the flux constraints of the standard formulation presented in eqn. (\ref{eq:Neq2BIsStandard}). Subsequently, using the flux conversion relations given in eqn. (\ref{eq:Neq2flux}) one can translate those standard Bianchi identities into cohomology form, which after some reshuffling can be presented by a set of flux constraints collected as follows, 
\bea
& {\bf (I)}  & H_0 \, \omega_A{}^0 = H^0 \, \omega_{A0} \qquad \forall A = 1, 2, 3. \\
& & \nonumber\\
& {\bf (II)}  & H_0\, Q^{A0} = H^0\, Q^A{}_0, \qquad \omega_{A0}\, \omega_B{}^0 = \omega_{B0}\, \omega_A{}^0; \qquad \forall A = 1, 2, 3. \nonumber\\
& & \omega_{A0}\, \omega_{B0} + \omega_{A0} \, \omega_{B0} = H_0\, Q^C{}_0 + H^0\, Q^{C0}; \quad A \neq B \neq C; \nonumber\\
& & \nonumber\\
& {\bf (III)}  & R_0\, H^0 = H_0\, R^0, \quad \omega_{A0} \,Q^{B0} = \omega_A{}^0\, Q^B{}_0, \quad \forall A, B; \\
& & H_0 R_0 + H^0 R^0 = \omega_{10}\,Q^{10} + \omega_1{}^0\, Q^1{}_0 = \omega_{20}\,Q^{20} + \omega_2{}^0\, Q^2{}_0= \omega_{30}\,Q^{30} + \omega_3{}^0\, Q^3{}_0, \nonumber\\
& & \nonumber\\
& {\bf (IV)}  & R_0\, \omega_A{}^0 = R^0\, \omega_{A0}, \qquad Q^{A0}\, Q^B{}_0 = Q^{B0}\, Q^A{}_0; \qquad \forall A = 1, 2, 3. \nonumber\\
& & Q^A{}_0\, Q^B{}_0 + Q^{A0} \, Q^{B0} = R_0\, \omega_{C0} + R^0\, \omega_C{}^0; \quad A \neq B \neq C, \nonumber\\
& & \nonumber\\
& {\bf (V)}  & R_0 \, Q^{A0} = R^0 \, Q^A{}_0 \qquad \forall A = 1, 2, 3. \nonumber
\eea
In addition, the `extra constraint' presented for the standard formulation as given in eqn. (\ref{eq:Neq2BIsStandard}) translates into the following cohomology form,
\bea
& & R_0 \, H^0 + \omega_{A0}\, Q^{A0} = H_0\, R^0 + \omega_A{}^0\, Q^A{}_0\,.
\eea
Therefore comparing the constraints arising in the two formulations, we obtain the following nine what we call ``missing" Bianchi identities,
\bea
& & \omega_{A0}\, \omega_{B0} + \omega_A{}^{0} \, \omega_{B}{}^{0} = H_0\, Q^C{}_0 + H^0\, Q^{C0}, \quad A \neq B \neq C;\\
& & Q^A{}_0\, Q^B{}_0 + Q^{A0} \, Q^{B0} = R_0\, \omega_{C0} + R^0\, \omega_C{}^0; \quad A \neq B \neq C; \nonumber\\
& & H_0 R_0 + H^0 R^0 = \omega_{10}\,Q^{1}{}_0 + \omega_{1}{}^{0}\, Q^{10} = \omega_{20}\,Q^{2}{}_0 + \omega_{2}{}^{0}\, Q^{20} = \omega_{30}\,Q^{3}{}_0 + \omega_{3}{}^{0}\, Q^{30}\,. \nonumber
\eea
These identities certainly constrain the flux choice allowed by the usual cohomology formulation, and hence can have the potential to significantly affect any ${\cal N} =2$ phenomenological conclusion derived from the scalar potential. Moreover, we find that using  the triple intersection numbers $\kappa_{ABC}$ such that $\kappa_{123} = 1$, these missing identities can be expressed in the following manner,
\bea
& & H_{(\underline 0} \, Q^A{}_{\underline 0)} + H^{(\underline 0} \, Q^{A \underline 0)}= \frac{1}{2}\, \kappa_{ABC}^{-1} \, \omega_{B \, (\underline 0} \, \omega_{C \, \underline 0)} + \frac{1}{2}\, \kappa_{ABC}^{-1} \, \omega_{B}{}^{(\underline 0} \, \omega_{C}{}^{\underline 0)}, \, \quad \forall \, A\, \in h^{1,1}\,;\nonumber\\
& & 3\, H_{(\underline 0} \, R_{\underline 0)} + 3\, H^{(\underline 0} \, R^{\underline 0)}  = \, \omega_{A (\underline 0} \, Q^A{}_{\underline 0)} + \, \omega_{A}{}^{(\underline 0} \, Q^{A \underline 0)}\,; \\
& & H_{(\underline 0} \, R_{\underline 0)} + \, H^{(\underline 0} \, R^{\underline 0)}  = \, \omega_{A^\prime (\underline 0} \, Q^{A^\prime}{}_{\underline 0)} + \, \omega_{A^\prime}{}^{(\underline 0} \, Q^{A^\prime \underline 0)}, \quad \forall \, A^\prime \in h^{1,1} \, \, {\rm and \, \, A^\prime \, \, not \, \, summed \, \, over}\,; \nonumber\\
& & R_{(\underline 0} \, \omega_{A \, \underline 0)} + R^{(\underline 0} \, \omega_{A}{}^{\underline 0)}= \frac{1}{2} \, \kappa_{ABC} \, Q^{B}{}_{(\underline 0} \, Q^{C}{}_{\underline 0)} + \frac{1}{2} \, \kappa_{ABC} \, Q^{B (\underline 0} \, Q^{C \underline 0)}, \,  \quad \forall \, A\, \in h^{1,1} \,.\nonumber
\eea
In the above, $\kappa_{ABC}^{-1} = 1/\kappa_{ABC}$ for fixed $A, B, C$ whenever $\kappa_{ABC} \neq 0$. It is important to mention that the above set of identities are well along the line with the conjectured form of the ${\cal N} = 1$ missing Bianchi identities presented in \cite{Gao:2018ayp}.

\subsection{Bianchi identities for the type IIA orientifold}
For the type IIA orientifold, the various flux action inherited from the ${\cal N}=2$ theory can be given as follows,
\bea
\label{eq:fluxActions0}
& & \hskip-0.3cm H \wedge {\bf 1} = H_K\, \beta^K, \qquad H \wedge \alpha_K = - H_K \,  \Phi_6, \qquad H\wedge \beta^K = 0, \\
& & \hskip-0.3cm \omega \triangleleft \nu_a = -\, \omega_{a K}\, \beta^K, \quad \omega \triangleleft \mu_\alpha = \hat{\omega}_{\alpha}^K\, \alpha_K, \qquad \omega \triangleleft \alpha_K = -\, \omega_{a K} \, \tilde\nu^a, \quad \omega \triangleleft \beta^K = -\, \hat\omega_{\alpha}^K \, \tilde\mu^\alpha \,,\nonumber\\
& & \hskip-0.3cm Q \triangleright \tilde\nu^a = Q^{a}_{K}\, \beta^K, \qquad Q \triangleright  \tilde\mu^\alpha = -\, \hat{Q}^{\alpha K}\, \alpha_K, \quad Q \triangleright \alpha_K = - Q^{a}_{K} \, \nu_a, \quad Q \triangleright \beta^K =-\, \hat{Q}^{\alpha K} \, \mu_\alpha\,, \nonumber\\
& & \hskip-0.3cm R \bullet {\Phi_6} = -\, R_K\, \beta^K, \qquad R \bullet \alpha_K = -\, R_K \,  {\bf 1}, \qquad R \bullet \beta^K =0\,. \nonumber
\eea
In addition, we get the following set of Bianchi identities in the cohomology formulation,
\bea
& & H_K\, \hat{\omega}_\alpha{}^K = 0, \quad H_K\, \hat{Q}^{\alpha K} = 0, \quad R_K\, \hat{Q}^{\alpha K} = 0, \quad R_K \, \hat{\omega}_\alpha{}^K = 0, \nonumber\\
& & \omega_{aK}\, \hat{\omega}_\alpha{}^K = 0, \quad \omega_{aK}\, \hat{Q}^{\alpha K} =0, \quad Q^a{}_K \, \hat{Q}^{\alpha K} = 0, \quad \hat{\omega}_\alpha{}^{K} \,Q^a{}_K=0,\\
& & \hat{\omega}_\alpha{}^{[K}\, \hat{Q}^{\alpha J]} = 0, \quad H_{[K} \, R_{J]} = \omega_{a[K}\, Q^a{}_{J]}\,, \nonumber
\eea
where the bracket $[..]$ appearing in the last two identities denote the anti-symmetrization of $J$ and $K$ indices.

The various fluxes which appear in the ${\cal N}=1$ type IIA orientifold framework are given, along with their respective conversion relations to the cohomology formulation, as follows:
\bea
\label{eq:Neq1IIAflux}
& & H_0 = H_{135}, \qquad  \omega_{10} = -\, \omega_{46}{}^2, \qquad \omega_{20} = -\, \omega_{62}{}^4 \qquad  \omega_{30} = -\, \omega_{24}{}^6, \\
& & Q^1{}_0 = Q_2{}^{45}, \quad \, \, \, Q^2{}_0 = Q_4{}^{61}, \qquad \, \, \, Q^3{}_0 = Q_6{}^{23}, \qquad \, \, \, \, R_0 = R^{246}\,. \nonumber
\eea
Let us mention that for our type IIA setup with ${\mathbb T}^6/({\mathbb Z}_3 \times {\mathbb Z}_3)$ orientifold using the standard involution $\sigma_{IIA}: z^i \to - \ov{z}^i$, the even (1,1)-cohomology and its dual odd (2,2)-cohomology are trivial, i.e. $h^{1,1}_+ = 0$, and therefore all the `hatted' fluxes in eqn. (\ref{eq:fluxActions0}) which are counted via $\alpha\in h^{1,1}_+$ are projected out. Such a situation provides a strong constraint on the cohomology formulation of the Bianchi identities as it suggests that all the Bianchi identities of the class {\bf (I)}, {\bf (II)}, {\bf (IV)} and {\bf (V)} are identically trivial ! In fact, the only Bianchi identities which could be non-trivial turns out to be the following one,
\bea
& & H_{[K} \, R_{J]} = \omega_{a[K}\, Q^a{}_{J]}\,.
\eea
However, for the case of frozen complex structure moduli, even the above identity is trivially satisfied. {\bf Thus, in our current type IIA construction all the cohomology formulation identities are trivial.} However translating the standard formulation identities using the conversion relations in eqn. (\ref{eq:Neq1IIAflux}), one ends up having the following status on the Bianchi identities,
\bea
\label{eq:BIs-IIAZ3xZ3O}
& {\bf (I)}  \quad & Trivial \\
& {\bf (II)}  \quad & \omega_{10}\, \omega_{20} = H_0\, Q^3{}_0, \quad \omega_{20}\, \omega_{30} = H_0\, Q^1{}_0, \quad \omega_{30}\, \omega_{10} = H_0\, Q^2{}_0, \nonumber\\
& {\bf (III)}  \quad &  H_0 R_0 = \omega_{10}\, Q^1{}_0 = \omega_{20}\, Q^2{}_0= \omega_{30}\, Q^3{}_0, \nonumber\\
& {\bf (IV)} \quad & Q^1{}_0\, Q^2{}_0 = R_0\, \omega_{30}, \quad Q^2{}_0\, Q^3{}_0 = R_0\, \omega_{10},, \quad Q^3{}_0\, Q^1{}_0 = R_0\, \omega_{20},\nonumber\\
& {\bf (V)}  \quad & Trivial \nonumber\\
& & Trivial \quad ({\rm extra \, \, constraint}). \nonumber
\eea
After looking at the index structure, it is clear that these identities correspond to the missing class, and do not arise from the list of cohomology identities. Moreover these can be also written in the following form,
\bea
& & H_{(\underline 0} \, Q^a{}_{\underline 0)} = \frac{1}{2}\, \kappa_{abc}^{-1} \, \omega_{b \, (\underline 0} \, \omega_{c \, \underline 0)}, \, \quad \forall \, \, a: \hat{\kappa}_{a\alpha\beta} = 0\,; \nonumber\\
& & 3\, H_{(\underline 0} \, R_{\underline 0)}  = \, \omega_{a (\underline 0} \, Q^a{}_{\underline 0)} + \hat\omega_\alpha{}^{(\underline 0} \, \hat{Q}^{\alpha \underline 0)}\,; \nonumber\\
& & R_{(\underline 0} \, \omega_{a \, \underline 0)} = \frac{1}{2} \, \kappa_{abc} \, Q^{b}{}_{(\underline 0} \, Q^{c}{}_{\underline 0)}, \, \quad \forall \, \, a: \hat{\kappa}_{a\alpha\beta} = 0\,, \nonumber
\eea
which matches with the conjecture given in \cite{Gao:2018ayp}. Here we used $\kappa_{abc}^{-1} = 1/\kappa_{abc}$ for fixed $a, b, c$ whenever $\kappa_{abc}$ is non-zero. Recall that for the current toroidal model, the intersection numbers of type $\hat{\kappa}_{a\alpha\beta} $ are absent due to orientifold projection, and for this reason, the flux components $\hat\omega_\alpha{}^{0}$ and $\hat{Q}^{\alpha 0}$ are also projected out. The various possible solutions of these Bianchi identities presented in eqn. (\ref{eq:BIs-IIAZ3xZ3O}) can be expressed in the following eight classes,
\bea
\label{eq:solBIs-IIAZ3xZ3O}
& {\bf S1:} & \omega_{10} = \omega_{20} = \omega_{30} = 0, \quad {Q}^1{}_0 = {Q}^2{}_0 = {Q}^3{}_0 =0, \quad {R}_0 = 0\,;\\ 
& {\bf S2:} & {H}_0 = 0, \quad \omega_{10} = \omega_{20} = \omega_{30} = 0, \quad {Q}^1{}_0 = {Q}^2{}_0 = {Q}^3{}_0 =0\,; \nonumber\\ 
& {\bf S3:} & {R}_0 = 0, \quad {Q}^1{}_0 = {Q}^2{}_0 = {Q}^3{}_0 =0, \quad \omega_{20} = 0 = \omega_{30}, \quad \omega_{10} \neq 0\,; \nonumber\\
& & {R}_0 = 0, \quad {Q}^1{}_0 = {Q}^2{}_0 = {Q}^3{}_0 =0, \quad \omega_{10} = 0 = \omega_{30}, \quad \omega_{20} \neq 0\,; \nonumber\\
& & {R}_0 = 0, \quad {Q}^1{}_0 = {Q}^2{}_0 = {Q}^3{}_0 =0, \quad \omega_{10} = 0 = \omega_{20}\,, \quad \omega_{30} \neq 0\,; \nonumber\\ 
& {\bf S4:} & {H}_0 = 0, \quad \omega_{10} = \omega_{20} = \omega_{30} = 0, \quad {Q}^2{}_0 = 0 = {Q}^3{}_0\,, \quad {Q}^1{}_0 \neq 0\,; \nonumber\\
& & {H}_0 = 0, \quad \omega_{10} = \omega_{20} = \omega_{30} = 0, \quad {Q}^1{}_0 = 0 = {Q}^3{}_0\,,\quad {Q}^2{}_0 \neq 0\,; \nonumber\\
& & {H}_0 = 0, \quad \omega_{10} = \omega_{20} = \omega_{30} = 0, \quad {Q}^1{}_0 = 0 = {Q}^2{}_0\,, \quad {Q}^3{}_0 \neq 0\,;\nonumber\\
& {\bf S5:} & {H}_0 = 0, \quad \omega_{20} = 0 = \omega_{30}, \quad {Q}^1{}_0 = 0 = {Q}^2{}_0, \quad {R}_0 = 0, \quad \omega_{10} \neq 0, \quad {Q}^3{}_0 \neq 0\,;\nonumber\\ 
& & {H}_0 = 0, \quad \omega_{10} = 0 = \omega_{30}, \quad {Q}^1{}_0 = 0 = {Q}^2{}_0, \quad {R}_0 = 0, \quad \omega_{20} \neq 0, \quad {Q}^3{}_0 \neq 0\,;\nonumber\\ 
& & {H}_0 = 0, \quad \omega_{20} = 0 = \omega_{30}, \quad {Q}^1{}_0 = 0 = {Q}^3{}_0, \quad {R}_0 = 0\,, \quad \omega_{10} \neq 0, \quad  {Q}^2{}_0 \neq 0\,;\nonumber\\ 
& & {H}_0 = 0, \quad \omega_{10} = 0 = \omega_{20}, \quad {Q}^1{}_0 = 0 = {Q}^3{}_0, \quad {R}_0 = 0\,, \quad \omega_{30} \neq 0, \quad {Q}^2{}_0 \neq 0\,;\nonumber\\ 
& & {H}_0 = 0, \quad \omega_{10} = 0 = \omega_{20}, \quad {Q}^2{}_0 = 0 = {Q}^3{}_0, \quad {R}_0 = 0\,, \quad \omega_{30} \neq 0, \quad {Q}^1{}_0 \neq 0\,;\nonumber\\ 
& & {H}_0 = 0, \quad \omega_{10} = 0 = \omega_{30}, \quad {Q}^2{}_0 = 0 = {Q}^3{}_0, \quad {R}_0 = 0\,, \quad \omega_{20} \neq 0, \quad  {Q}^1{}_0 \neq 0\,;\nonumber
\eea
\bea
& {\bf S6:} & {H}_0 = 0\,, \, {Q}^1{}_0 = 0, \, \omega_{20} = 0 = \omega_{30}, \, {R}_0 = \frac{{Q}^2{}_0\, {Q}^3{}_0}{\omega_{10}}, \, \omega_{10} \neq 0, \quad {Q}^2{}_0\, {Q}^3{}_0 \neq 0\,;\nonumber\\ 
& & {H}_0 = 0\,, \, {Q}^2{}_0 = 0, \, \omega_{10} = 0 = \omega_{30}, \, {R}_0 = \frac{{Q}^1{}_0\, {Q}^3{}_0}{\omega_{20}}, \, \omega_{20} \neq 0, \quad {Q}^1{}_0\, {Q}^3{}_0 \neq 0\,; \nonumber\\ 
& & {H}_0 = 0\,, \, {Q}^3{}_0 = 0, \, \omega_{10} = 0 = \omega_{20}, \, {R}_0 = \frac{{Q}^1{}_0\, {Q}^2{}_0}{\omega_{30}}, \, \omega_{30} \neq 0, \quad {Q}^1{}_0\, {Q}^2{}_0 \neq 0\,; \nonumber\\ 
& {\bf S7:} & {R}_0 = 0\,, \, \omega_{10} = 0, \, {Q}^2{}_0 = 0 = {Q}^3{}_0, \, {H}_0 = \frac{\omega_{20} \, \omega_{30}}{{Q}^1{}_0}\,, \, {Q}^1{}_0 \neq 0, \quad \omega_{20}\, \omega_{30} \neq 0\,;\nonumber\\ 
& & {R}_0 = 0\,, \, \omega_{20} =0, \, {Q}^1{}_0 = 0 = {Q}^3{}_0, \, {H}_0 = \frac{\omega_{10} \, \omega_{30}}{{Q}^2{}_0}\,, \, {Q}^2{}_0 \neq 0, \quad \omega_{10}\, \omega_{30} \neq 0\,;\nonumber\\ 
& & {R}_0 = 0\,, \, \omega_{30} =0, \, {Q}^1{}_0 = 0 = {Q}^2{}_0, \, {H}_0 = \frac{\omega_{10} \, \omega_{20}}{{Q}^3{}_0}\,, \, {Q}^3{}_0 \neq 0, \quad \omega_{20}\, \omega_{10} \neq 0\,;\nonumber\\ 
& {\bf S8:} & \biggl\{{Q}^1{}_0 \neq 0, \, {Q}^2{}_0 \neq 0, \, {Q}^3{}_0\neq 0, \,  \omega_{10} = \frac{\omega_{30}\, {Q}^3{}_0}{{Q}^1{}_0}\, , \, \omega_{20} = \frac{\omega_{30}\, {Q}^3{}_0}{{Q}^2{}_0}\,, \omega_{30} \neq 0, \,\nonumber\\
& & \hskip2cm  {H}_0 = \frac{\omega_{10} \, \omega_{20}}{{Q}^3{}_0} \, , \, {R}_0 = \frac{{Q}^1{}_0\, {Q}^2{}_0}{\omega_{30}} \biggr\} \,; \nonumber
\eea
where in the above solutions if some flux components are not mentioned then it means that they are not constrained in the given solution of identities. We will utilise these flux constraints for exploring the de-Sitter (no-go) scenarios in one of the upcoming sections. 

\subsection{Lessons from the missing Bianchi identities in the type IIB orientifold}
The various fluxes which appear in the ${\cal N}=1$ type IIB orientifold framework are given, along with their respective conversion relations to the cohomology formulation, as follows:
\bea
\label{eq:Neq1IIBflux}
& & \hskip-0.5cm H_0 = H_{135}, \qquad Q^1{}_0 = Q_2{}^{45}, \qquad  \, \, \, \, \, Q^2{}_0 = Q_4{}^{61}, \qquad \, \, \, \, \, \, Q^3{}_0 = Q_6{}^{23}, \\
& & \hskip-0.5cm H^0 = H_{246}, \qquad Q^{10} = -\,Q_2{}^{46}, \qquad  Q^{20} = -\,Q_4{}^{62}, \qquad \, Q^{30} = -\,Q_6{}^{24}\,. \nonumber
\eea
Thus we have 8 independent NS-NS flux components that survive the orbifold twist, and they are further constrained by the set of Bianchi identities which turns out to be of the following form,
\bea
\label{eq:IIBZ3O-allBIs}
& {\bf (I)}  \qquad & Trivial \\
& {\bf (II)}  \qquad & H_0\, Q^{10} = H^0\, Q^1{}_0, \quad  H_0\, Q^{20} = H^0\, Q^2{}_0, \quad H_0\, Q^{30} = H^0\, Q^3{}_0, \nonumber\\
& & H_0\, Q^1{}_0 + H^0\, Q^{10} =0, \quad H_0\, Q^2{}_0 + H^0\, Q^{20} =0, \quad H_0\, Q^3{}_0 + H^0\, Q^{30} =0,\nonumber\\
& {\bf (III)} \qquad & Trivial \nonumber\\
& {\bf (IV)}  \qquad & Q^{10}\, Q^2{}_0 = Q^{20}\, Q^1{}_0, \quad Q^{10}\, Q^3{}_0 = Q^{30}\, Q^1{}_0, \quad Q^{20}\, Q^3{}_0 = Q^{30}\, Q^2{}_0, \nonumber\\
& & Q^1{}_0\, Q^2{}_0 + Q^{10} \, Q^{20} = 0, \quad Q^2{}_0\, Q^3{}_0 + Q^{2} \, Q^{30} = 0, \quad Q^1{}_0\, Q^3{}_0 + Q^{10} \, Q^{30} = 0, \nonumber\\
& {\bf (V)}  \qquad & Trivial \nonumber\\
& & Trivial \quad ({\rm extra \, \, constraint}). \nonumber
\eea
Let us first analyse the cohomology formulation identities pretending that those are the only ones to worry about. In this regard, there are six such flux constraints which arise from the two classes of identities, namely $H^\Lambda \, Q^A{}_\Lambda = H_\Lambda\, Q^{A\Lambda}$ and $Q^{A\Lambda} \, Q^B{}_\Lambda = Q^A{}_\Lambda\, Q^{B\Lambda}$. These can be explicitly given as follows,
\bea
\label{eq:IIBZ3O-cohomBIs}
& & H_0\, Q^{10} = H^0\, Q^1{}_0, \qquad  H_0\, Q^{20} = H^0\, Q^2{}_0, \qquad H_0\, Q^{30} = H^0\, Q^3{}_0, \\
& & Q^{10}\, Q^2{}_0 = Q^{20}\, Q^1{}_0, \qquad Q^{10}\, Q^3{}_0 = Q^{30}\, Q^1{}_0, \qquad Q^{20}\, Q^3{}_0 = Q^{30}\, Q^2{}_0\,. \nonumber
\eea
These constraint are indeed  a part of the standard set of Bianchi identities computed for this model as given in eqn. (\ref{eq:IIBZ3O-allBIs}). Assuming the case for $H\neq0$ and $Q\neq0$, i.e. when both the kinds of fluxes have at least one non-zero component, we find that there are four classes of solutions of the above set of Bianchi identities in eqn. (\ref{eq:IIBZ3O-cohomBIs}), which can be given as follows, 
\bea
& {\bf (S1):} \quad & H^0 = Q^{10} = Q^{20} = Q^{30} = 0, \\
& {\bf (S2):} \quad & Q^{20} = Q^{30} = Q^2{}_0 = Q^3{}_0= 0, \quad Q^{10} \neq 0, \quad H_0 = \frac{Q^1{}_0\, H^0}{Q^{10}}, \nonumber\\
& {\bf (S3):} \quad & Q^{30} = Q^3{}_0= 0, \quad Q^{20} \neq 0, \quad Q^1{}_0 = \frac{Q^{10}\, Q^2{}_0}{Q^{20}}, \quad H_0 = \frac{Q^2{}_0\, H^0}{Q^{20}}, \nonumber\\
& {\bf (S4):} \quad & Q^{30} \neq 0, \quad Q^1{}_0 = \frac{Q^{10}\, Q^3{}_0}{Q^{30}}, \quad Q^2{}_0 = \frac{Q^{20}\, Q^3{}_0}{Q^{30}},\quad H_0 = \frac{Q^3{}_0\, H^0}{Q^{30}}. \nonumber
\eea
Now we can observe that if one does not consider the set of missing identities, there are indeed some possible non-trivial solutions. In fact the solution {\bf (S1)} is quite popular one in type IIB orientifold compactifications with nongeometric fluxes as the cohomology formulation identities given in eqn. (\ref{eq:IIBZ3O-cohomBIs}) can be trivially satisfied by switching-off half of the fluxes, either with the lower or with the upper $h^{2,1}$-indices \cite{Robbins:2007yv, Blumenhagen:2015kja, Shukla:2016xdy}. 

However, imposing the additional (missing) identities from the full set of constraints given in eqn. (\ref{eq:IIBZ3O-allBIs}) leads to vanishing of the $Q$-flux whenever we assume that $H$-flux has at least one non-zero component in order to stabilize the axion-dilaton $(\tau)$ which is a universal chiral variable in type IIB setting. This observation should be also crucial for the recent phenomenological studies made for the case of frozen complex structure moduli, as these are the simplest possible setups which could have some non-trivial applications for moduli stabilization and breaking of the no-scale structure using nongeometric fluxes. So the main observation or the lesson which we get after analysing this type IIB compactification on the orientifold of a ${\mathbb T}^6/({\mathbb Z}_3 \times {\mathbb Z}_3)$-orbifold is the fact that $H \neq 0$ implies $Q = 0$, i.e. no nongeometric flux can be turned-on in this rigid orientifold without switching-off the NS-NS three-form flux $H_3$. This observation is apparently quite strong and a peculiar one in its own!

In that regard, let us also consider even a simpler case of the isotropic limit for the current example. This leads to the following identifications of the flux components,
\bea
& & \hskip-1.5cm H_0 = h_0, \, \, \,  H^0 = \tilde{h}^0, \quad Q^1{}_0 = Q^2{}_0 = Q^3{}_0 = q^1{}_0, \quad Q^{10} = Q^{20} = Q^{30} = \tilde{q}^{10},
\eea
which are constrained by the following three NS-NS Bianchi identities,
\bea
& & \hskip-1cm h_0\, \tilde{q}^{10} = \tilde{h}^0 \, q^1{}_0, \quad \quad \quad h_0\, q^1{}_0 + \tilde{h}^0 \,\tilde{q}^{10} = 0, \quad \quad \quad {(q^1{}_0)}^2 + {(\tilde{q}^{10})}^2 = 0.
\eea
Note that only the first identity corresponds to the second formulation while the later two are the missing ones. It is now obvious that setting $\tilde{h}^0 = 0 = \tilde{q}^{10}$ which are usually sufficient to satisfy the first identity does not trivially satisfy the remaining two, and in fact leads to vanishing of $Q$-flux. 
On these lines, let is consider the following superpotential which can be generically written for a setting (beyond the toroidal example such as Calabi Yau) having the frozen complex structure moduli and a single K\"ahler modulus $T$,
\bea
& & \hskip-1cm W = \left(f_0 + \tau\, h_0 + T\, q^1{}_0 \right)+ i\, \left(\tilde{f}^0\,+ \tau\, \tilde{h}^0 + T\, \tilde{q}^{10} \right),
\eea
in which the fluxes need to respect the single Bianchi identity $\tilde{h}^0 \, q^1{}_0 = h_0 \, \tilde{q}^{10}$ as known from the cohomology formulation. Therefore, for studying the phenomenology, what one usually does is that to set $\tilde{h}^0= 0 = \tilde{q}^{10}$ which leads to the well studied superpotential given as follows \cite{Blumenhagen:2015jva, Blumenhagen:2015qda, Blumenhagen:2015xpa, Damian:2018tlf},
\bea
& & \hskip-1cm W = f + \tau\, h_0 + T\, q^1{}_0 + i\, \tilde{f}^0 \,,
\eea
and in addition to that one has to impose the RR tadpole cancellation conditions as well. In this case, the no-scale structure is broken by the nongeometric $(q^1{}_0)$-flux fixing the $T$ modulus, and results in a set of $AdS$ vacua, which would disappear if the missing Bianchi identities anticipated from the explicit computations for the rigid toroidal example continue to hold for the case of the rigid Calabi Yau orientifolds. However, one still needs to check if the missing identities we have anticipated to hold beyond the toroidal cases are indeed true for the generic rigid Calabi Yau compactifications. 

With the reasons/observations elaborated as above, for the current article we will not consider the study of nongeometric type IIB scalar potentials arising in the rigid orientifold flux compactifications, though the interested readers can read-off the scalar potential for rigid compactifications from the generic results in \cite{Shukla:2015hpa, Shukla:2015rua}.

\section{Type IIA de-Sitter no-go scenarios}
\label{sec_IIA}
The nongeometric type IIA scalar potential has been generically computed in \cite{Gao:2017gxk}, and we review the relevant pieces of information in the appendix \ref{sec_appendix2}. Given that the explicit expressions of the scalar potential in terms of various moduli and axions are known, we are not only in a position to re-derive the various previously proposed no-go theorems regarding the $dS$/inflationary realizations but also we can explore more such conditions in a given specific setting with ${\mathbb T}^6/({\mathbb Z}_3 \times {\mathbb Z}_3)$ orientifolds as we plan to present in this section. 

\subsection{Two known no-go conditions without nongeometric fluxes}
To begin with, using our generic results for the type IIA scalar potential we will re-derive the de-Sitter no-go conditions for two particular models which have been presented in \cite{DeWolfe:2005uu, Hertzberg:2007wc, Flauger:2008ad, Haque:2008jz, Caviezel:2008tf}. The first no-go condition corresponds to the case of having no geometric and nongeometric fluxes \cite{DeWolfe:2005uu, Hertzberg:2007wc}, while the second case includes geometric flux but does not include the nongeometric flux \cite{Flauger:2008ad, Haque:2008jz, Caviezel:2008tf}. 

In order to explore the possible de-Sitter no-go scenarios, let us begin with introducing the notations of \cite{Flauger:2008ad} by using the following redefinitions,
\bea
\label{eq:IIAvolgamma}
& & {\cal V} = \rho^3, \qquad \qquad t^a = \rho\, \gamma^a, \qquad \qquad \kappa_{abc}\, \gamma^a\, \gamma^b\, \gamma^c = 6\,,
\eea
where $\rho$ measures the overall volume of the internal manifold while $\gamma^a$'s correspond to the angular K\"ahler moduli. For invoking the ``swampland-like inequalities", we focus only on the dynamics in the volume-dilaton ($\rho$ -$D$) plane of the moduli space and ignore the other moduli dependences which otherwise would heavily complicate the setup. Given that the scalar potential can be expressed in terms of polynomials in $\rho$ and $e^D$ variables, this effectively ``two-field scenario" will not only keep the setup simple but will also impose quite interesting generic restrictions on the existence of de-Sitter vacua. Being simple, this two-field model has been often found to be attractive and studying this model can also provide the ``large volume" and ``weak coupling" insights relevant for moduli stabilization and the other phenomenological aspects. However, our current aim is limited to study these models in the lights of swampland inequalities given in eqns. (\ref{eq:old-swamp}) and (\ref{eq:new-swamp}).

Using the new redefinitions in eqn. (\ref{eq:IIAvolgamma}), the generic type IIA scalar potential pieces given in eqns. (\ref{eq:typeIIA-genpot1})-(\ref{eq:typeIIA-genpot2}) can be rewritten in the following manner,
\bea
\label{eq:typeIIA-genpot3}
&& \hskip-0cm V = V_{h} + V_{\omega} + V_{q}+ V_{r} + V_{f_0} + V_{f_2} + V_{f_4} + V_{f_6} + V_{\rm loc}\,\, ; \\
& & \nonumber\\
& & V_h = \frac{e^{2D}}{\rho^3} \, A_h\,, \quad V_\omega = \frac{e^{2D}}{\rho} \, A_\omega\, , \quad V_q = \, {e^{2D}\, \rho}\, A_q\,, \quad V_r  =\, {e^{2D} \, \rho^3} \, A_r\,,\nonumber\\
& & V_{f_0} = \, {e^{4D}\, \rho^3}\, A_{f_0}\,, \quad V_{f_2} =  \, {e^{4D}\, \rho}\, A_{f_2}\,, \quad V_{f_4} = \, \frac{e^{4D}}{\rho} \, A_{f_4}\,, \quad V_{f_6} = \, \frac{e^{4D}}{\rho^3} \,A_{f_6}\,, \nonumber\\
& & V_{\rm loc} = \, e^{3D} \,A_{\rm loc}\,, \nonumber
\eea  
where the various flux/moduli dependent quantities $A_i$'s are explicitly given by the following expressions,
\bea
\label{eq:typeIIA-genpot4}
& & A_h = \, \frac{1}{2}\, {\mathbb H}_I \, {\cal M}^{I J} \, {\mathbb H}_J \,, \\
& & A_\omega = \, \frac{1}{2} \biggl[{\mathbb \mho}_{aI} \, {\cal M}^{I J} \, {\mathbb \mho}_{bJ} \, \gamma^a \, \gamma^b  + 4 \, \, {\mathbb \mho}_{aI} \, {\mathbb \mho}_{bJ} \, {\cal X}^I\, {\cal X}^J \, \left(\tilde{K}^{ab} -\, \gamma^a \, \gamma^b \right) \nonumber\\
& & \hskip0.5cm -2\, \left({\mathbb H}_I \, {\cal M}^{I J} \, {\mathbb Q}^a{}_J \, - \, 4 \, {\mathbb H}_I \, {\cal X}^I \, {\cal X}^J \, {\mathbb Q}^a{}_J \right)\, \tilde\sigma_a +  4\, {\cal F}_I \, {\cal F}_J \, \, \hat{\mathbb \mho}_{\alpha}{}^I \,\left(\hat{\kappa}_{a\alpha\beta}\, \gamma^a\right)^{-1}\, \hat{\mathbb \mho}_{\beta}{}^J \biggr]\,, \nonumber\\
& & A_q = \, \frac{1}{2} \biggl[{\mathbb Q}^{a}{}_I \,{\cal M}^{I J} \, {\mathbb Q}^{b}{}_J  \, \tilde\sigma_a \, \tilde\sigma_b + 4\, \,{\mathbb Q}^a{}_I \, {\mathbb Q}^{b}{}_{J} \, {\cal X}^I\, {\cal X}^J \, \left(\tilde{K}_{ab} - \, \tilde\sigma_a \, \tilde\sigma_b \right) \nonumber\\
& & \hskip1.5cm -2\, \left({\mathbb R}_I \, {\cal M}^{I J} \, {\mathbb \mho}_{aJ}\, - \,4 \, {\mathbb R}_I \, {\cal X}^I \, {\cal X}^J \, {\mathbb \mho}_{aJ} \right)\, \gamma^a + 4\, {\cal F}_I \, {\cal F}_J \,\hat{\mathbb Q}^{\alpha I} \, \left(\hat{\kappa}_{a\alpha\beta}\, \gamma^a\right) \, \hat{\mathbb Q}^{\beta J} \biggr]\,, \nonumber\\
& & A_r = \, \frac{1}{2} \, {\mathbb R}_I \, {\cal M}^{I J} \, {\mathbb R}_J \,, \nonumber\\
& & A_{f_0} = \,\frac{1}{2}\, \left({\mathbb G}_0 \right)^2, \quad A_{f_2} = \,\frac{1}{2}\, {\mathbb G}^a \,{\tilde K}_{ab}\, {\mathbb G}^b, \quad A_{f_4} = \, \frac{1}{2}\, {\mathbb G}_a \,\,{\tilde K}^{ab} \, {\mathbb G}_b, \quad A_{f_6} = \,\frac{1}{2} \, \left({\mathbb G}^0 \right)^2 \,, \nonumber\\
& & A_{\rm loc} = -\, 2 \, \left({\mathbb H}_K \, {\mathbb G}_{0} - \, {\mathbb \mho}_{a K}\, {\mathbb G}^a + \, {\mathbb Q}^a{}_K \, {\mathbb G}_a - \, {\mathbb R}_K \, {\mathbb G}^0 \right) {\cal X}^K\,. \nonumber
\eea
In the above collection, the axionic flux quantities $\{{\mathbb H}_I, {\mathbb \mho}_{aI}, {\mathbb Q}^a{}_I, {\mathbb R}_I\}$ and $\{{\mathbb G}_0, {\mathbb G}^a, {\mathbb G}_a, {\mathbb G}^0\}$ are defined in eqn. (\ref{eq:TypeIIAfluxOrbits0}) of the appendix.  Also note that in the above collection $A_i$'s, there are quantities involving only the angular K\"ahler moduli and not the $\rho$ modulus, as they have been obtained after extracting the volume modulus via using $t^a = \rho \, \gamma^a$. Similarly the moduli space metric and its inverse can be given after extracting the powers of $\rho$ in the following manner,
\bea
\label{eq:tilde-modulispace}
& & \hskip-1cm \tilde\sigma_a = \frac{1}{2} \, \kappa_{abc} \, \gamma^b\, \gamma^c, \quad \tilde{K}^{ab} = \frac{1}{2} \, \gamma^a \, \gamma^b - \left(\kappa_{abc}\, \gamma^c\right)^{-1}, \quad \tilde{K}_{ab} = \, \tilde\sigma_a\, \tilde\sigma_b - \, \kappa_{abc}\, \gamma^c\,.
\eea
Note that none of the above quantities explicitly involve the $\rho$ modulus. One can easily convince that our scalar potential in eqns. (\ref{eq:typeIIA-genpot3})-(\ref{eq:typeIIA-genpot4}) reduces into the ones presented in \cite{Flauger:2008ad, Caviezel:2008tf} when the nongeometric $Q$ and $R$ fluxes are turned-off but the geometric flux contributions are present; for example, see eqn. (2.34) of \cite{Flauger:2008ad} and eqn. (3.13) of \cite{Caviezel:2008tf} where apart from the nongeometric fluxes, the $D$-term geometric fluxes $\hat{\mathbb \omega}_{\alpha}{}^I$ are also switched-off. 

In the context of extending the models with de-Sitter solution to building inflationary scenarios, let us make a side remark that using the new basis of moduli through the relation given in eqn. (\ref{eq:IIAvolgamma}), the slow-roll parameter $\epsilon_V$ can be simplified to 
satisfy the following bound \cite{Flauger:2008ad},
\bea
\label{eq:epsilon-bound}
& & \epsilon_V \geq \frac{1}{V^2} \biggl[\frac{\rho^2}{3} \, \left(\frac{\partial V}{\partial \rho}\right)^2 + \frac{1}{4} \left(\frac{\partial V}{\partial D}\right)^2 \biggr]\,.
\eea
In fact, it has been shown in \cite{Flauger:2008ad} that the slow-roll parameter $\epsilon_V$ can be expressed in terms of six non-negative pieces including the two of the above eqn.(\ref{eq:epsilon-bound}) which corresponds to derivatives of $\rho$ and $D$ moduli. So if there exists a set of parameters $\{\alpha > 0, \, \beta >0\}$ such that one has an inequality of the form: $\left(\alpha \, \partial_D V - \rho \, \partial_\rho V \right) \geq \beta \, V$, then the generic extremum cannot be a de-Sitter one because for this case $ \partial_D V = 0 =  \partial_\rho V$ would imply $V_{\rm ext} \leq 0$. Therefore, finding such what we call ``swampland-like inequalities", namely $\left(\alpha \, \partial_D V - \rho \, \partial_\rho V \right) \geq \beta \, V$ for some positive quantities $\alpha \simeq {\cal O}(1)$ and $\beta \simeq {\cal O}(1)$ helps in determining the sign of potential at the extremum. Moreover, if such an inequality exits, using eqn. (\ref{eq:epsilon-bound}) one can convince that it leads to the following lower bound on the slow-roll parameter,
\bea
\label{eq:epsion-cond}
& & \hskip-0.3cm \epsilon_V \geq V^{-2}\biggl[\frac{\rho^2}{3} \, (\partial_\rho V)^2 + \frac{1}{4} \, (\partial_D V)^2 \biggr] \\
& & \hskip0cm = \frac{V^{-2}}{3 + 4\, \alpha^2} \biggl[\left(\alpha \, (\partial_D V) - \rho \, (\partial_\rho V)\right)^2 + \frac{1}{12} \, (3\, (\partial_D V) + 4\, \alpha\, \rho \, (\partial_\rho V))^2 \biggr]\,. \nonumber\\
& & \geq \, \frac{\beta^2}{3 + 4\, \alpha^2}\,. \nonumber
\eea
This relation can be used for reading off the lower bound on $\epsilon_V$ parameter in several models as we willl see in the upcoming subsections. 

\subsubsection*{NoGo-1: without (non-)geometric flux, i.e. $V_\omega = V_q = V_r = 0$}
In the absence of any geometric and nongeometric fluxes, the scalar potential takes the following form
\bea
V = V_{h} + V_{f_0} + V_{f_2} + V_{f_4} + V_{f_6} + V_{\rm loc}.
\eea
Recall that the generic expressions for the various scalar potential pieces such as $V_h, V_{{\mathbb H} {\mathbb H}}$ etc. are presented through the eqns. (\ref{eq:IIA-pot-gen-all}), (\ref{eq:TypeIIAfluxOrbits0}), (\ref{eq:typeIIA-genpot1}) and (\ref{eq:typeIIA-genpot2}) of the appendix.. Here in this particular setting the term $V_h$ arises from $V_{{\mathbb H} {\mathbb H}}$ after setting the $\omega, Q$ and $R$ fluxes to zero. Similarly, the RR potential terms $V_{f_p}$, for $p = 0, 2, 4, 6$, arise from their respective pieces after setting the $\omega, Q$ and $R$ fluxes to zero. This leads to the following form of the scalar potential pieces,
\bea
& & \hskip-1cm V_h = \frac{e^{2 D}}{\rho^3} \, A_{h}\,, \quad V_{\rm loc} = e^{3 D} \, A_{\rm loc}\,, \quad V_{f_p} = \frac{e^{4 D}}{\rho^{(p-3)}}\, A_{f_p}\,,
\eea
where $A_h$ and $A_{f_p}$'s are flux dependent non-negative quantities which do not depend on the volume moduli $\rho$ and the dilaton $D$, while the local piece $A_{\rm loc}$ can a priory have any sign, however for avoiding the runaway in the dilatonic direction $D$, one needs $A_{\rm loc} < 0$. Now it is easy to observe that the following relations hold,
\bea
& & \hskip-0.5cm \partial_D \, V_h = 2\, V_h\,, \quad \partial_D \, V_{\rm loc} = 3\, V_{\rm loc}\,, \quad \partial_D \, V_{f_p} = 4\, V_{f_p}, \quad \forall p \in \{0, 2, 4, 6\}\,, \nonumber\\
& & \hskip-0.5cm \partial_\rho \, V_h = -\frac{3}{\rho}\, V_h\,, \quad \partial_\rho \, V_{\rm loc} = 0\,, \quad \partial_\rho \, V_{f_p} = \frac{3-p}{\rho}\, V_{f_p}, \quad \forall p \in \{0, 2, 4, 6\}\,. 
\eea
Using these partial derivatives results in the following inequality,
\bea
& & 3\, \partial_D \, V - \rho \, \partial_\rho\, V = 9\, V + \sum_{p=2,4,6}\, p \, V_{f_p} > 9 \, V\,.
\eea
Finding this information is very crucial for two reasons. First, it immediately shows that at the extremum in the two modulus $(\rho, D)$ plane, the scalar potential cannot be positive as one has,
\bea
& & V_{\rm ext} = -\, \frac{1}{9}\, \sum_{p=2,4,6}\, p \, V_{f_p}\,,
\eea
which therefore forbids de-Sitter solution in this minimal setting. Second, using the eqn. (\ref{eq:epsilon-bound}), the inequality also leads to the following lower bound on the $\epsilon_V$ parameter,
\bea
& & \epsilon_V \geq V^{-2}\biggl[\frac{\rho^2}{3} \, (\partial_\rho V)^2 + \frac{1}{4} \, (\partial_D V)^2 \biggr] \geq \frac{27}{13}\,,
\eea
which forbids any inflationary realization in this model.

\subsubsection*{NoGo-2: without nongeometric flux, i.e. $V_q = 0 = V_r$}
The previous framework is very minimal and now we plan to include more ingredients to look for the possibility of avoiding the de-Sitter no-go condition. After including the so-called geometric flux $\omega$, and  still having no nongeometric fluxes, namely for the case when $Q = 0 = R$, the scalar potential takes the following form\footnote{Recall that the generic expressions for the various scalar potential pieces such as $V_h, V_{{\mathbb H} {\mathbb H}}$ etc. are presented in eqns. (\ref{eq:IIA-pot-gen-all}), (\ref{eq:TypeIIAfluxOrbits0}), (\ref{eq:typeIIA-genpot1}) and (\ref{eq:typeIIA-genpot2}) of the appendix.},
\bea
&& \hskip-1cm V = V_{h} + V_{\omega} + V_{f_0} + V_{f_2} + V_{f_4} + V_{f_6} + V_{\rm loc}\,,
\eea
where as before $V_{h}$ arises from $V_{{\mathbb H} {\mathbb H}}$ but now only setting the $Q$ and $R$ fluxes to zero. Similarly, the RR potential terms $V_{f_p}$, for $p = 0, 2, 4, 6$, arise from their respective pieces after setting the $Q$ and $R$ fluxes to zero. In addition, now we have the $V_{\omega}$ piece which can be generically sourced from both the $F$-term as well as the $D$-term contributions as seen from the eqn. (\ref{eq:typeIIA-genpot2}). Let us recall here again that we are only setting the $Q$ and $R$ fluxes to zero, and not the geometric flux, and therefore $V_h$ will involve geometric flux contributions through the `generalized flux orbits' we have defined in eqn. (\ref{eq:TypeIIAfluxOrbits0}) to compactly write down the scalar potential. This is the case for the pieces $V_{f_p}$'s and the local piece $V_{\rm loc}$ which involve the flux orbits as defined in eqn. (\ref{eq:TypeIIAfluxOrbits0}). With this in mind, we arrive at the following form of the scalar potential from our generic results,
\bea
& & \hskip-1cm V_h = \frac{e^{2 D}}{\rho^3} \, A_{h}\,, \quad V_\omega = \frac{e^{2 D}}{\rho} \, A_{\omega}\,, \quad V_{\rm loc} = e^{3 D} \, A_{D6}\,, \quad V_{f_p} = \frac{e^{4 D}}{\rho^{(p-3)}}\, A_{f_p}\,,
\eea
where as can be directly read-off from the collection given in eqn. (\ref{eq:typeIIA-genpot4}), one finds that $A_h, \, A_\omega, \, A_{\rm loc}$ and $A_{f_p}$ are flux dependent quantities which do not depend on the volume moduli $\rho$ and the dilaton $D$. Further, it is true that except the term $A_\omega$ and the local term $A_{\rm loc}$, all the remaining ones are guaranteed to be positive semidefinite. Although a couple of pieces within $A_\omega$ can be guaranteed to be positive semidefinite (e.g. the one coming from the $D$-term contributions), but the overall sign of $A_\omega$ can not be a priory fixed. Moreover, all the terms have positive powers of $e^D$, and therefore to avoid a runaway in that direction one needs $A_{\rm loc} < 0$ in this case. Now it is easy to observe that,
\bea
& & \partial_D \, V_h = 2\, V_h\,, \qquad \partial_D \, V_\omega = 2\, V_\omega\,, \qquad \partial_D \, V_{\rm loc} = 3\, V_{\rm loc}\,, \\
& & \partial_\rho \, V_h = -\frac{3}{\rho}\, V_h\,, \qquad \partial_\rho \, V_\omega = -\frac{1}{\rho}\, V_\omega\,, \qquad \partial_\rho \, V_{\rm loc} = 0\,, \nonumber\\
& & \partial_D \, V_{f_p} = 4\, V_{f_p}, \qquad \partial_\rho \, V_{f_p} = \frac{3-p}{\rho}\, V_{f_p}; \quad \forall p \in \{0, 2, 4, 6\}\,. \nonumber
\eea
Using these partial derivatives results in the following relation,
\bea
& & \partial_D \, V - \rho \, \partial_\rho\, V = 3\, V + 2 V_{h} + \, 2 \, V_{f_4} + \, 4 \, V_{f_6} - \, 2 \, V_{f_0}.
\eea
This presents an interesting fact. In the absence of Romans mass term, i.e. for $V_{f_0} =0$, one arrives at the following relations,
\bea
& & \partial_D \, V - \rho \, \partial_\rho\, V = 3\, V + 2 V_{h} + \, 2 \, V_{f_4} + \, 4 \, V_{f_6} \geq 3 V,
\eea
which not only forbids the existence of de-Sitter solution but also leads to the following lower bound on the $\epsilon_V$ parameter using eqn. (\ref{eq:epsilon-bound}),
\bea
& & \epsilon_V \geq V^{-2}\biggl[\frac{\rho^2}{3} \, (\partial_\rho V)^2 + \frac{1}{4} \, (\partial_D V)^2 \biggr] \geq \frac{9}{7}\,.
\eea
This observation tells us that in the absence of any nongeometric flux ($Q$ and $R$), one needs the non-zero Romans mass term along with $A_{\rm loc} < 0$ and $A_\omega > 0$ in order to avoid the no-go condition \cite{Flauger:2008ad, Haque:2008jz, Caviezel:2008tf}. Note that the need of $V_\omega > 0$ is not manifested from the above combination $(\partial_D \, V - \rho \, \partial_\rho\, V)$ but can be seen from the previous one as given under,
\bea
& & 3\, \partial_D \, V - \rho \, \partial_\rho\, V = 9\, V  - 2\, V_\omega + \sum_{p=2,4,6}\, p \, V_{f_p} \,,
\eea
and hence if $V_\omega < 0$, then the previous {\bf NoGo-1} condition would continue to hold leading to $(3\, \partial_D \, V - \rho \, \partial_\rho\, V) > 9 \, V$ and hence $\epsilon_V > 27/13$. Thus we conclude that merely including the geometric flux does not help for the purpose, and there are other things to engineer to avoid the no-go condition. However, let us recall that merely not finding a no-go condition does not guaranteed for the existence of de-Sitter vacua as breaking the known no-go is only a necessary but not a sufficient condition for the de-Sitter existence. 

\subsection{Evading the no-go conditions with nongeometric fluxes}
On the similar lines of analysis we have discussed and continuing with the focus on the dependence in the $(D, \rho)$ plane only, the generic scalar potential can be clubbed in the following form,
\bea
&& \hskip-1cm V = V_{h} + V_{\omega} + V_{q}+ V_{r} + V_{f_0} + V_{f_2} + V_{f_4} + V_{f_6} + V_{\rm loc}\,,
\eea
where we have
\bea
& & \hskip-1cm V_h = \frac{e^{2 D}}{\rho^3} \, A_{h}\,, \quad V_\omega = \frac{e^{2 D}}{\rho} \, A_{\omega}\,, \quad V_q = e^{2 D} \, \rho \, A_{q}\,, \quad V_r = e^{2 D} \, \rho^3 \, A_{r}\,, \\
& & \hskip-1cm V_{\rm loc} = e^{3 D} \, A_{\rm loc}\,, \quad V_{f_p} = \frac{e^{4 D}}{\rho^{(p-3)}}\, A_{f_p}\,.\nonumber
\eea
Here as before $A_h, \, A_\omega, \, A_q, \, A_r, \, A_{\rm loc}$ and $A_{f_p}$'s are flux dependent quantities which do not depend on the volume moduli $\rho$ and the dilaton $D$. Now having full generalities at hand, the following conditions hold on the signs of these coefficients,
\bea
\label{eq:no-go2sign}
& & A_h  \geq 0, \qquad A_r \geq 0, \qquad A_{f_p} \geq 0, \qquad  \forall p = 0, 2, 4, 6; \\
& & {\rm sign}(A_\omega), \quad {\rm sign}(A_q), \quad {\rm and} \quad {\rm sign}(A_{\rm loc}) \quad {\rm not \, \, fixed}\,. \nonumber
\eea 
Subsequently it is easy to observe that the following relations hold,
\bea
& & \partial_D \, V_h = 2\, V_h\,, \qquad \partial_D \, V_\omega = 2\, V_\omega\,, \qquad \partial_D \, V_q = 2\, V_q\,, \quad \partial_D \, V_r= 2\, V_r\,, \\
& & \partial_D \, V_{\rm loc} = 3\, V_{\rm loc}\,, \qquad \partial_D \, V_{f_p} = 4\, V_{f_p}, \qquad \forall p \in \{0, 2, 4, 6\};  \nonumber\\
& & \partial_\rho \, V_h = -\frac{3}{\rho}\, V_h\,, \qquad \partial_\rho \, V_\omega = -\frac{1}{\rho}\, V_\omega\,, \qquad \partial_\rho \, V_q = \frac{1}{\rho}\, V_q\,, \quad \partial_\rho \, V_r = \frac{3}{\rho}\, V_r\,, \nonumber\\
& & \partial_\rho \, V_{\rm loc} = 0\,, \qquad \partial_\rho \, V_{f_p} = \frac{3-p}{\rho}\, V_{f_p}, \quad \forall p \in \{0, 2, 4, 6\}\,. \nonumber
\eea
For studying the swampland conditions for generic cases having nongeometric fluxes, we take the following combination of the scalar potential derivatives,
\bea
\label{eq:cond-del-rho-D}
& & \hskip-0.5cm \left(\alpha \, \partial_D \, V - \rho \, \partial_\rho\, V \right) = (2 \alpha + 3) \, V_{h} + (2 \alpha + 1) \, V_{\omega} + (2 \alpha - 1) \, V_{q} + (2 \alpha - 3) \, V_{r} \\
& & \hskip0cm  + (4 \alpha - 3) \, V_{f_0} + \, (4 \alpha - 1) \, V_{f_2} + \, (4 \alpha + 1) \, V_{f_4} + \, (4 \alpha + 3) \, V_{f_6} + 3 \alpha \, V_{\rm loc}\,.\nonumber\\
& & = 3\,\alpha\, V + (3 -\alpha)\, V_h + (1 - \alpha) \, V_\omega - (1 + \alpha)\, V_q - (3 + \alpha)\, V_r \, + \sum_{p=0}^{2,4,6} \,(\alpha - 3 + p) V_{f_p}\,,\nonumber
\eea
where in the second equality we have eliminated the $V_{\rm loc}$ piece as it is (one of) the possible negative piece on the right hand side of the equality. Now, given the uncertainties in the signs of various terms as mentioned in eqn. (\ref{eq:no-go2sign}) it turns out that unlike the cases we have analysed without having any nongeometric flux, now it is not obvious if RHS in the second line of eqn. (\ref{eq:cond-del-rho-D}) can be guaranteed to be positive semidefnite for some $\alpha > 0$. In particular, one can easily verify how the two $dS$/inflationary no-go conditions revisited earlier can be evaded in the presence of the nongeometric fluxes, e.g. using $\alpha = 3$ we get,
\bea
& & \hskip-1cm \left(3\, \partial_D \, V - \rho \, \partial_\rho\, V \right) = 9\, V - 2 \, V_{\omega} - 4 \, V_{q} - 6 \, V_{r} + \, 2 \, V_{f_2} + \, 4 \, V_{f_4} + \,6 \, V_{f_6}\,.
\eea
Without knowing the explicit form of the full scalar potential depending on all the moduli, axions and fluxes, and using schematic dependence on the $D$ and $\rho$ moduli it has been mentioned in \cite{Hertzberg:2007wc}, that if $V_{\omega}, \, V_q$ and $V_r$ all are negative, then the earlier condition {\bf NoGo-1} continues to hold even after including the (non-)geometric fluxes
. However, from our explicit computations in appendix \ref{sec_appendix2} now we have seen that $V_r \geq 0$ and therefore the condition {\bf NoGo-1} is guaranteed to be evaded after including the nongeometric $R$-flux while keeping the $V_\omega$ and $V_q$ pieces to zero, otherwise one would need to ensure that $V_{\omega} + 2 \, V_{q} + 3 \, V_{r} > 0$ holds in order to break the no-go condition. 

Similarly if we consider the second case with $\alpha = 1$ corresponding to the condition {\bf NoGo-2} we get the following,
\bea
& & \hskip-1cm \left(\partial_D \, V - \rho \, \partial_\rho\, V \right) =  3\, V + 2\, V_{h} - 2 \, V_{q} - 4\, V_{r}  - 2 \, V_{f_0} + 2 \, V_{f_4} + 4 \, V_{f_6} \,.\nonumber
\eea
Therefore it shows that the condition {\bf NoGo-2} would hold as long as $V_{q} + 2\, V_{r}  +\, V_{f_0} < 0$ is satisfied. This would be though unlikely as $V_r \geq 0$ and $V_{f_0} \geq 0$ and so $V_q$ has to be negative and compensate these two positive contributions. Moreover, this analysis also suggests that unlike the earlier proposal of \cite{Flauger:2008ad}, now one may not need to demand the vanishing of the Romans mass term $V_{f_0}$, and the condition {\bf NoGo-2} can be also evaded by appropriately turning-on the nongeometric $R$-flux. 

As a side remark, it is easy to observe that the two cases discussed in the previous subsections can be directly read-off from the eqn. (\ref{eq:cond-del-rho-D}) and eqn. (\ref{eq:epsion-cond}) leading to the following values of the parameters $\alpha, \, \beta$ along with a bound on the $\epsilon_V$ parameter,
\bea
& & {\rm \bf NoGo1:} \qquad \alpha = 3, \quad \beta = 9 \qquad \Longrightarrow \qquad \epsilon_V \, \geq \, \frac{27}{13}\,, \\
& & {\rm \bf NoGo2:} \qquad \alpha = 1, \quad \beta = 3 \qquad \Longrightarrow \qquad \epsilon_V \, \geq \, \frac{9}{7}\,. \nonumber
\eea

\subsection{New dS no-go scenarios for rigid nongeometric compactifications}
In this subsection we will explore the possibility of finding some new de-Sitter no-go conditions in the context of type IIA setup with frozen complex structure moduli. This corresponds to having the following period vectors,
\bea
& & {\cal X}^0 = 1\, \quad \quad \quad {\cal F}_0 = - \frac{i}{2}; \quad \quad \quad  i \, \int_{X_3} \, \Omega \wedge \ov\Omega = 1\,,
\eea
which subsequently leads to constant components in the moduli space metric given as follows,
\bea
& & {\cal M}^{00} = 1 = {\cal M}_{00}\,.
\eea
It is straight forward to convince that this choice of normalization satisfy all the relations presented in eqn. (\ref{eq:IIAcsmoduli}). Subsequently, the total $F$-term and $D$-term contributions to the scalar potential given in eqns. (\ref{eq:typeIIA-genpot1})-(\ref{eq:typeIIA-genpot2}) takes the following simpler form,
\bea
\label{eq:typeIIArigid}
& & \hskip-1cm V = \frac{e^{2D}}{2 \, {\cal V}} \biggl[{\mathbb H}_0^2 \, + \, {\mathbb \mho}_{a0} \, {\mathbb \mho}_{b0} \left(4\,{\cal V} {\cal G}^{ab} - 3\, t^a\, t^b \right) + \, {\mathbb Q}^{a}{}_0\, {\mathbb Q}^{b}{}_0 \, \left(4\,{\cal V}\, {\cal G}_{ab} - 3\, \sigma_a\, \sigma_b \right) \\
& & \qquad \qquad + \, {\cal V}^2 \, {\mathbb R}_0^2 + \,  6  \, {\mathbb H}_0 \, {\mathbb Q}^{a}{}_0 \,\sigma_a + 6\, {\cal V} \, {\mathbb \mho}_{a0}\, {\mathbb R}_0 \, t^a \biggr] \nonumber\\
& & + \, \frac{e^{4D}}{2 \, {\cal V}} \biggl[\left({\mathbb G}^0\right)^2 + \, {\mathbb G}_a \, {\cal V}\, {\cal G}^{ab}\, {\mathbb G}_b + \, {\mathbb G}^a \, {\cal V}\, {\cal G}_{ab}\, {\mathbb G}^b + {\cal V}^2 \, {\mathbb G}_0^2 \biggr] \nonumber\\
& & -\, 2\, e^{3D} \,  \left({\mathbb H}_0 \, {\mathbb G}_{0} - \, {\mathbb \mho}_{a 0}\, {\mathbb G}^a + \, {\mathbb Q}^a{}_0 \, {\mathbb G}_a - \, {\mathbb R}_0 \, {\mathbb G}^0 \right)\,, \nonumber \\
& & + \frac{e^{2D}}{2 \, {\cal V}} \biggl[\, \hat{\mathbb \mho}_{\alpha}{}^0 \, {\cal V} \,\left(\hat{\kappa}_{a\alpha\beta} t^a \right)^{-1}\, \hat{\mathbb \mho}_{\beta}{}^0 + \, {\cal V} \, \hat{\mathbb Q}^{\alpha0} \, \left(\hat{\kappa}_{a\alpha\beta} t^a \right) \, \hat{\mathbb Q}^{\beta0}\biggr]\,, \nonumber
\eea
where the `generalized' flux orbits contain the NS-NS and R-R axionic moduli, namely the $b^a$ and $\xi^K$ as seen from eqn. (\ref{eq:TypeIIAfluxOrbits0}). Let us emphasize here that the above formulation for the scalar potential given in the eqn. (\ref{eq:typeIIArigid}) is generic for the type IIA setups with frozen complex structure moduli, and once a complete set of Bianchi identities is known, one can utilize them for studying the possibilities about having or not having the de-Sitter solutions in generic scenario of the rigid type IIA flux compactifications. 

Rather than studying it in full generality, we investigate the possibilities for our toroidal setup with ${\mathbb T}^6/({\mathbb Z}_3 \times {\mathbb Z}_3)$-orientifold, where e.g. the full set of Bianchi identities are explicitly known to us, and then one may extrapolate our results and speculate for beyond toroidal situations. For our current type IIA setup with ${\mathbb T}^6/{({\mathbb Z}_3 \times {\mathbb Z}_3)}$ orientifold compactification, we have $h^{1,1}_+ = 0$ and hence no fluxes with indices $\alpha \in h^{1,1}_+$ are allowed to be present, and subsequently there are no $D$-terms generated, $V_D = 0$. Further, for our toroidal case, $h^{1,1}_- = 3$ and so $a \in \{1, 2, 3\}$. In addition, we have the following relations arising from the only non-vanishing triple intersection number being $\kappa_{123} = 1$,
\bea
\label{eq:kaehler-modulispace}
& & {\cal V} = t^1 \, t^2 \, t^3\,, \quad \sigma_1 = \frac{1}{2}\, \kappa_1 = t^2 \, t^3 , \quad \sigma_2 = \frac{1}{2}\, \kappa_2 = t^1 \, t^3, \quad \sigma_3 = \frac{1}{2}\, \kappa_3 = t^1 \, t^2\,, \nonumber\\
& & \\
& & \kappa_{ab} = \begin{pmatrix} 
0 & \, \, t^3 & \, \, t^2 \\
t^3 & \, \, 0 & \, \, t^1 \\
t^2 & \, \ t^1 & \, \, 0 \\
\end{pmatrix}, \quad -4\, {\cal V}\,\kappa^{ab} = \begin{pmatrix} 
2\, (t^1)^2 & \quad - \, 2\, t^1 \, t^2 & \quad - \, 2\, t^1 \, t^3 \\
- \, 2\, t^1 \, t^2 & \quad 2\, (t^2)^2 & \quad - \, 2\, t^2 \, t^3 \\
- \, 2\, t^1 \, t^3 & \quad - \, 2\, t^2 \, t^3 & \quad 2\, (t^3)^2 \\
\end{pmatrix} \, , \nonumber\\
& & \nonumber\\
& & K^{a \ov b} = \begin{pmatrix} 
4\, (t^1)^2 & 0 & 0 \\
0 & 4\, (t^2)^2 & 0 \\
0 & 0 & 4\, (t^3)^2 \\
\end{pmatrix} = 4\, {\cal V} \, {\cal G}^{ab}\,, \quad
K_{a \ov b} = \begin{pmatrix} 
\frac{1}{4\, (t^1)^2} & 0 & 0 \\
0 & \frac{1}{4\, (t^2)^2} & 0 \\
0 & 0 & \frac{1}{4\, (t^3)^2} \\
\end{pmatrix} = \frac{1}{4\, {\cal V}} \, {\cal G}_{ab}\,. \nonumber
\eea
In the absence of any $D$-term contribution in this construction, the total scalar potential arising from the generic fluxes can be given as follows,
\bea
\label{eq:typeIIArigidsimp}
& & \hskip-1cm V = \frac{e^{2D}}{2 \, {\cal V}} \biggl[{\mathbb H}_0^2 \, + \, {\mathbb \mho}_{a0} \, {\mathbb \mho}_{b0} \left(4\,{\cal V} {\cal G}^{ab} - 3\, t^a\, t^b \right) + \, {\mathbb Q}^{a}{}_0\, {\mathbb Q}^{b}{}_0 \, \left(4\,{\cal V}\, {\cal G}_{ab} - 3\, \sigma_a\, \sigma_b \right) \\
& & \qquad \qquad + \, {\cal V}^2 \, {\mathbb R}_0^2 + \,  6  \, {\mathbb H}_0 \, {\mathbb Q}^{a}{}_0 \,\sigma_a + 6\, {\cal V} \, {\mathbb \mho}_{a0}\, {\mathbb R}_0 \, t^a \biggr] \nonumber\\
& & + \, \frac{e^{4D}}{2 \, {\cal V}} \biggl[\left({\mathbb G}^0\right)^2 + \, {\mathbb G}_a \, {\cal V}\, {\cal G}^{ab}\, {\mathbb G}_b + \, {\mathbb G}^a \, {\cal V}\, {\cal G}_{ab}\, {\mathbb G}^b + {\cal V}^2 \, {\mathbb G}_0^2 \biggr] \nonumber\\
& & -\, 2\, e^{3D} \,  \left({\mathbb H}_0 \, {\mathbb G}_{0} - \, {\mathbb \mho}_{a 0}\, {\mathbb G}^a + \, {\mathbb Q}^a{}_0 \, {\mathbb G}_a - \, {\mathbb R}_0 \, {\mathbb G}^0 \right)\,, \nonumber 
\eea
Thanks to our generalized flux-orbits and the generic formulation of the scalar potential using which we have read-off the scalar potential  in the above compact form as given in eqn. (\ref{eq:typeIIArigidsimp}). It is remarkable that the full scalar potential written in the above four lines has 391 terms, and to have a cross-check, the interested readers can compute the scalar potential in two ways; one from eqn. (\ref{eq:typeIIArigidsimp}) via using moduli space metric given in eqn. (\ref{eq:kaehler-modulispace}) along with the generalized flux orbits in eqn. (\ref{eq:TypeIIAfluxOrbits0}) while the other one via using the following K\"ahler and super-potentials,
\bea
& & K \equiv 4D - \ln(8\, {\cal V}) = -4\, \ln\Bigl[(-i (N^0 - \ov{N}^0)/2\Bigr] -  \sum_{a=1,2,3} \ln\Bigl[(-i (T^a - \ov{T}^a)\Bigr]\,, \\
& & W =  e_0 +  b^1\, e_1 + b^2\, e_2 + b^3\, e_3 + b^1\, b^2\, m^3 + b^2\, b^3\, m^1 + b^3\, b^1\, m^2 +  \, b^1 \, b^2 \, b^3\, m_0 \nonumber\\
& & \hskip0.5cm + 2\, N^0 \left(H_0 +  b^1 \omega_{10} + b^2 \omega_{20} + b^3 \omega_{30} + b^1\, b^2\, Q^3{}_0 + b^2\, b^3\, Q^1{}_0 + b^3\, b^1\, Q^2{}_0 + b^1 \, b^2 \, b^3\, R_0 \right), \nonumber
\eea
where the chiral variables are given as $T^a = b^a + i t^a$ and $N^0 = \frac{\xi^0}{2} + i\, e^{-D}$.  Using the particular details (e.g. the moduli space metrics etc.) about this model we can be more specific about the scalar potential terms; for example the generalized RR piece of the scalar potential takes the following form,
\bea
\label{eq:Vrr-Z3xZ3simp}
& & \hskip-2.0cm \sum_{p = 0, 2, 4, 6} V_{f_p} \equiv \, \frac{e^{4D}}{2 \, {\cal V}} \biggl[\left({\mathbb G}^0\right)^2 + \, {\mathbb G}_a \, {\cal V}\, {\cal G}^{ab}\, {\mathbb G}_b + \, {\mathbb G}^a \, {\cal V}\, {\cal G}_{ab}\, {\mathbb G}^b + {\cal V}^2 \, {\mathbb G}_0^2 \biggr] \\
& & = \frac{e^{4D}}{2 \, {\cal V}} \biggl[\left({\mathbb G}^0\right)^2 + \sum_{a=1}^{3} \, ({\mathbb G}_a)^2 \, (t^a)^2 + \sum_{a=1}^{3} \, ({\mathbb G}^a)^2 \, (\sigma_a)^2 + {\cal V}^2 \, {\mathbb G}_0^2 \biggr]\,\geq 0\,. \nonumber
\eea
Let us mention that the Bianchi identities satisfied by the usual fluxes continue to hold after promoting the various flux components to their respective generalized flux orbits as defined in eqn. (\ref{eq:TypeIIAfluxOrbits0}). Subsequently, the `generalized' flux constraints for the current type IIA setup can be given as follows,
\bea
\label{eq:IIA-genBIs}
& & \hskip-1cm \mho_{10}\, \mho_{20} = {\mathbb H}_0\, {\mathbb Q}^3{}_0, \quad \mho_{20}\, \mho_{30} = {\mathbb H}_0\, {\mathbb Q}^1{}_0, \quad \, \, \, \mho_{30}\, \mho_{10} = {\mathbb H}_0\, {\mathbb Q}^2{}_0, \\
& & \hskip-1cm {\mathbb Q}^1{}_0\, {\mathbb Q}^2{}_0 = {\mathbb R}_0\, \mho_{30}, \quad {\mathbb Q}^2{}_0\, {\mathbb Q}^3{}_0 = {\mathbb R}_0\, \mho_{10},, \quad {\mathbb Q}^3{}_0\, {\mathbb Q}^1{}_0 = {\mathbb R}_0\, \mho_{20}\,, \nonumber\\
& & \hskip-1cm {\mathbb H}_0 \, {\mathbb R}_0 \, = \, \mho_{10}\, {\mathbb Q}^1{}_0 \, = \, \mho_{20}\, {\mathbb Q}^2{}_0 \, = \, \mho_{30}\, {\mathbb Q}^3{}_0 \,. \nonumber
\eea
This helps us in significantly nullifying the various scalar potential terms. For example, let us consider all the terms of the $(\mho\mho + {\mathbb H}{\mathbb Q})$-type which we denote as $V_\omega$,
\bea
& & \hskip-1.75cm V_\omega \equiv \frac{e^{2D}}{2 \, {\cal V}} \, \biggl[{\mathbb \mho}_{a0} \, {\mathbb \mho}_{b0} \left(4\, {\cal V}\, {\cal G}^{ab} - 3\, t^a\, t^b \right) + \,  6 \, {\mathbb H}_0 \, {\mathbb Q}^{a}{}_0 \, \sigma_a \biggr] \\
& & = \frac{e^{2D}}{2} \biggl[\frac{t^1\, ({\mathbb \mho}_{10})^2}{t^2\, t^3} + \frac{t^2\, ({\mathbb \mho}_{20})^2}{t^1\, t^3} + \frac{t^3\, ({\mathbb \mho}_{30})^2}{t^1\, t^2} \biggr] \, \geq \, 0 \,. \nonumber
\eea
This not only shows that the ${\mathbb H}{\mathbb Q}$-terms are cancelled by the $\mho\mho$-terms but also fixes the sign of $V_\omega$ to be positive semidefinite. Similarly for the scalar potential terms of $(\mho {\mathbb R} + {\mathbb Q}{\mathbb Q})$-type clubbed in $V_q$, we get the following simplifications after nullifying many pieces out of the generalized Bianchi identities,
\bea
& & \hskip-1.5cm V_q \equiv  \frac{e^{2D}}{2 \, {\cal V}} \biggl[{\mathbb Q}^{a}{}_0\, {\mathbb Q}^{b}{}_0 \, \left(4\, {\cal V}\, {\cal G}_{ab} - 3\, \sigma_a\, \sigma_b \right) + 6 \, {\cal V}\, {\mathbb \mho}_{a0}\, {\mathbb R}_0 \, t^a \biggr] \\
& & \hskip-1cm = \frac{e^{2D}}{2} \biggl[\frac{t^2\, t^3}{t^1} \, ({\mathbb Q}^{1}{}_0)^2\, +\, \frac{t^1\, t^3}{t^2} \, \, ({\mathbb Q}^{2}{}_0)^2 + \frac{t^1\, t^2}{t^3} \, \, ({\mathbb Q}^{3}{}_0)^2 \biggr] \, \geq 0\,. \nonumber
\eea
Now let us define a new set of moduli $\{\rho^1, \rho^2, \rho^3\}$ out of the two-cycle moduli $\{t^1, t^2, t^3\}$ in the following manner,
\bea
& & \frac{t^2\, t^3}{t^1} = \rho^1\,, \quad \frac{t^1\, t^3}{t^2} = \rho^2, \quad \frac{t^1\, t^2}{t^3} = \rho^3 \quad \Longrightarrow \quad {\cal V} = t^1\, t^2\,t^3 = \rho^1\, \rho^2\, \rho^3
\eea
Subsequently we find that our type IIA scalar potential arising from the rigid compactification with (non-)geometric fluxes as given in eqn. (\ref{eq:typeIIArigidsimp}) reduces to the following form,
\bea
& & \hskip-1cm V = V_h + V_\omega + V_q + V_r + \sum_{p = 0, 2, 4, 6} V_{f_p} + V_{\rm loc} \,, 
\eea
where all pieces except the $V_{\rm loc}$ are non-negative with their explicit forms given as follows,
\bea
\label{eq:V-Z3xZ3simp0}
& & V_h = \frac{e^{2D}}{2 \, \rho^1\, \rho^2\,\rho^3} \, {\mathbb H}_0^2 \geq 0 \,, \\
& & V_\omega = \frac{e^{2D}}{2} \biggl[\frac{({\mathbb \mho}_{10})^2}{\rho^1} + \frac{({\mathbb \mho}_{20})^2}{\rho^2}  + \frac{({\mathbb \mho}_{30})^2}{\rho^3}  \biggr] \equiv V_{\omega 1} + V_{\omega 2} + V_{\omega 3} \, \geq 0, \nonumber\\
& & V_q = \frac{e^{2D}}{2} \biggl[({\mathbb Q}^{1}{}_0)^2\, \rho^1 +\, ({\mathbb Q}^{2}{}_0)^2 \, \rho^2 + \, ({\mathbb Q}^{3}{}_0)^2 \, \rho^3\biggr] \equiv V_{q1} + V_{q2} + V_{q3} \, \geq 0\,, \nonumber\\
& & V_r =  \frac{e^{2D}}{2} \, \rho^1\, \rho^2\,\rho^3 \, \, {\mathbb R}_0^2 \, \geq 0\,, \nonumber\\
& & V_{f_0} = \frac{e^{4D}}{2} \, \rho^1\, \rho^2\,\rho^3 \, {\mathbb G}_0^2 \, \geq \, 0\,, \nonumber\\
& & V_{f_2} = \frac{e^{4D}}{2} \, \biggl[({\mathbb G}^1)^2 \, \rho^1 +({\mathbb G}^2)^2 \, \rho^2 + ({\mathbb G}^3)^2 \, \rho^3 \biggr] \equiv V_{f_{21}} + V_{f_{22}} + V_{f_{23}} \, \geq \, 0\,,\nonumber\\
& & V_{f_4} = \frac{e^{4D}}{2} \, \biggl[\frac{({\mathbb G}_1)^2}{\rho^1} + \frac{({\mathbb G}_2)^2}{\rho^2} + \frac{({\mathbb G}_3)^2}{\rho^3} \biggr] \equiv V_{f_{41}} +  V_{f_{42}} + V_{f_{43}} \, \geq 0\,, \nonumber\\
& & V_{f_6} = \frac{e^{4D}}{2 \, \rho^1\, \rho^2\,\rho^3} \, \left({\mathbb G}^0\right)^2 \geq 0\,, \nonumber\\
& & V_{\rm loc} = -\, 2\, e^{3D} \,  \left({\mathbb H}_0 \, {\mathbb G}_{0} - \,\sum_{a=1}^{3} {\mathbb \mho}_{a 0}\, {\mathbb G}^a + \sum_{a=1}^{3}\, {\mathbb Q}^a{}_0 \, {\mathbb G}_a - \, {\mathbb R}_0 \, {\mathbb G}^0 \right) < 0. \nonumber
\eea
Now let us recall that in order to avoid  the de-Sitter no-go case we have invoked two conditions: $V_\omega + 2 \, V_q + 3 V_r > 0$ and the other one being $V_q + 2\, V_r + V_{f_0} > 0$. In our current example, we have all the pieces positive except the local term $V_{\rm loc}$, and therefore if moduli stabilization can be made in a trustworthy regime, there is a possibility of finding a de-Sitter minimum. On these lines, now our aim is to investigate if there exist some de-Sitter solution, and also the possible new de-Sitter no-go conditions with a given particular choice of the fluxes. 

After simplifying the scalar potential explicitly, one can convince that the Bianchi identities in eqn. (\ref{eq:IIA-genBIs}) written in terms of the generalized flux orbits are equivalent to the following correlations among the various pieces,
\bea
\label{eq:IIA-genBIs-pot}
& & \hskip-1cm V_{\omega 1}\, V_{\omega 2} = V_{h}\, V_{q3}, \qquad V_{\omega 2}\, V_{\omega 3} = V_{h}\, V_{q1}, \qquad V_{\omega 1}\, V_{\omega 3} = V_{h}\, V_{q2}, \\
& & \hskip-1cm V_{q1}\, V_{q2} = V_r\, V_{\omega3}, \qquad V_{q2}\, V_{q3} = V_r\, V_{\omega1}, \qquad V_{q1}\, V_{q3} = V_r\, V_{\omega2}\,, \nonumber\\
& & \hskip-1cm V_h\, V_r \, = \, V_{\omega1}\, V_{q1} \, = \, V_{\omega2}\, V_{q2} \, = \, V_{\omega3}\, V_{q3} \,. \nonumber
\eea
This is quite peculiar simplification for this model. 
Although we have nullified some terms in the scalar potential via directly using the Bianchi identities, it is still possible that there would be further simplifications once we consider the explicit flux solutions allowed by the Bianchi identities given in eqn. (\ref{eq:BIs-IIAZ3xZ3O}). We have classified the solutions of these identities into eight types as presented in eqn. (\ref{eq:solBIs-IIAZ3xZ3O}). Moreover, let us also mention that as we have argued, the Bianchi identities with usual fluxes presented in eqn.  (\ref{eq:BIs-IIAZ3xZ3O}) can be simply promoted to the ones given in eqn. (\ref{eq:IIA-genBIs}) using generalized flux orbits. These can be further used in simplifying the potential and for any phenomenological purpose because of the scalar potential being compactly written out in terms of the generalized flux orbits. In addition, given that these generalized flux orbits depend only on fluxes and axions and not on the saxions, we can simply use the solutions of Bianchi identities given in eqn. (\ref{eq:solBIs-IIAZ3xZ3O}) by promoting them to the generalized flux orbits. As axions do not interfere with our saxionic analysis and the approach to invoke or evade de-Sitter no-go conditions via looking at some inequalities arising from the extremization of the saxions, it is justified to take the axionic flux orbits just as fluxes satisfying the relations in (\ref{eq:IIA-genBIs}) and having solutions as follows,
\bea
\label{eq:solBIs-IIAZ3xZ3Ogen}
& {\bf S1:} & \mho_{10} = \mho_{20} = \mho_{30} = 0, \quad {\mathbb Q}^1{}_0 = {\mathbb Q}^2{}_0 = {\mathbb Q}^3{}_0 =0, \quad {\mathbb R}_0 = 0\,;\\ 
& {\bf S2:} & {\mathbb H}_0 = 0, \quad \mho_{10} = \mho_{20} = \mho_{30} = 0, \quad {\mathbb Q}^1{}_0 = {\mathbb Q}^2{}_0 = {\mathbb Q}^3{}_0 =0\,; \nonumber\\ 
& {\bf S3:} & {\mathbb R}_0 = 0, \quad {\mathbb Q}^1{}_0 = {\mathbb Q}^2{}_0 = {\mathbb Q}^3{}_0 =0, \quad \mho_{20} = 0 = \mho_{30}, \quad \mho_{10} \neq 0\,; \nonumber\\
& & {\mathbb R}_0 = 0, \quad {\mathbb Q}^1{}_0 = {\mathbb Q}^2{}_0 = {\mathbb Q}^3{}_0 =0, \quad \mho_{10} = 0 = \mho_{30}, \quad \mho_{20} \neq 0\,; \nonumber\\
& & {\mathbb R}_0 = 0, \quad {\mathbb Q}^1{}_0 = {\mathbb Q}^2{}_0 = {\mathbb Q}^3{}_0 =0, \quad \mho_{10} = 0 = \mho_{20}\,, \quad \mho_{30} \neq 0\,; \nonumber\\ 
& {\bf S4:} & {\mathbb H}_0 = 0, \quad \mho_{10} = \mho_{20} = \mho_{30} = 0, \quad {\mathbb Q}^2{}_0 = 0 = {\mathbb Q}^3{}_0\,, \quad {\mathbb Q}^1{}_0 \neq 0\,; \nonumber\\
& & {\mathbb H}_0 = 0, \quad \mho_{10} = \mho_{20} = \mho_{30} = 0, \quad {\mathbb Q}^1{}_0 = 0 = {\mathbb Q}^3{}_0\,,\quad {\mathbb Q}^2{}_0 \neq 0\,; \nonumber\\
& & {\mathbb H}_0 = 0, \quad \mho_{10} = \mho_{20} = \mho_{30} = 0, \quad {\mathbb Q}^1{}_0 = 0 = {\mathbb Q}^2{}_0\,, \quad {\mathbb Q}^3{}_0 \neq 0\,;\nonumber\\
& {\bf S5:} & {\mathbb H}_0 = 0, \quad \mho_{20} = 0 = \mho_{30}, \quad {\mathbb Q}^1{}_0 = 0 = {\mathbb Q}^2{}_0, \quad {\mathbb R}_0 = 0, \quad \mho_{10} \neq 0, \, \, {\mathbb Q}^3{}_0 \neq 0\,;\nonumber\\ 
& & {\mathbb H}_0 = 0, \quad \mho_{10} = 0 = \mho_{30}, \quad {\mathbb Q}^1{}_0 = 0 = {\mathbb Q}^2{}_0, \quad {\mathbb R}_0 = 0, \quad \mho_{20} \neq 0, \, \, {\mathbb Q}^3{}_0 \neq 0\,;\nonumber\\ 
& & {\mathbb H}_0 = 0, \quad \mho_{20} = 0 = \mho_{30}, \quad {\mathbb Q}^1{}_0 = 0 = {\mathbb Q}^3{}_0, \quad {\mathbb R}_0 = 0\,, \quad \mho_{10} \neq 0, \, \,  {\mathbb Q}^2{}_0 \neq 0\,;\nonumber\\ 
& & {\mathbb H}_0 = 0, \quad \mho_{10} = 0 = \mho_{20}, \quad {\mathbb Q}^1{}_0 = 0 = {\mathbb Q}^3{}_0, \quad {\mathbb R}_0 = 0\,, \quad \mho_{30} \neq 0, \, \, {\mathbb Q}^2{}_0 \neq 0\,;\nonumber\\ 
& & {\mathbb H}_0 = 0, \quad \mho_{10} = 0 = \mho_{20}, \quad {\mathbb Q}^2{}_0 = 0 = {\mathbb Q}^3{}_0, \quad {\mathbb R}_0 = 0\,, \quad \mho_{30} \neq 0, \, \, {\mathbb Q}^1{}_0 \neq 0\,;\nonumber\\ 
& & {\mathbb H}_0 = 0, \quad \mho_{10} = 0 = \mho_{30}, \quad {\mathbb Q}^2{}_0 = 0 = {\mathbb Q}^3{}_0, \quad {\mathbb R}_0 = 0\,, \quad \mho_{20} \neq 0, \, \,  {\mathbb Q}^1{}_0 \neq 0\,;\nonumber\\ 
& {\bf S6:} & {\mathbb H}_0 = 0\,, \, {\mathbb Q}^1{}_0 = 0, \, \mho_{20} = 0 = \mho_{30}, \, {\mathbb R}_0 = \frac{{\mathbb Q}^2{}_0\, {\mathbb Q}^3{}_0}{\mho_{10}}, \, \mho_{10} \neq 0, \quad {\mathbb Q}^2{}_0\, {\mathbb Q}^3{}_0 \neq 0\,;\nonumber\\ 
& & {\mathbb H}_0 = 0\,, \, {\mathbb Q}^2{}_0 = 0, \, \mho_{10} = 0 = \mho_{30}, \, {\mathbb R}_0 = \frac{{\mathbb Q}^1{}_0\, {\mathbb Q}^3{}_0}{\mho_{20}}, \, \mho_{20} \neq 0, \quad {\mathbb Q}^1{}_0\, {\mathbb Q}^3{}_0 \neq 0\,; \nonumber\\ 
& & {\mathbb H}_0 = 0\,, \, {\mathbb Q}^3{}_0 = 0, \, \mho_{10} = 0 = \mho_{20}, \, {\mathbb R}_0 = \frac{{\mathbb Q}^1{}_0\, {\mathbb Q}^2{}_0}{\mho_{30}}, \, \mho_{30} \neq 0, \quad {\mathbb Q}^1{}_0\, {\mathbb Q}^2{}_0 \neq 0\,; \nonumber\\ 
& {\bf S7:} & {\mathbb R}_0 = 0\,, \, \mho_{10} =0, \, {\mathbb Q}^2{}_0 = 0 = {\mathbb Q}^3{}_0, \, {\mathbb H}_0 = \frac{\omega_{20} \, \mho_{30}}{{\mathbb Q}^1{}_0}\,, \, {\mathbb Q}^1{}_0 \neq 0, \quad \mho_{20}\, \mho_{30} \neq 0\,;\nonumber\\ 
& & {\mathbb R}_0 = 0\,, \, \mho_{20} =0, \, {\mathbb Q}^1{}_0 = 0 = {\mathbb Q}^3{}_0, \, {\mathbb H}_0 = \frac{\mho_{10} \, \mho_{30}}{{\mathbb Q}^2{}_0}\,, \, {\mathbb Q}^2{}_0 \neq 0, \quad \mho_{10}\, \mho_{30} \neq 0\,;\nonumber\\ 
& & {\mathbb R}_0 = 0\,, \, \mho_{30} =0, \, {\mathbb Q}^1{}_0 = 0 = {\mathbb Q}^2{}_0, \, {\mathbb H}_0 = \frac{\mho_{10} \, \mho_{20}}{{\mathbb Q}^3{}_0}\,, \, {\mathbb Q}^3{}_0 \neq 0, \quad \mho_{20}\, \mho_{10} \neq 0\,;\nonumber\\ 
& {\bf S8:} & \biggl\{{\mathbb Q}^1{}_0 \neq 0, \, {\mathbb Q}^2{}_0 \neq 0, \, {\mathbb Q}^3{}_0\neq 0, \,  \mho_{10} = \frac{\mho_{30}\, {\mathbb Q}^3{}_0}{{\mathbb Q}^1{}_0}\, , \, \mho_{20} = \frac{\mho_{30}\, {\mathbb Q}^3{}_0}{{\mathbb Q}^2{}_0}\,, \mho_{30} \neq 0, \,\nonumber\\
& & \hskip2cm  {\mathbb H}_0 = \frac{\mho_{10} \, \mho_{20}}{{\mathbb Q}^3{}_0} \, , \, {\mathbb R}_0 = \frac{{\mathbb Q}^1{}_0\, {\mathbb Q}^2{}_0}{\mho_{30}} \biggr\}. \nonumber
\eea
We can see that except the class of solutions {\bf S8}, all the other ones have at least one axionic flux orbit set to zero. Let us make a few comments and observations on the generic solutions of Bianchi identities in eqn. (\ref{eq:IIA-genBIs}) when none of the eight NS-NS fluxes, namely $\{\mathbb H_0, \mho_{10}, \mho_{20}, \mho_{30}, {\mathbb Q}^1{}_0, {\mathbb Q}^2{}_0, {\mathbb Q}^3{}_0, {\mathbb R}_0 \}$, are zero. In this case, all the nine Bianchi identities in eqn. (\ref{eq:IIA-genBIs}) can ``effectively" reduce to satisfying the following four constraints,
\bea
\label{eq:solBIsnew1}
& & \hskip-1cm {\mathbb Q}^1{}_0 = \frac{\mho_{20}\, \mho_{30}}{{\mathbb H}_0}, \quad {\mathbb Q}^2{}_0 = \frac{\mho_{30}\, \mho_{10}}{{\mathbb H}_0}, \quad {\mathbb Q}^3{}_0 = \frac{\mho_{10}\, \mho_{20}}{{\mathbb H}_0}, \quad {\mathbb R}_0 = \frac{\mho_{10}\, \mho_{20}\, \mho_{30}}{{\mathbb H}_0^2}.
\eea
Note that this choice of solution is made to show that one could get rid of the purely nongeometric (${\mathbb Q}^a{}_0$ and ${\mathbb R}_0$) flux orbits  in terms of ${\mathbb H}_0$ and $\mho_{a0}$ fluxes. Given that the Bianchi identities satisfied by the set of standard fluxes $\{H, \omega, Q, R\}$ remain to be true after promoting these fluxes to their respective ``axionic flux" orbits $\{{\mathbb H}, \mho, {\mathbb Q}, {\mathbb R}\}$ defined in eqn. (\ref{eq:TypeIIAfluxOrbits0}), we have preferred to work with the latter in which the scalar potential pieces are compactly rewritten as in eqn. (\ref{eq:V-Z3xZ3simp0}). Having this in mind, the solution in eqn. (\ref{eq:solBIsnew1}) translates into the following correlation among the scalar potential pieces,
\bea
\label{eq:solBIsnew2}
& & \hskip-1cm V_{q1} = \frac{V_{\omega_2}V_{\omega_3}}{V_h}, \quad V_{q2} = \frac{V_{\omega_3}V_{\omega_1}}{V_h}, \quad V_{q3} = \frac{V_{\omega_1}V_{\omega_2}}{V_h}, \quad V_{r} = \frac{V_{\omega_1}V_{\omega_2}V_{\omega_3}}{V_h^2}.
\eea
Subsequently the NS-NS pieces can be expressed in the following simplified form,
\bea
& & \hskip-1cm V_{\rm NS} \equiv V_h + V_{\omega} + V_q + V_r \\
& & \hskip-0.25cm = V_h + V_{\omega_1}+ V_{\omega_2} + V_{\omega_3} + V_{q1} + V_{q2} + V_{q3} + V_r \nonumber\\
& & \hskip-0.25cm = V_h \left(1 + \frac{V_{\omega_1}}{V_h} \right)\left(1 + \frac{V_{\omega_2}}{V_h} \right)\left(1 + \frac{V_{\omega_3}}{V_h} \right). \nonumber
\eea
So we observe that there can be indeed some interesting reshuffling/simplification of the scalar potential pieces induced by the set of Bianchi identities, which could help in the search for an analytic study of the possibility of finding the de-Sitter (no-go) solutions.

\subsubsection*{Exploring new de-Sitter no-go scenarios}
Now we will separately take each of the eight solutions of the Bianchi identities in eqn. (\ref{eq:solBIs-IIAZ3xZ3Ogen}) to check if there exist a no-go condition for realizing de-Sitter vacua. For that purpose, we will simply consider the following three steps:
\begin{itemize}
\item{{\bf step1:} first, we solve for extremization conditions for all the four saxionic moduli, namely $\{\rho^1, \rho^2, \rho^3\}$ and $D$ via,
\bea
\label{eq:derViia}
& & \hskip-1cm \partial_D V = 0, \qquad  \rho^1\, \partial_{\rho^1} V = 0, \qquad \rho^2\, \partial_{\rho^2} V = 0, \qquad  \rho^3\, \partial_{\rho^3} V = 0\,,
\eea
which leads to identification of some of the scalar potential pieces when evaluated at the extremum.}
\item{{\bf step2:} we evaluate the potential at the extremum using the identification of pieces from step 1, and we check if that is non-positive or not ! In case it is, we are done, otherwise we further go to the step 3.}
\item{{\bf step3:} we evaluate the trace and the determinant of the $4\times4$ Hessian at the critical point to check if they are positive or not !}
\end{itemize}
Let us demonstrate this procedure for one of the cases in detail, say with the third solution of {\bf S3} to illustrate the steps explicitly and then we will directly tabulated the results for all the solutions. For this case the only non-zero NS-NS flux orbit we have is $\mho_{30} \neq 0$ which leads to,
\bea
& & V_{\omega 1} = 0 = V_{\omega 2}, \qquad V_q = 0, \qquad V_r = 0\,, 
\eea
This can be also understood from the relations in eqn. (\ref{eq:IIA-genBIs-pot}) which follow from the Bianchi identities given in eqn. (\ref{eq:IIA-genBIs}). Subsequently, the extremization conditions can be translated into the following four constraints,
\bea
& & V_{\rm loc} = - \frac{2}{3}\, \left(4 \, V_{f_{22}} + 4 \,V_{f_6} + 4 \,V_{f_{41}} + 4 \, V_{f_{43}} + 3\, V_h + 3 \, V_{\omega 3} \right) \\
& & V_{f_0} = - \, V_{f_{22}} + V_{f_6} + V_{f_{42}} + V_h, \,\, V_{f_{21}} = V_{f_{22}} + V_{f_{41}} - V_{f_{42}}, \nonumber\\
& & V_{f_{23}} = V_{f_{22}} - V_{f_{42}} + V_{f_{43}} + V_{\omega3}\,. \nonumber
\eea
Using the above conditions in the scalar potential leads to the following form at the extremum,
\bea
& & \hskip-2cm V_{\rm ext} = -\frac{2}{3} \left(V_{f_{22}} + V_{f_6} + V_{f_{41}} + V_{f_{43}} \right) \leq 0\,,
\eea
which ensures the no-go for the de-Sitter vacua. We have applied this technique to conclude that all of the flux solutions except the {\bf S8}  result in the no-go scenarios for de-Sitter realization. Moreover one can also derive the swampland inequalities using the derivatives of the potential which are given as follows,
\bea
\label{eq:derV-Z3xZ3}
& & \partial_D V = 2\, (V_h + V_\omega + V_q + V_r) + 3\, V_{\rm loc} + 4 \, V_f \,,\\
& & \hskip-0.8cm -\, \rho^1\, \partial_{\rho^1} V = V_h + V_{\omega 1} - V_{q1} - V_r - V_{f_0} - V_{f_{21}} + V_{f_{41}} + V_{f_6}\,,\nonumber\\
& & \hskip-0.8cm -\, \rho^2\, \partial_{\rho^2} V = V_h + V_{\omega 2} - V_{q2} - V_r - V_{f_0} - V_{f_{22}} + V_{f_{42}} + V_{f_6}\,,\nonumber\\
& & \hskip-0.8cm -\, \rho^3\, \partial_{\rho^3} V = V_h + V_{\omega 3} - V_{q3} - V_r - V_{f_0} - V_{f_{23}} + V_{f_{43}} + V_{f_6}\,. \nonumber
\eea
Given that all the pieces in our scalar potential are non-negative except for the piece $V_{\rm loc}$, and therefore we can cook many possible scenarios in which de-Sitter no-go conditions can be derived for some particular set of flux choices. For example, we can take the following combinations of the scalar potential derivatives which would be useful,
\bea
& & 3\, \partial_D V - \sum_{a=1,2,3} \rho^a\, \partial_{\rho^a} V = 9 V - 2 V_\omega - 4 V_q - 6 V_r + \sum_{p=0,2,4,6} \, p\, V_{f_p} \,,\\
& & 3\, \partial_D V + \sum_{a=1,2,3} \rho^a\, \partial_{\rho^a} V = 9 V - 2 V_q - 4 V_\omega - 6 V_h + \sum_{p=0,2,4,6} \, (6 - p)\, V_{f_p} \,.\nonumber
\eea
With some efforts, one leads to the following inequalities resulting into the de-Sitter no-go conditions for their respective flux solutions,
\bea
\label{eq:swamp-ineq}
& {\bf S1:} & \quad 3\, \partial_D V - \rho^1\, \partial_{\rho^1} V - \rho^2\, \partial_{\rho^2} V - \rho^3\, \partial_{\rho^3} V  = 9 \,V + \sum_{p=0,2,4,6} \, p\, V_{f_p} \, > 9 V\,,\\
& & \nonumber\\
& {\bf S2:} & \quad 3\, \partial_D V + \rho^1\, \partial_{\rho^1} V + \rho^2\, \partial_{\rho^2} V + \rho^3\, \partial_{\rho^3} V  = 9 \, V + \sum_{p=0,2,4,6} \, (6 - p)\, V_{f_p} \, > 9 V\,, \nonumber\\
& & \nonumber\\
& {\bf S3:} & \quad \, \partial_D V - \rho^1\, \partial_{\rho^1} V  = 3\, V + \,V_{f_{22}} + V_{f_{23}} + V_{f_{4}} + V_{f_{41}} + 2\,V_{f_6} > \, 3 V\,, \nonumber\\
& & \quad \, \partial_D V - \rho^2\, \partial_{\rho^2} V  = 3\, V + \,V_{f_{21}} + V_{f_{23}} + V_{f_{4}} + V_{f_{42}} + 2\,V_{f_6} > \, 3 V\,, \nonumber\\
& & \quad \, \partial_D V - \rho^3\, \partial_{\rho^3} V  = 3\, V + \,V_{f_{21}} + V_{f_{22}} + V_{f_{4}} + V_{f_{43}} + 2\,V_{f_6} > \, 3 V\,, \nonumber\\
& & \nonumber\\
& {\bf S4:} & \quad \, \partial_D V + \rho^1\, \partial_{\rho^1} V  = 3\, V + 2\,V_{f_0} + \,V_{f_{2}} + V_{f_{21}} + V_{f_{42}} + V_{f_{43}}  > \, 3 V\,, \nonumber\\
& & \quad \, \partial_D V + \rho^2\, \partial_{\rho^2} V  = 3\, V + 2\,V_{f_0} + \,V_{f_{2}} + V_{f_{22}} + V_{f_{41}} + V_{f_{43}}  > \, 3 V\,, \nonumber\\
& & \quad \, \partial_D V + \rho^3\, \partial_{\rho^3} V  = 3\, V + 2\,V_{f_0} + \,V_{f_{2}} + V_{f_{23}} + V_{f_{41}} + V_{f_{42}}  > \, 3 V\,, \nonumber
\eea
\bea
& {\bf S5:} & \quad \, \partial_D V + \rho^3\, \partial_{\rho^3} V - \rho^1\, \partial_{\rho^1} V = 3 V + V_{f_0} + 2 V_{f_{22}} + V_{f_{23}} + 2 V_{f_{41}} + V_{f_{43}} + V_{f_6} > \, 3 V\,, \nonumber\\ 
& & \quad \, \partial_D V + \rho^3\, \partial_{\rho^3} V - \rho^2\, \partial_{\rho^2} V = 3 V + V_{f_0} + V_{f_{21}} + 2 V_{f_{22}} +  V_{f_{41}} + 2 V_{f_{43}} + V_{f_6} > \, 3 V\,, \nonumber\\
& & \quad \, \partial_D V + \rho^2\, \partial_{\rho^2} V - \rho^1\, \partial_{\rho^1} V = 3 V + V_{f_0} + V_{f_{21}} + 2 V_{f_{23}} +  V_{f_{41}} + 2 V_{f_{42}} + V_{f_6} > \, 3 V\,, \nonumber\\
& & \quad \, \partial_D V + \rho^2\, \partial_{\rho^2} V - \rho^3\, \partial_{\rho^3} V = 3 V + V_{f_0} + V_{f_{22}} + 2 V_{f_{23}} + 2 V_{f_{41}} + V_{f_{42}} + V_{f_6} > \, 3 V\,, \nonumber\\
&  & \quad \, \partial_D V + \rho^1\, \partial_{\rho^1} V - \rho^3\, \partial_{\rho^3} V = 3 V + V_{f_0} + 2 V_{f_{21}} + V_{f_{22}} + V_{f_{42}} + 2 V_{f_{43}} + V_{f_6} > \, 3 V\,, \nonumber\\
& & \quad \, \partial_D V + \rho^1\, \partial_{\rho^1} V - \rho^2\, \partial_{\rho^2} V = 3 V + V_{f_0} + 2 V_{f_{21}} + V_{f_{23}} + 2 V_{f_{42}} + V_{f_{43}} + V_{f_6} > \, 3 V\,, \nonumber\\
& & \nonumber\\
& {\bf S6:} & \quad \, \partial_D V - \rho^1\, \partial_{\rho^1} V + \rho^2\, \partial_{\rho^2} V + \rho^3\, \partial_{\rho^3} V = 3 V + 2 V_{f_{0}} + 2 V_{f_{22}} + 2 V_{f_{23}} + 2 V_{f_{41}} > \, 3 V\,, \nonumber\\
& & \quad \, \partial_D V + \rho^1\, \partial_{\rho^1} V - \rho^2\, \partial_{\rho^2} V + \rho^3\, \partial_{\rho^3} V = 3 V + 2 V_{f_{0}} + 2 V_{f_{21}} + 2 V_{f_{23}} + 2 V_{f_{42}} > \, 3 V\,, \nonumber\\
& & \quad \, \partial_D V + \rho^1\, \partial_{\rho^1} V + \rho^2\, \partial_{\rho^2} V - \rho^3\, \partial_{\rho^3} V = 3 V + 2 V_{f_{0}} + 2 V_{f_{21}} + 2 V_{f_{22}} + 2 V_{f_{43}} > \, 3 V\,, \nonumber\\
& & \nonumber\\
& {\bf S7:} & \quad \, \partial_D V + \rho^1\, \partial_{\rho^1} V - \rho^2\, \partial_{\rho^2} V - \rho^3\, \partial_{\rho^3} V = 3 V + 2 V_{f_{21}} + 2 V_{f_{42}} + 2 V_{f_{43}} + 2 V_{f_6} > \, 3 V\,, \nonumber\\
& & \quad \, \partial_D V - \rho^1\, \partial_{\rho^1} V + \rho^2\, \partial_{\rho^2} V - \rho^3\, \partial_{\rho^3} V = 3 V + 2 V_{f_{22}} + 2 V_{f_{41}} + 2 V_{f_{43}} + 2 V_{f_6} > \, 3 V\,, \nonumber\\
& & \quad \, \partial_D V - \rho^1\, \partial_{\rho^1} V - \rho^2\, \partial_{\rho^2} V + \rho^3\, \partial_{\rho^3} V = 3 V + 2 V_{f_{23}} + 2 V_{f_{41}} + 2 V_{f_{42}} + 2 V_{f_6} > \, 3 V\,. \nonumber
\eea
It is remarkable that for all the solutions in eqn. (\ref{eq:solBIs-IIAZ3xZ3Ogen}) except the last one, namely the one denoted as {\bf S8}, we  find de-Sitter no-go conditions. For the solution {\bf S8}, we have performed some numerical investigations with random choice of allowed fluxes, but we could not manage to find any de-Sitter vacua using integral fluxes. 
It would be interesting to perform a comprehensive and systematic numerical analysis to investigate the possibility of the existence of a viable de-Sitter solution or an extension of the no-go results for the remaining generic case.


\section{Conclusions}
\label{sec_conclusions}
In this article we have studied the effects of including nongeometric fluxes in the four-dimensional type II models which arise with compactifications using rigid (Calabi Yau) threefolds. In this regard we have considered an explicit construction using the ${\mathbb T}^6/({\mathbb Z}_3 \times {\mathbb Z}_3)$ toroidal orbifold. First we have presented the so-called missing Bianchi identities in the four-dimensional ${\cal N} =2$ theory before orientifolding the setup to be later studied in explicit type IIA and type IIB context. In type IIB orientifold we have observed that the missing Bianchi identities are quite strong and whenever one wants to have at least one non-trivial component for the $H_3$-flux, say for example needed to fix the universal axio-dilaton modulus, there remains no scope for turning-on any nongeometric $Q$-flux, and hence getting back to the no-scale-structure in which the three K\"ahler moduli remain unstabilized. Based on these observations we also argue that if these could be promoted to the case of rigid CY threefolds, then several minimal nongeometric setups with rigid CYs, for example (some) models proposed in \cite{Blumenhagen:2015kja, Blumenhagen:2015qda, Blumenhagen:2015jva, Damian:2018tlf} might be in problem, though one would certainly need to ensure that the conjectured form of the identities in \cite{Gao:2018ayp}, which is further supported by the current analysis, is indeed true for the beyond toroidal cases. 

On the type IIA side, we have investigated the solutions of Bianchi identities in some good detail and have found that seven out of eight classes of solutions result in de-Sitter no-go scenarios despite having nongeometric fluxes turned-on, though a no-go for the most generic case could not be found. However for this generic case also, in our (limited) numerical search we could not manage to find any de-Sitter vacua, and 
it could be conjectured that finding de-Sitter vacua with integer fluxes satisfying all the Bianchi identities should not be possible in models of rigid compactifications with (non-)geometric fluxes.  We plan to get back to addressing this issue in a comprehensive numerical search in near future.
 
\section*{Acknowledgments}
We are grateful to Fernando Quevedo for his kind support and encouragements throughout. We would like to thank David Andriot, Erik Plauschinn, Thomas Van Riet and Timm Wrase for useful discussions and communications. In addition, we are very thankful to the referee for her/his useful comments and suggestions which have helped in improving the manuscript.

\newpage
\appendix
\setcounter{equation}{0}

\section{Ingredients for the ${\mathbb T}^6/({\mathbb Z}_3 \times {\mathbb Z}_3)$ orbifold and its orientifolds}
\label{sec_appendix1}
In this section we present the relevant pieces of information on the explicit construction of the ${\mathbb T}^6/({\mathbb Z}_3\times {\mathbb Z}_3)$ orbifold \cite{DeWolfe:2005uu}. Our aim is to first present the computations about (nongeometric) flux components and moduli surviving under the orbifold action regarding the ${\cal N}=2$ nongeometric construction, and subsequently we will provide those explicit details for the type IIA and type IIB orientifolds as well.

\subsection{Fluxes and moduli in the ${\cal N} = 2$ orbifold compactifications}
Let us start by considering the complexified coordinates on the six torus ${\mathbb T}^6$ to be defined as follows,
\bea
& & \hskip-0.5cm z^1 = x^1 + i\,\, x^2, \qquad z^2 = x^3 + i\, x^4, \qquad z^3 = \, x^5 + i\,\, x^6 \,,
\eea
subject to the periodic conditions,
\bea
& & z^i \sim z^i + 1 \sim z^i + \theta, \quad {\rm where } \quad \theta = e^{i\, \pi/3}; \qquad \forall \, i = 1, 2, 3.
\eea
This six-torus has a ${\mathbb Z}_3$ symmetry $\Theta_1$ acting in the following manner,
\bea
& & \Theta_1: \left( z^1, \, z^2, \, z^3 \right) \to \left( \theta^2 \, z^1, \, \theta^2 \, z^2, \, \theta^2 \, z^3 \right).
\eea
This $\Theta_1$ action has 27 fixed points and the subsequent orbifold is a singular limit of a Calabi Yau threefold with Euler character $\chi = 72$. Moreover, as analysed in \cite{Dixon:1985jw, Strominger:1985ku} the resulting space has an additional ${\mathbb Z}_3$ symmetry acting in the following manner,
\bea
& & \Theta_2: \left( z^1, \, z^2, \, z^3 \right) \to \left( \theta^2 \, z^1 + \frac{\theta+1}{3}, \, \theta^4 \, z^2 + \frac{\theta+1}{3}, \, \, z^3 \right),
\eea
which doesn't have any fixed point. Further modding out the six-torus with the second ${\mathbb Z}_3$ action leads to a singular limit of a Calabi Yau threefold with $\chi =24$ having 9 ${\mathbb Z}_3$ singularities. To be more specific, the ${\mathbb T}^6/({\mathbb Z}_3\times {\mathbb Z}_3)$ orbifold compactification results in a construction with frozen complex structure moduli, i.e. having $h^{2,1} = 0$ and $h^{1,1} = 12$ which corresponds to 3 standard K\"ahler moduli and 9 blow-up modes. However, in our current study we will focus only on the untwisted sector and hence for us,
\bea
& & h^{1,1} = 3, \qquad \qquad \qquad h^{2,1} = 0\,.
\eea

\subsubsection*{Twist invariant forms and moduli}
Now it is easy to construct the twist invariant forms to represent the various moduli and fluxes present in the theory. For that purpose let us first note that 
\bea
& & \Theta_1: dz^i \to \theta^2 \, dz^i, \quad \quad \quad  \Theta_2: dz^i \to \theta^{2i} \, dz^i\, ; \qquad \qquad \forall \, i = 1, \, 2, \, 3.
\eea
This leads to the following twist invariant two-forms $\mu_A$,
\bea
& & \mu_1 = \left(\kappa \, \sqrt{3}\right)^{1/3}\, i\, \, dz^1 \wedge \ov{dz^1} = 2\,\left(\kappa \, \sqrt{3}\right)^{1/3}\, \, dx^1 \wedge dx^2\, , \\
& & \mu_2 = \left(\kappa \, \sqrt{3}\right)^{1/3}\, i\, \, dz^2 \wedge \ov{dz^2} = 2\,\left(\kappa \, \sqrt{3}\right)^{1/3}\, \, dx^3 \wedge dx^4 \, , \nonumber\\
& & \mu_3 = \left(\kappa \, \sqrt{3}\right)^{1/3}\, i\, \, dz^3 \wedge \ov{dz^3} = 2\,\left(\kappa \, \sqrt{3}\right)^{1/3}\, \, dx^5 \wedge dx^6 \, , \nonumber
\eea
and the four-forms $\tilde\mu^A$'s which are dual to the two-forms ($\mu_A$) are defined as follows,
\bea
& & \tilde\mu^1 = \left(\frac{3}{\kappa}\right)^{1/3} \, \left(i\, \, dz^2 \wedge \ov{dz^2}\right) \, \left(i\, \, dz^3 \wedge \ov{dz^3}\right) = \kappa^{-1} \, \mu_2 \wedge \mu_3\,,\\
& & \tilde\mu^2 = \left(\frac{3}{\kappa}\right)^{1/3} \, \left(i\, \, dz^3 \wedge \ov{dz^3}\right) \, \left(i\, \, dz^1 \wedge \ov{dz^1}\right) = \kappa^{-1} \, \mu_3 \wedge \mu_1\,,\nonumber\\
& & \tilde\mu^3 = \left(\frac{3}{\kappa}\right)^{1/3} \, \left(i\, \, dz^1 \wedge \ov{dz^1}\right) \, \left(i\, \, dz^2 \wedge \ov{dz^2}\right) = \kappa^{-1} \, \mu_1 \wedge \mu_2\,.\nonumber
\eea
Here the overall normalization has been fixed as follows,
\bea
& & \int_{{\mathbb T}^6/({\mathbb Z}_3\times {\mathbb Z}_3)} \, \mu_A \wedge \tilde\mu^B = \delta_A{}^B\,, \qquad \int_{{\mathbb T}^6/({\mathbb Z}_3\times {\mathbb Z}_3)} \, \mu_1 \wedge \mu_2 \wedge \mu_3 = \kappa\,.
\eea
Further, the twist-invariant three-form $\Omega$ can be defined as follows,
\bea
& & \Omega = 3^{1/4} \, i \, \, dz^1 \wedge dz^2 \wedge dz^3\,,
\eea
where the normalization in the above unique holomorphic three-form $\Omega$ has been fixed via the following constraint,
\bea
& & i \, \int_{{\mathbb T}^6/({\mathbb Z}_3\times {\mathbb Z}_3)} \Omega \wedge \ov\Omega = 1\,,
\eea
where we have used $i\, \int_{{\mathbb T}^6/({\mathbb Z}_3\times {\mathbb Z}_3)} dz^i \wedge \ov{dz^i} = \sqrt{3}$. In terms of real cohomology, the holomorphic three-form $\Omega$ can be represented as,
\bea
& & \Omega = \frac{1}{\sqrt{2}} \left({\cal A}_0 + i \, {\cal B}^0 \right), \qquad \int_{{\mathbb T}^6/({\mathbb Z}_3\times {\mathbb Z}_3)} {\cal A}_0 \wedge {\cal B}^0 = 1\,.
\eea
With a slight difference to the conventions of  \cite{DeWolfe:2005uu}, for our convenience, we want to get rid of normalization factors to write the volume as ${\cal V} = t^1\, t^2\, t^3$, i.e. the triple intersection number $\kappa_{123} \equiv \kappa = 1$. For that let us rescale the complex coordinates by $dz^i \to 3^{-\frac{1}{12}}\, dz^i; \, \, \forall \, \, i=\{1, 2, 3\}$, which would lead to $i\, \int_{{\mathbb T}^6/({\mathbb Z}_3\times {\mathbb Z}_3)} dz^i \wedge \ov{dz^i} = 3^{2/3}$, and subsequently the normalizations can be set as follows,
\bea
& & \Theta_1: dz^i \to \theta^2 \, dz^i, \quad \quad \quad  \Theta_2: dz^i \to \theta^{2i} \, dz^i\, ; \qquad \qquad \forall \, i = 1, \, 2, \, 3.
\eea
This leads to the following twist invariant two-forms $\mu_A$,
\bea
& & \mu_1 = dx^1 \wedge dx^2 = \frac{i}{2}\, \, dz^1 \wedge \ov{dz^1} \, , \\
& & \mu_2 = dx^3 \wedge dx^4 = \frac{i}{2}\, \, dz^2 \wedge \ov{dz^2} \, , \nonumber\\
& & \mu_3 = dx^5 \wedge dx^6 = \frac{i}{2}\, \, dz^3 \wedge \ov{dz^3} \, , \nonumber
\eea
and the four-forms $\tilde\mu^A$'s which are dual to the two-forms ($\mu_A$) are defined as follows,
\bea
& & \tilde\mu^1 = \frac{1}{4}\, \left(i\, \, dz^2 \wedge \ov{dz^2}\right) \, \left(i\, \, dz^3 \wedge \ov{dz^3}\right) = \, \mu_2 \wedge \mu_3\,,\\
& & \tilde\mu^2 = \frac{1}{4} \, \left(i\, \, dz^3 \wedge \ov{dz^3}\right) \, \left(i\, \, dz^1 \wedge \ov{dz^1}\right) =  \, \mu_3 \wedge \mu_1\,,\nonumber\\
& & \tilde\mu^3 = \frac{1}{4} \, \left(i\, \, dz^1 \wedge \ov{dz^1}\right) \, \left(i\, \, dz^2 \wedge \ov{dz^2}\right) =  \, \mu_1 \wedge \mu_2\,.\nonumber
\eea
Now the overall normalization has been fixed as follows,
\bea
& & \int_{{\mathbb T}^6/({\mathbb Z}_3\times {\mathbb Z}_3)} \, \mu_A \wedge \tilde\mu^B = \delta_A{}^B\,, \qquad \int_{{\mathbb T}^6/({\mathbb Z}_3\times {\mathbb Z}_3)} \, \mu_1 \wedge \mu_2 \wedge \mu_3 = 1\,.
\eea
Further, the twist-invariant three-form $\Omega$ can be defined as follows,
\bea
& & \Omega = \, i \, \, dz^1 \wedge dz^2 \wedge dz^3\, = \sum_{\Lambda=0}^3\, \left(\alpha_\Lambda + i\, \beta^\Lambda \right)
\eea
where the real cohomology bases are represented as follows,
\bea 
& & \hskip-0.5cm \alpha_0 = 2 \wedge 4 \wedge 6, \quad  \alpha_1 = - 2 \wedge 3 \wedge 5, \quad \alpha_2 = - 1 \wedge 4 \wedge 5, \quad  \alpha_3 = - 1 \wedge 3 \wedge 6, \\
& & \hskip-0.5cm \beta^0 = 1 \wedge 3 \wedge 5, \quad \beta^1 = - 1 \wedge 4 \wedge 6, \quad \beta^2 = - 2 \wedge 3 \wedge 6, \quad \beta^3 = - 2 \wedge 4 \wedge 5\,. \nonumber
\eea
Here the shorthand notations $2\wedge 4 \wedge 6 = dx^2 \wedge dx^4 \wedge dx^6$ etc. have been used, for which the normalization is given as $\int \alpha_\Lambda \wedge \beta^\Delta = \delta_\Lambda{}^\Delta$. Moreover under the orbifold twists we have only two components of the three-form to be non-zero as there are following constraints to be imposed,
\bea
& & \alpha_0 = \alpha_1= \alpha_2 =\alpha_3, \qquad \beta^0 = \beta^1 = \beta^2 = \beta^3 \,.
\eea
where the normalization in the unique holomorphic three-form $\Omega$ has been fixed via,
\bea
& & i \, \int_{{\mathbb T}^6/({\mathbb Z}_3\times {\mathbb Z}_3)} \Omega \wedge \ov\Omega = 1\,.
\eea
In terms of real cohomology, the holomorphic three-form $\Omega$ can be represented as,
\bea
& & \Omega = \frac{1}{\sqrt{2}} \left({\cal A}_0 + i \, {\cal B}^0 \right), \qquad \int_{{\mathbb T}^6/({\mathbb Z}_3\times {\mathbb Z}_3)} {\cal A}_0 \wedge {\cal B}^0 = 1\,.
\eea

\subsubsection*{Twist invariant fluxes}
For our particular ${\mathbb T}^6/({\mathbb Z}_3\times {\mathbb Z}_3)$ orbifold construction, we now come to the discussion on the various flux components which survive under the two orbifold twists,
\begin{itemize}
\item{Out of 20 flux components of $H_{ijk}$, there are only 8 components which are allowed to be non-zero and, are further constrained by the following 6 relations,
\bea
& & \hskip-1cm H_{235} = H_{145} = H_{136} = - \, H_{246}, \qquad H_{146} = H_{236} = H_{245} = -\,H_{135},
\eea
which lead to only two independent flux components, namely $H_{246}$ and $H_{135}$.}
\item{Out of 90 flux components of $\omega_{ij}{}^k$, there are only 24 components which are allowed to be non-zero and, are further constrained by the following 18 relations,
\bea
& & \hskip-0.8cm \omega_{23}{}^6 = \omega_{14}{}^6 = -\, \omega_{13}{}^5 = \omega_{24}{}^5, \qquad \qquad \omega_{23}{}^5 = \omega_{14}{}^5 = \omega_{13}{}^6 = -\,\omega_{24}{}^6,\\
& & \hskip-0.8cm  \omega_{45}{}^2 = \omega_{36}{}^2 = -\, \omega_{35}{}^1 = \omega_{46}{}^1, \qquad \qquad \omega_{45}{}^1= \omega_{36}{}^1 = \omega_{35}{}^2 = - \, \omega_{46}{}^2, \nonumber\\
& & \hskip-0.8cm \omega_{61}{}^4 = \omega_{52}{}^4 = - \, \omega_{51}{}^3 = \omega_{62}{}^3, \qquad \qquad \omega_{61}{}^3 = \omega_{52}{}^3 = \omega_{51}{}^4 = -\,\omega_{62}{}^4, \nonumber
\eea
which lead to the following six ``independent'" flux components denoted as,
\bea
& & \hskip-1cm \omega_{46}{}^1, \quad \omega_{62}{}^3, \quad \omega_{24}{}^5, \quad \omega_{24}{}^6, \quad \omega_{46}{}^2, \quad   \omega_{62}{}^4.
\eea}
\item{Out of 90 flux components of $Q_i{}^{jk}$, there are only 24 components which are allowed to be non-zero and, are further constrained by the following 18 relations,
\bea
& & \hskip-1cm Q_2{}^{36} = Q_1{}^{46} = -\, Q_1{}^{35} = Q_2{}^{45}, \quad \qquad Q_2{}^{35} = Q_1{}^{45} = Q_1{}^{36} = -\, Q_2{}^{46}, \\
& & \hskip-1cm Q_4{}^{52} = Q_3{}^{62} = -\, Q_3{}^{51} = Q_4{}^{61}, \quad \qquad Q_4{}^{51} = Q_3{}^{61} = Q_3{}^{52} = -\,Q_4{}^{62}, \nonumber\\
& & \hskip-1cm Q_6{}^{14} = Q_5{}^{24} = -\, Q_5{}^{13} = Q_6{}^{23}, \quad \qquad Q_6{}^{13} = Q_5{}^{23} = Q_5{}^{14} = -\,Q_6{}^{24}\,,\nonumber
\eea
which lead to the following six ``independent'" flux components denoted as,
\bea
& & \hskip-1cm Q_4{}^{61}, \quad Q_6{}^{23}, \quad Q_2{}^{45}, \quad Q_2{}^{46}, \quad Q_4{}^{62}, \quad Q_6{}^{24}\,.
\eea}
\item{Out of 20 flux components of $R^{ijk}$, there are only 8 components which are  allowed to be non-zero and, are further constrained by the following 6 relations,
\bea
& & \hskip-1cm R^{235} = R^{145} = R^{136} = - \, R^{246}, \qquad R^{146} = R^{236} = R^{245} = - \, R^{135},
\eea
which leads to only two independent flux components, namely $R^{246}$ and $R^{135}$.}
\end{itemize}
Moreover the flux conversion relations between the standard and cohomology formulation representations are given as follows,
\bea
& & H_0 = H_{135}, \qquad \qquad \quad H^0 = H_{246}\,,\\
& & \omega_{10} = -\, \omega_{46}{}^2, \qquad \qquad \omega_{20} = -\, \omega_{62}{}^4, \qquad \qquad \omega_{30} = -\, \omega_{24}{}^6, \nonumber\\
& & \omega_{1}{}^0 = -\, \omega_{46}{}^1, \qquad \qquad \omega_{2}{}^0 = -\, \omega_{62}{}^3, \qquad \qquad \omega_{3}{}^0 = -\, \omega_{24}{}^5, \nonumber\\
& & Q^1{}_0 = Q_2{}^{45}, \qquad \qquad \, \, \, Q^2{}_0 = Q_4{}^{61}, \qquad \qquad \, \, \, \, Q^3{}_0 = Q_6{}^{23}, \nonumber\\
& & Q^{10} = -\,Q_2{}^{46}, \qquad \quad \, \, \, \, Q^{20} = -\,Q_4{}^{62}, \qquad \quad \, \, \, \, \, Q^{30} = -\,Q_6{}^{24}, \nonumber\\
& & R_0 = R^{246}, \qquad \qquad \quad \, \, \, R^0 = - R^{135}\,. \nonumber
\eea

\subsection{Fluxes and moduli in the ${\cal N} = 1$ orientifold compactifications}
\subsubsection*{Type IIA using a ${\mathbb T}^6/({\mathbb Z}_3 \times {\mathbb Z}_3)$ orientifold:}
The four-dimensional ${\cal N}=1$ type IIA model is constructed with the orientifold of a ${\mathbb T}^6/({\mathbb Z}_3\times {\mathbb Z}_3)$ orbifold. For that purpose, we further quotient the orbifold by ${\cal O} = \Omega_p \, (-1)^{F_L} \, \sigma$, where $\Omega_p$ is the worldsheet parity while $(-1)^{F_L}$ corresponds to the left moving fermion number, and the anti-holomorphic involution $\sigma$ is defined by the following action on the complex coordinates:
\bea
& & \hskip-2cm \sigma: \, \left(z^1, \, z^2, \, z^3 \right) \quad \to \quad  \left( -\ov z^1, \, -\ov z^2, \, -\ov z^3 \right).
\eea
Under the orientifold action $\sigma$, the $(1,1)$-cohomology splits into a trivial even sector and a non-trivial odd sector, i.e. we have $h^{1,1} = h^{1,1}_- = 3$ and $h^{1,1}_+ = 0$. In our conventions $A = \{\alpha \in h^{1,1}_+, a \in h^{1,1}_-\}$. Further one has $\sigma: \Omega \to \ov \Omega$, and therefore ${\cal A}_0$ is even while ${\cal B}^0$ is odd under the involution. With some computations it turns out that the surviving NS-NS flux components and their respective cohomology versions are given as follows,
\bea
& & \hskip-1cm H_0 = H_{135}, \qquad  \omega_{10} = -\, \omega_{46}{}^2, \qquad \omega_{20} = -\, \omega_{62}{}^4 \qquad  \omega_{30} = -\, \omega_{24}{}^6, \\
& & \hskip-1cm Q^1{}_0 = Q_2{}^{45}, \qquad Q^2{}_0 = Q_4{}^{61}, \qquad Q^3{}_0 = Q_6{}^{23}, \qquad \, \, \, \, R_0 = R^{246}\,. \nonumber
\eea

\subsubsection*{Type IIB using a ${\mathbb T}^6/({\mathbb Z}_3 \times {\mathbb Z}_3)$ orientifold:}
The four-dimensional ${\cal N}=1$ type IIB model is constructed with the orientifold of a ${\mathbb T}^6/({\mathbb Z}_3\times {\mathbb Z}_3)$ orbifold. For that purpose, we further quotient the orbifold by ${\cal O} = \Omega_p \, (-1)^{F_L} \, \sigma$, where $\Omega_p$ is the worldsheet parity while $(-1)^{F_L}$ corresponds to the left moving fermion number, and the holomorphic involution $\sigma$ is defined by the following action:
\bea
& & \sigma: \, \left(z^1, \, z^2, \, z^3 \right) \quad \to \quad  \left( -\,z^1, \, -\,z^2, \, -\,z^3 \right).
\eea
Under the orientifold action $\sigma$, the $(1,1)$-cohomology splits into a trivial odd sector and a non-trivial even sector, i.e. we have $h^{1,1} = h^{1,1}_+ = 3$ and $h^{1,1}_- = 0$ which means that there are three K\"ahler moduli $T_\alpha$ and no odd-moduli $G^a$ being present in this construction. Recall that in our conventions $A = \{\alpha \in h^{1,1}_+, a \in h^{1,1}_-\}$. Further one has $\sigma: \Omega \to -\, \Omega$, and therefore ${\cal A}_0$ as well as ${\cal B}^0$ both are odd under the involution. Also given that the NS-NS flux $H_3$ is odd, and therefore with some computations it turns out that the surviving NS-NS flux components are given as follows,
\bea
& & \hskip-0.5cm H_0 = H_{135}, \qquad Q^1{}_0 = Q_2{}^{45}, \qquad  \, \, \, \, \, Q^2{}_0 = Q_4{}^{61}, \qquad \, \, \, \, \, \, Q^3{}_0 = Q_6{}^{23}, \\
& & \hskip-0.5cm H^0 = H_{246}, \qquad Q^{10} = -\,Q_2{}^{46}, \qquad  Q^{20} = -\,Q_4{}^{62}, \qquad  Q^{30} = -\,Q_6{}^{24}\,. \nonumber
\eea

\section{Details on the Type IIA scalar potential}
\label{sec_appendix2}
Within the two-derivative approximations, the four-dimensional ${\cal N} =1$ type IIA scalar potential receives two types of contributions; one arising from the $F$-term while the other arising from the $D$-term effects, which have been computed for the generic case in \cite{Gao:2017gxk}, and can be expressed in the following collection,
\bea
\label{eq:IIA-pot-gen}
& & \hskip-2cm V_{\rm IIA} \equiv V_F + V_D \\
& &  \hskip-1.2cm = V_{{\mathbb H} {\mathbb H}} + V_{{\mathbb \mho} {\mathbb \mho}} + V_{{\mathbb Q} {\mathbb Q}} + V_{{\mathbb R} {\mathbb R}}  + V_{{\mathbb H} {\mathbb Q}} + V_{{\mathbb R} {\mathbb \mho}} \nonumber\\
& & \hskip-0.5cm + V_{{\mathbb G}^0 {\mathbb G}^0} + V_{{\mathbb G}^a {\mathbb G}^a} + V_{{\mathbb G}_a {\mathbb G}_a} + V_{{\mathbb G}_0 {\mathbb G}_0} + V_{D6/O6} + V_D\, , \nonumber
\eea
where explicit expressions for the various pieces are summarized as follows, 
\bea
\label{eq:IIA-pot-gen-all}
& & V_{{\mathbb H} {\mathbb H}} = \frac{e^{2D}}{2 \, {\cal V}} \, \left[{\mathbb H}_I \, {\cal M}^{I J} \, {\mathbb H}_J \right] \\
& & V_{{\mathbb \mho} {\mathbb \mho}} =\frac{e^{2D}}{2 \, {\cal V}} \, \, \left[{\mathbb \mho}_{aI} \, {\cal M}^{I J} \, {\mathbb \mho}_{bJ} \, t^a \, t^b  + 4 \, {\mathbb \mho}_{aI} \, {\mathbb \mho}_{bJ} \, {\cal X}^I\, {\cal X}^J \, \left({\cal V}\, {\cal G}^{ab} - \, t^a t^b \right)\right] \nonumber\\
& & V_{{\mathbb Q} {\mathbb Q}} = \frac{e^{2D}}{2 \, {\cal V}} \,\, \left[{\mathbb Q}^{a}{}_I \,{\cal M}^{I J} \, {\mathbb Q}^{b}{}_J  \, \sigma_a \, \sigma_b + \, 4 \,{\mathbb Q}^a{}_I \, {\mathbb Q}^{b}{}_{J} \, {\cal X}^I\, {\cal X}^J \, \left({\cal V} \, {\cal G}_{ab} - \, \sigma_a \, \sigma_b \right)\right] \nonumber\\
& & V_{{\mathbb R} {\mathbb R}} =\frac{e^{2D}}{2 \, {\cal V}} \, \, \left[{\cal V}^2 \, \, \, {\mathbb R}_I \, {\cal M}^{I J} \, {\mathbb R}_J \right] \nonumber\\
& & V_{{\mathbb H} {\mathbb Q}} = \frac{e^{2D}}{2 \, {\cal V}} \times (-2) \,\, \left[ {\mathbb H}_I \, {\cal M}^{I J} \, {\mathbb Q}^a{}_J \, \sigma_a \, - \, 4 \, {\mathbb H}_I \, {\cal X}^I \, {\cal X}^J \, {\mathbb Q}^a{}_J \, \sigma_a \right] \nonumber\\
& & V_{{\mathbb R} {\mathbb \mho}} = \frac{e^{2D}}{2 \, {\cal V}} \times (-2) \, \, {\cal V}\, \left[ {\mathbb R}_I \, {\cal M}^{I J} \, {\mathbb \mho}_{aJ} \, t^a \, - \,4 \, {\mathbb R}_I \, {\cal X}^I \, {\cal X}^J \, {\mathbb \mho}_{aJ} \, t^a \right] \nonumber
\eea
\bea
& & V_{{\mathbb G}_0 {\mathbb G}_0} = \frac{e^{4D}}{2 \, {\cal V}} \, \left[{\cal V}^2 \, \left({\mathbb G}_0 \right)^2\right], \nonumber\\
& & V_{{\mathbb G}^a {\mathbb G}^a} =  \frac{e^{4D}}{2\, {\cal V}} \, \, \left[{\cal V}\, {\mathbb G}^a \,{\cal G}_{ab}\, {\mathbb G}^b \right] \, , \\
& & V_{{\mathbb G}_a {\mathbb G}_a}= \,  \frac{e^{4D}}{2\, {\cal V}} \, \, \left[{\cal V}\, {\mathbb G}_a \,\,{\cal G}^{ab} \, {\mathbb G}_b \right]\, , \nonumber\\
& & V_{{\mathbb G}^0 {\mathbb G}^0} = \, \frac{e^{4D}}{2 \, {\cal V}} \,\left({\mathbb G}^0 \right)^2 \,, \nonumber\\
& & V_{D6/O6} = -\, 2\, e^{3D} \left({\mathbb H}_K \, {\mathbb G}_{0} - \, {\mathbb \mho}_{a K}\, {\mathbb G}^a + \, {\mathbb Q}^a{}_K \, {\mathbb G}_a - \, {\mathbb R}_K \, {\mathbb G}^0 \right) {\cal X}^K . \nonumber\\
& & V_D = - \, 2\, e^{2D} \, {\cal F}_I \, {\cal F}_J \biggl[\hat{\mathbb \mho}_{\alpha}{}^I \, {\cal G}^{\alpha\beta}\, \hat{\mathbb \mho}_{\beta}{}^J + \,\hat{\mathbb Q}^{\alpha I} \, {\cal G}_{\alpha\beta} \, \hat{\mathbb Q}^{\beta J}\biggr] = V_D^{(1)} + V_D^{(2)}\,. \nonumber
\eea
Here $D$ denotes the four-dimensional dilaton which is related to the ten-dimensional dilaton $\phi$ via $e^{-D} = e^{-\phi}\, \sqrt{\cal V}$ where ${\cal V}$ is the volume of the complex threefold. In addition, the various flux orbits involving the axionic moduli $b^a$ and $\xi^K$ are given as follows,
\bea
\label{eq:TypeIIAfluxOrbits0}
& & {\mathbb H}_K \, \, = H_K + \omega_{a K}\, b^a + \frac{1}{2} \kappa_{abc} \, b^b \, b^c \, Q^a{}_K + \frac{1}{6} \kappa_{abc} \, b^a \, b^b \, b^c \, R_K \, ,\nonumber\\
& & {\mathbb \mho}_{a K} = \omega_{aK} + \kappa_{abc} b^b \, Q^c{}_K + \frac{1}{2} \kappa_{abc} \, b^b\, b^c \, \, R_K, \nonumber\\
& & {\mathbb Q}^a{}_K = Q^a{}_K + \, b^a\, R_K \, , \nonumber\\
& & {\mathbb R}_K \, \,\,= \, R_K \,.\nonumber\\
& & \nonumber\\
& & {\mathbb G}^0 = {\mathbb F}^0\, + \xi^K\, {\mathbb H}_K, \qquad \, \, \, \qquad {\mathbb G}_a = {\mathbb F}_a\, + \xi^K \, {\mathbb \mho}_{a K}, \nonumber\\
& & {\mathbb G}^a = {\mathbb F}^a\, + \xi^K\, {\mathbb Q}^a{}_K \,, \qquad \qquad {\mathbb G}_0 = {\mathbb F}_0 \, +  \xi^K \, {\mathbb R}_K . \nonumber\\
& & {\mathbb F}^0 = e^0 +\, b^a\, e_a + \frac{1}{2} \, \kappa_{abc} \, b^a\, b^b \,m^c + \frac{1}{6}\, \kappa_{abc}\, b^a \,b^b\,b^c \, m_0\, , \\
& & {\mathbb F}_a = e_a + \, \kappa_{abc} \, b^b \,m^c + \frac{1}{2}\, \kappa_{abc}\, b^b\,b^c \, m_0\,, \nonumber\\
& & {\mathbb F}^a = m^a + m_0\, b^a\,, \nonumber\\
& & {\mathbb F}_0 = m_0 \,. \nonumber
\eea
Moreover, the moduli space metrics follow from \cite{Grimm:2004uq}, and are obtained by orientifolding their parental ${\cal N}=2$ metrics given in eqns. (\ref{eq:Neq2Kmatrices1})-(\ref{eq:Neq2Kmatrices2}),
\bea
\label{eq:KabIIA}
& \hskip-1cm K_{a \ov b} = \frac{\kappa_a \, \kappa_b - 4\, {\cal V} \, \kappa_{ab}}{16\, {\cal V}^2}, \quad & K_{a \ov b} = \frac{1}{4{\cal V}} \, \int_{X_3} \nu_a \wedge \ast \nu_b :=  \frac{1}{4{\cal V}} \, {\cal G}_{ab}\, , \\
& \hskip-1cm K^{a \ov b} = 2 \, \, t^a \, t^b - 4\, {\cal V} \, \, \kappa^{ab}\,, \quad & K^{a \ov b} = \, 4 \, {\cal V} \,  \int_{X_3} \, \tilde{\nu}^a \wedge \ast \tilde{\nu}^b :=  {4{\cal V}} \, {\cal G}^{ab}\,, \nonumber\\
& \hskip-1cm K_{\alpha \ov{\beta}} = - \, \frac{\hat\kappa_{\alpha\beta}}{4 \, {\cal V}}, \quad & {K}_{\alpha \ov \beta} = \frac{1}{4{\cal V}} \, \int_{X_3} \mu_\alpha \wedge \ast \mu_\beta :=  \frac{{\cal G}_{\alpha\beta}}{4{\cal V}}\,, \nonumber\\
& \hskip-1cm K^{\alpha \ov{\beta}} = - \, 4\, {\cal V} \, \hat\kappa^{\alpha\beta}, \quad & {K}^{\alpha \ov \beta} = 4 \, {\cal V} \,  \int_{X_3} \, \tilde{\mu}^\alpha \wedge \ast \tilde{\mu}^\beta :=  {4{\cal V}} \, {\cal G}^{\alpha\beta}\,. \nonumber\\
& \hskip-1cm K_{I \ov J} = \, e^{2 \, D} \, {\cal M}_{IJ}, \quad & K^{I \ov J} = \, e^{-2D}\, {\cal M}^{IJ} \,,\nonumber
\eea
where the shorthand notations such as $\kappa_a\, t^a = 6\, {\cal V}, \, \kappa_{ab} = \kappa_{abc}\, t^c, \, \kappa_{a} = \kappa_{abc} \, t^b \, t^c = 2 \, \sigma_a$ along with $\kappa^{ab}$ and $\kappa_{ab}$, are used whenever needed. In addition, we have chosen the normalization of the three-form such that the following relations are satisfied \cite{Flauger:2008ad},
\bea
\label{eq:IIAcsmoduli}
& & \hskip-1cm {\cal X}^K \, {\cal F}_K = - \, \frac{i}{2}, \qquad {\cal M}^{IJ} \, {\cal F}_{J} = -\, \frac{i}{2}\, {\cal X}^I, \qquad {\cal M}_{IJ} \,{\cal X}^J = 2\, i\, {\cal F}_I\, , \\
& & \hskip-1cm {\cal M}^{IJ}\, {\cal F}_I \, {\cal F}_J = -\, \frac{1}{4}, \qquad  {\cal M}_{IJ}\, {\cal X}^I \, {\cal X}^J = 1, \qquad \partial_J \, K = -4\, e^{D} \, {\cal F}_J \,. \nonumber
\eea
Notice that in the absence of (non-)geometric fluxes, the collection of scalar potential pieces in eqns. (\ref{eq:IIA-pot-gen-all}) reduces into the ones presented, e.g. in eqn. (3.15) of \cite{Carta:2016ynn} and eqn. (4.26) of \cite{Farakos:2017jme}. In addition, using the moduli space metric relations in eqn. (\ref{eq:KabIIA}), it is easy to observe that our scalar potential in eqn. (\ref{eq:IIA-pot-gen-all}) reduces into the one presented in eqn. (2.34) of \cite{Flauger:2008ad} when the nongeometric $Q$ and $R$ fluxes are absent but the geometric flux contributions are included.

For the current interest in this work regarding the de-Sitter no-go scenarios, we recollect the scalar potential pieces in eqn. (\ref{eq:IIA-pot-gen}) in some particular manner given as follows,
 \bea
\label{eq:typeIIA-genpot1}
&& \hskip-2cm V_{\rm IIA} = V_{h} + V_{\omega} + V_{q}+ V_{r} + V_{f_0} + V_{f_2} + V_{f_4} + V_{f_6} + V_{\rm loc}\,,
\eea
where the various pieces can be read-off from the eqn. (\ref{eq:IIA-pot-gen-all}) in the following way,
\bea
\label{eq:typeIIA-genpot2}
& & \hskip-1cm V_h \equiv V_{{\mathbb H} {\mathbb H}}\,, \quad \, \, \, \, V_\omega \equiv V_{{\mathbb \mho} {\mathbb \mho}} + V_{{\mathbb H} {\mathbb Q}} + V_D^{(1)}\, \quad V_q \equiv V_{{\mathbb Q} {\mathbb Q}} + V_{{\mathbb R} {\mathbb \mho}} + V_D^{(2)}\,, \quad V_r \equiv V_{{\mathbb R} {\mathbb R}}\,,\\
& & \hskip-1cm V_{f_0} \equiv V_{{\mathbb G}_0 {\mathbb G}_0}\,, \quad V_{f_2} \equiv V_{{\mathbb G}^a {\mathbb G}^a}\,, \quad V_{f_4} \equiv V_{{\mathbb G}_a {\mathbb G}_a}\,, \quad V_{f_6} \equiv V_{{\mathbb G}^0 {\mathbb G}^0}\,, \quad V_{\rm loc} \equiv V_{D6/O6}\,. \nonumber
\eea 
Here let us mention that we have clubbed the ``$HQ$-type" and the ``$\omega R$-type" cross-terms into what we call $V_\omega$ and $V_q$ pieces respectively. As may be obvious from eqn. (\ref{eq:IIA-pot-gen-all}), this has been done because the ``$HQ$-terms" scale similar to ``$\omega^2$-terms" in two-cycle volume moduli $t^a$ and the dilaton $D$, and similarly the ``$\omega R$-terms" scale as those of the ``$Q^2$-terms" in these two moduli. Moreover, the two positive definite $D$-term contributions, which we have denoted as $V_D^{(1)}$ and $V_{D}^{(2)}$, are also clubbed along with their respective pieces into $V_\omega$ and $V_q$  respectively. 

Also note the fact that the various pieces in the scalar potential given in eqns. (\ref{eq:typeIIA-genpot1})-(\ref{eq:typeIIA-genpot2}) involve the generalized flux orbits as defined in eqn.  (\ref{eq:TypeIIAfluxOrbits0}), and therefore assuming that $V_h$ piece involves only the NS-NS $H_3$-flux components would be misleading as the flux orbit ${\mathbb H}$ can generically have all the NS-NS fluxes, namely the $H, \omega, Q$ and $R$ fluxes. This argument holds true for the other pieces as well. Our observation from the explicit computations of the scalar potential also tells us that the naive form of the scalar potential as assumed in \cite{deCarlos:2009fq} can be only valid for the cases when the NS-NS axionic moduli $b^a$'s are set/stabilized to zero as in that case the generalized flux orbits would reduce into the usual flux components. 

We use this scalar potential given in eqns. (\ref{eq:typeIIA-genpot1})- (\ref{eq:typeIIA-genpot2}) for our current analysis in this work. Our generic scalar potential with explicit dependence on all the moduli and the fluxes should enable one to explore more possibilities either for evading or finding new no-go conditions with different set of flux choices considered in a given scenario.


\newpage
\bibliographystyle{utphys}
\bibliography{reference}

\providecommand{\href}[2]{#2}\begingroup\raggedright\begin{thebibliography}{100}

\bibitem{Kachru:2003aw}
S.~Kachru, R.~Kallosh, A.~D. Linde, and S.~P. Trivedi, ``{De Sitter vacua in
  string theory},'' {\em Phys. Rev.} {\bf D68} (2003) 046005,
\href{http://www.arXiv.org/abs/hep-th/0301240}{{\tt hep-th/0301240}}.

\bibitem{Taylor:1999ii}
T.~R. Taylor and C.~Vafa, ``{R R flux on Calabi-Yau and partial supersymmetry
  breaking},'' {\em Phys.Lett.} {\bf B474} (2000) 130--137,
\href{http://www.arXiv.org/abs/hep-th/9912152}{{\tt hep-th/9912152}}.

\bibitem{Blumenhagen:2003vr}
R.~Blumenhagen, D.~Lust, and T.~R. Taylor, ``{Moduli stabilization in chiral
  type IIB orientifold models with fluxes},'' {\em Nucl.Phys.} {\bf B663}
  (2003) 319--342,
\href{http://www.arXiv.org/abs/hep-th/0303016}{{\tt hep-th/0303016}}.

\bibitem{Grimm:2004ua}
T.~W. Grimm and J.~Louis, ``{The Effective action of type IIA Calabi-Yau
  orientifolds},'' {\em Nucl. Phys.} {\bf B718} (2005) 153--202,
\href{http://www.arXiv.org/abs/hep-th/0412277}{{\tt hep-th/0412277}}.

\bibitem{Grimm:2004uq}
T.~W. Grimm and J.~Louis, ``{The Effective action of N = 1 Calabi-Yau
  orientifolds},'' {\em Nucl.Phys.} {\bf B699} (2004) 387--426,
\href{http://www.arXiv.org/abs/hep-th/0403067}{{\tt hep-th/0403067}}.

\bibitem{Denef:2005mm}
F.~Denef, M.~R. Douglas, B.~Florea, A.~Grassi, and S.~Kachru, ``{Fixing all
  moduli in a simple f-theory compactification},'' {\em Adv. Theor. Math.
  Phys.} {\bf 9} (2005) 861--929,
\href{http://www.arXiv.org/abs/hep-th/0503124}{{\tt hep-th/0503124}}.

\bibitem{Grana:2005jc}
M.~Grana, ``{Flux compactifications in string theory: A Comprehensive
  review},'' {\em Phys. Rept.} {\bf 423} (2006) 91--158,
\href{http://www.arXiv.org/abs/hep-th/0509003}{{\tt hep-th/0509003}}.

\bibitem{Balasubramanian:2005zx}
V.~Balasubramanian, P.~Berglund, J.~P. Conlon, and F.~Quevedo, ``{Systematics
  of moduli stabilisation in Calabi-Yau flux compactifications},'' {\em JHEP}
  {\bf 03} (2005) 007,
\href{http://www.arXiv.org/abs/hep-th/0502058}{{\tt hep-th/0502058}}.

\bibitem{Blumenhagen:2006ci}
R.~Blumenhagen, B.~Kors, D.~Lust, and S.~Stieberger, ``{Four-dimensional String
  Compactifications with D-Branes, Orientifolds and Fluxes},'' {\em Phys.Rept.}
  {\bf 445} (2007) 1--193,
\href{http://www.arXiv.org/abs/hep-th/0610327}{{\tt hep-th/0610327}}.

\bibitem{Douglas:2006es}
M.~R. Douglas and S.~Kachru, ``{Flux compactification},'' {\em Rev. Mod. Phys.}
  {\bf 79} (2007) 733--796,
\href{http://www.arXiv.org/abs/hep-th/0610102}{{\tt hep-th/0610102}}.

\bibitem{Blumenhagen:2007sm}
R.~Blumenhagen, S.~Moster, and E.~Plauschinn, ``{Moduli Stabilisation versus
  Chirality for MSSM like Type IIB Orientifolds},'' {\em JHEP} {\bf 01} (2008)
  058,
\href{http://www.arXiv.org/abs/0711.3389}{{\tt 0711.3389}}.

\bibitem{Aldazabal:2006up}
G.~Aldazabal, P.~G. Camara, A.~Font, and L.~Ibanez, ``{More dual fluxes and
  moduli fixing},'' {\em JHEP} {\bf 0605} (2006) 070,
\href{http://www.arXiv.org/abs/hep-th/0602089}{{\tt hep-th/0602089}}.

\bibitem{Ihl:2006pp}
M.~Ihl and T.~Wrase, ``{Towards a Realistic Type IIA T**6/Z(4) Orientifold
  Model with Background Fluxes. Part 1. Moduli Stabilization},'' {\em JHEP}
  {\bf 07} (2006) 027,
\href{http://www.arXiv.org/abs/hep-th/0604087}{{\tt hep-th/0604087}}.

\bibitem{Ihl:2007ah}
M.~Ihl, D.~Robbins, and T.~Wrase, ``{Toroidal orientifolds in IIA with general
  NS-NS fluxes},'' {\em JHEP} {\bf 0708} (2007) 043,
\href{http://www.arXiv.org/abs/0705.3410}{{\tt 0705.3410}}.

\bibitem{Font:2008vd}
A.~Font, A.~Guarino, and J.~M. Moreno, ``{Algebras and non-geometric flux
  vacua},'' {\em JHEP} {\bf 0812} (2008) 050,
\href{http://www.arXiv.org/abs/0809.3748}{{\tt 0809.3748}}.

\bibitem{Guarino:2008ik}
A.~Guarino and G.~J. Weatherill, ``{Non-geometric flux vacua, S-duality and
  algebraic geometry},'' {\em JHEP} {\bf 0902} (2009) 042,
\href{http://www.arXiv.org/abs/0811.2190}{{\tt 0811.2190}}.

\bibitem{Aldazabal:2008zza}
G.~Aldazabal, P.~G. Camara, and J.~Rosabal, ``{Flux algebra, Bianchi identities
  and Freed-Witten anomalies in F-theory compactifications},'' {\em Nucl.Phys.}
  {\bf B814} (2009) 21--52,
\href{http://www.arXiv.org/abs/0811.2900}{{\tt 0811.2900}}.

\bibitem{deCarlos:2009qm}
B.~de~Carlos, A.~Guarino, and J.~M. Moreno, ``{Complete classification of
  Minkowski vacua in generalised flux models},'' {\em JHEP} {\bf 1002} (2010)
  076,
\href{http://www.arXiv.org/abs/0911.2876}{{\tt 0911.2876}}.

\bibitem{Danielsson:2012by}
U.~Danielsson and G.~Dibitetto, ``{On the distribution of stable de Sitter
  vacua},'' {\em JHEP} {\bf 1303} (2013) 018,
\href{http://www.arXiv.org/abs/1212.4984}{{\tt 1212.4984}}.

\bibitem{Blaback:2013ht}
J.~Blåbäck, U.~Danielsson, and G.~Dibitetto, ``{Fully stable dS vacua from
  generalised fluxes},'' {\em JHEP} {\bf 1308} (2013) 054,
\href{http://www.arXiv.org/abs/1301.7073}{{\tt 1301.7073}}.

\bibitem{Damian:2013dwa}
C.~Damian and O.~Loaiza-Brito, ``{More stable de Sitter vacua from S-dual
  nongeometric fluxes},'' {\em Phys.Rev.} {\bf D88} (2013), no.~4, 046008,
\href{http://www.arXiv.org/abs/1304.0792}{{\tt 1304.0792}}.

\bibitem{Damian:2013dq}
C.~Damian, L.~R. Diaz-Barron, O.~Loaiza-Brito, and M.~Sabido, ``{Slow-Roll
  Inflation in Non-geometric Flux Compactification},'' {\em JHEP} {\bf 1306}
  (2013) 109,
\href{http://www.arXiv.org/abs/1302.0529}{{\tt 1302.0529}}.

\bibitem{Hassler:2014mla}
F.~Hassler, D.~Lust, and S.~Massai, ``{On Inflation and de Sitter in
  Non‐Geometric String Backgrounds},'' {\em Fortsch. Phys.} {\bf 65} (2017),
  no.~10-11, 1700062,
\href{http://www.arXiv.org/abs/1405.2325}{{\tt 1405.2325}}.

\bibitem{Blumenhagen:2014gta}
R.~Blumenhagen and E.~Plauschinn, ``{Towards Universal Axion Inflation and
  Reheating in String Theory},'' {\em Phys.Lett.} {\bf B736} (2014) 482--487,
\href{http://www.arXiv.org/abs/1404.3542}{{\tt 1404.3542}}.

\bibitem{Blumenhagen:2015qda}
R.~Blumenhagen, A.~Font, M.~Fuchs, D.~Herschmann, and E.~Plauschinn, ``{Towards
  Axionic Starobinsky-like Inflation in String Theory},'' {\em Phys. Lett.}
  {\bf B746} (2015) 217--222,
\href{http://www.arXiv.org/abs/1503.01607}{{\tt 1503.01607}}.

\bibitem{Blumenhagen:2015jva}
R.~Blumenhagen, A.~Font, M.~Fuchs, D.~Herschmann, and E.~Plauschinn, ``{Large
  field inflation and string moduli stabilization},'' {\em PoS} {\bf
  PLANCK2015} (2015) 021,
\href{http://www.arXiv.org/abs/1510.04059}{{\tt 1510.04059}}.

\bibitem{Li:2015taa}
T.~Li, Z.~Li, and D.~V. Nanopoulos, ``{Helical Phase Inflation via
  Non-Geometric Flux Compactifications: from Natural to Starobinsky-like
  Inflation},'' {\em JHEP} {\bf 10} (2015) 138,
\href{http://www.arXiv.org/abs/1507.04687}{{\tt 1507.04687}}.

\bibitem{Blumenhagen:2015kja}
R.~Blumenhagen, A.~Font, M.~Fuchs, D.~Herschmann, E.~Plauschinn, Y.~Sekiguchi,
  and F.~Wolf, ``{A Flux-Scaling Scenario for High-Scale Moduli Stabilization
  in String Theory},'' {\em Nucl. Phys.} {\bf B897} (2015) 500--554,
\href{http://www.arXiv.org/abs/1503.07634}{{\tt 1503.07634}}.

\bibitem{Blumenhagen:2015xpa}
R.~Blumenhagen, C.~Damian, A.~Font, D.~Herschmann, and R.~Sun, ``{The
  Flux-Scaling Scenario: De Sitter Uplift and Axion Inflation},'' {\em Fortsch.
  Phys.} {\bf 64} (2016), no.~6-7, 536--550,
\href{http://www.arXiv.org/abs/1510.01522}{{\tt 1510.01522}}.

\bibitem{Blaback:2015zra}
J.~Blåbäck, U.~H. Danielsson, G.~Dibitetto, and S.~C. Vargas, ``{Universal dS
  vacua in STU-models},'' {\em JHEP} {\bf 10} (2015) 069,
\href{http://www.arXiv.org/abs/1505.04283}{{\tt 1505.04283}}.

\bibitem{Hellerman:2002ax}
S.~Hellerman, J.~McGreevy, and B.~Williams, ``{Geometric constructions of
  nongeometric string theories},'' {\em JHEP} {\bf 0401} (2004) 024,
\href{http://www.arXiv.org/abs/hep-th/0208174}{{\tt hep-th/0208174}}.

\bibitem{Dabholkar:2002sy}
A.~Dabholkar and C.~Hull, ``{Duality twists, orbifolds, and fluxes},'' {\em
  JHEP} {\bf 0309} (2003) 054,
\href{http://www.arXiv.org/abs/hep-th/0210209}{{\tt hep-th/0210209}}.

\bibitem{Hull:2004in}
C.~Hull, ``{A Geometry for non-geometric string backgrounds},'' {\em JHEP} {\bf
  0510} (2005) 065,
\href{http://www.arXiv.org/abs/hep-th/0406102}{{\tt hep-th/0406102}}.

\bibitem{Derendinger:2004jn}
J.-P. Derendinger, C.~Kounnas, P.~M. Petropoulos, and F.~Zwirner,
  ``{Superpotentials in IIA compactifications with general fluxes},'' {\em
  Nucl.Phys.} {\bf B715} (2005) 211--233,
\href{http://www.arXiv.org/abs/hep-th/0411276}{{\tt hep-th/0411276}}.

\bibitem{Derendinger:2005ph}
J.-P. Derendinger, C.~Kounnas, P.~Petropoulos, and F.~Zwirner, ``{Fluxes and
  gaugings: N=1 effective superpotentials},'' {\em Fortsch.Phys.} {\bf 53}
  (2005) 926--935,
\href{http://www.arXiv.org/abs/hep-th/0503229}{{\tt hep-th/0503229}}.

\bibitem{Shelton:2005cf}
J.~Shelton, W.~Taylor, and B.~Wecht, ``{Nongeometric flux compactifications},''
  {\em JHEP} {\bf 0510} (2005) 085,
\href{http://www.arXiv.org/abs/hep-th/0508133}{{\tt hep-th/0508133}}.

\bibitem{Wecht:2007wu}
B.~Wecht, ``{Lectures on Nongeometric Flux Compactifications},'' {\em Class.
  Quant. Grav.} {\bf 24} (2007) S773--S794,
\href{http://www.arXiv.org/abs/0708.3984}{{\tt 0708.3984}}.

\bibitem{Dall'Agata:2009gv}
G.~Dall'Agata, G.~Villadoro, and F.~Zwirner, ``{Type-IIA flux compactifications
  and N=4 gauged supergravities},'' {\em JHEP} {\bf 0908} (2009) 018,
\href{http://www.arXiv.org/abs/0906.0370}{{\tt 0906.0370}}.

\bibitem{Aldazabal:2011yz}
G.~Aldazabal, D.~Marques, C.~Nunez, and J.~A. Rosabal, ``{On Type IIB moduli
  stabilization and N = 4, 8 supergravities},'' {\em Nucl. Phys.} {\bf B849}
  (2011) 80--111,
\href{http://www.arXiv.org/abs/1101.5954}{{\tt 1101.5954}}.

\bibitem{Aldazabal:2011nj}
G.~Aldazabal, W.~Baron, D.~Marques, and C.~Nunez, ``{The effective action of
  Double Field Theory},'' {\em JHEP} {\bf 1111} (2011) 052,
\href{http://www.arXiv.org/abs/1109.0290}{{\tt 1109.0290}}.

\bibitem{Geissbuhler:2011mx}
D.~Geissbuhler, ``{Double Field Theory and N=4 Gauged Supergravity},'' {\em
  JHEP} {\bf 1111} (2011) 116,
\href{http://www.arXiv.org/abs/1109.4280}{{\tt 1109.4280}}.

\bibitem{Grana:2012rr}
M.~Gra\~{n}a and D.~Marques, ``{Gauged Double Field Theory},'' {\em JHEP} {\bf
  1204} (2012) 020,
\href{http://www.arXiv.org/abs/1201.2924}{{\tt 1201.2924}}.

\bibitem{Dibitetto:2012rk}
G.~Dibitetto, J.~Fernandez-Melgarejo, D.~Marques, and D.~Roest, ``{Duality
  orbits of non-geometric fluxes},'' {\em Fortsch.Phys.} {\bf 60} (2012)
  1123--1149,
\href{http://www.arXiv.org/abs/1203.6562}{{\tt 1203.6562}}.

\bibitem{Andriot:2013xca}
D.~Andriot and A.~Betz, ``{$\beta$-supergravity: a ten-dimensional theory with
  non-geometric fluxes, and its geometric framework},'' {\em JHEP} {\bf 1312}
  (2013) 083,
\href{http://www.arXiv.org/abs/1306.4381}{{\tt 1306.4381}}.

\bibitem{Andriot:2014qla}
D.~Andriot and A.~Betz, ``{Supersymmetry with non-geometric fluxes, or a
  $\beta$-twist in Generalized Geometry and Dirac operator},'' {\em JHEP} {\bf
  04} (2015) 006,
\href{http://www.arXiv.org/abs/1411.6640}{{\tt 1411.6640}}.

\bibitem{Blair:2014zba}
C.~D.~A. Blair and E.~Malek, ``{Geometry and fluxes of SL(5) exceptional field
  theory},'' {\em JHEP} {\bf 03} (2015) 144,
\href{http://www.arXiv.org/abs/1412.0635}{{\tt 1412.0635}}.

\bibitem{Andriot:2012an}
D.~Andriot, O.~Hohm, M.~Larfors, D.~Lust, and P.~Patalong, ``{Non-Geometric
  Fluxes in Supergravity and Double Field Theory},'' {\em Fortsch.Phys.} {\bf
  60} (2012) 1150--1186,
\href{http://www.arXiv.org/abs/1204.1979}{{\tt 1204.1979}}.

\bibitem{Geissbuhler:2013uka}
D.~Geissbuhler, D.~Marques, C.~Nunez, and V.~Penas, ``{Exploring Double Field
  Theory},'' {\em JHEP} {\bf 06} (2013) 101,
\href{http://www.arXiv.org/abs/1304.1472}{{\tt 1304.1472}}.

\bibitem{Blumenhagen:2013hva}
R.~Blumenhagen, X.~Gao, D.~Herschmann, and P.~Shukla, ``{Dimensional Oxidation
  of Non-geometric Fluxes in Type II Orientifolds},'' {\em JHEP} {\bf 1310}
  (2013) 201,
\href{http://www.arXiv.org/abs/1306.2761}{{\tt 1306.2761}}.

\bibitem{Villadoro:2005cu}
G.~Villadoro and F.~Zwirner, ``{N=1 effective potential from dual type-IIA
  D6/O6 orientifolds with general fluxes},'' {\em JHEP} {\bf 0506} (2005) 047,
\href{http://www.arXiv.org/abs/hep-th/0503169}{{\tt hep-th/0503169}}.

\bibitem{Robbins:2007yv}
D.~Robbins and T.~Wrase, ``{D-terms from generalized NS-NS fluxes in type
  II},'' {\em JHEP} {\bf 0712} (2007) 058,
\href{http://www.arXiv.org/abs/0709.2186}{{\tt 0709.2186}}.

\bibitem{Lombardo:2016swq}
D.~M. Lombardo, F.~Riccioni, and S.~Risoli, ``{$P$ fluxes and exotic branes},''
  {\em JHEP} {\bf 12} (2016) 114,
\href{http://www.arXiv.org/abs/1610.07975}{{\tt 1610.07975}}.

\bibitem{Lombardo:2017yme}
D.~M. Lombardo, F.~Riccioni, and S.~Risoli, ``{Non-geometric fluxes \& tadpole
  conditions for exotic branes},'' {\em JHEP} {\bf 10} (2017) 134,
\href{http://www.arXiv.org/abs/1704.08566}{{\tt 1704.08566}}.

\bibitem{Ceresole:1995ca}
A.~Ceresole, R.~D'Auria, and S.~Ferrara, ``{The Symplectic structure of N=2
  supergravity and its central extension},'' {\em Nucl.Phys.Proc.Suppl.} {\bf
  46} (1996) 67--74,
\href{http://www.arXiv.org/abs/hep-th/9509160}{{\tt hep-th/9509160}}.

\bibitem{D'Auria:2007ay}
R.~D'Auria, S.~Ferrara, and M.~Trigiante, ``{On the supergravity formulation of
  mirror symmetry in generalized Calabi-Yau manifolds},'' {\em Nucl. Phys.}
  {\bf B780} (2007) 28--39,
\href{http://www.arXiv.org/abs/hep-th/0701247}{{\tt hep-th/0701247}}.

\bibitem{Shukla:2015hpa}
P.~Shukla, ``{A symplectic rearrangement of the four dimensional non-geometric
  scalar potential},'' {\em JHEP} {\bf 11} (2015) 162,
\href{http://www.arXiv.org/abs/1508.01197}{{\tt 1508.01197}}.

\bibitem{Gao:2017gxk}
X.~Gao, P.~Shukla, and R.~Sun, ``{Symplectic formulation of the type IIA
  nongeometric scalar potential},'' {\em Phys. Rev.} {\bf D98} (2018), no.~4,
  046009,
\href{http://www.arXiv.org/abs/1712.07310}{{\tt 1712.07310}}.

\bibitem{Shukla:2016hyy}
P.~Shukla, ``{Reading off the nongeometric scalar potentials via the
  topological data of the compactifying Calabi-Yau manifolds},'' {\em Phys.
  Rev.} {\bf D94} (2016), no.~8, 086003,
\href{http://www.arXiv.org/abs/1603.01290}{{\tt 1603.01290}}.

\bibitem{Gao:2015nra}
X.~Gao and P.~Shukla, ``{Dimensional oxidation and modular completion of
  non-geometric type IIB action},'' {\em JHEP} {\bf 1505} (2015) 018,
\href{http://www.arXiv.org/abs/1501.07248}{{\tt 1501.07248}}.

\bibitem{Shukla:2015rua}
P.~Shukla, ``{On modular completion of generalized flux orbits},'' {\em JHEP}
  {\bf 11} (2015) 075,
\href{http://www.arXiv.org/abs/1505.00544}{{\tt 1505.00544}}.

\bibitem{Shukla:2015bca}
P.~Shukla, ``{Implementing odd-axions in dimensional oxidation of 4D
  non-geometric type IIB scalar potential},'' {\em Nucl. Phys.} {\bf B902}
  (2016) 458--482,
\href{http://www.arXiv.org/abs/1507.01612}{{\tt 1507.01612}}.

\bibitem{Andriot:2012wx}
D.~Andriot, O.~Hohm, M.~Larfors, D.~Lust, and P.~Patalong, ``{A geometric
  action for non-geometric fluxes},'' {\em Phys.Rev.Lett.} {\bf 108} (2012)
  261602,
\href{http://www.arXiv.org/abs/1202.3060}{{\tt 1202.3060}}.

\bibitem{Andriot:2011uh}
D.~Andriot, M.~Larfors, D.~Lust, and P.~Patalong, ``{A ten-dimensional action
  for non-geometric fluxes},'' {\em JHEP} {\bf 1109} (2011) 134,
\href{http://www.arXiv.org/abs/1106.4015}{{\tt 1106.4015}}.

\bibitem{Blumenhagen:2015lta}
R.~Blumenhagen, A.~Font, and E.~Plauschinn, ``{Relating double field theory to
  the scalar potential of N = 2 gauged supergravity},'' {\em JHEP} {\bf 12}
  (2015) 122,
\href{http://www.arXiv.org/abs/1507.08059}{{\tt 1507.08059}}.

\bibitem{Plauschinn:2018wbo}
E.~Plauschinn, ``{Non-geometric backgrounds in string theory},'' {\em Phys.
  Rept.} {\bf 798} (2019) 1--122,
\href{http://www.arXiv.org/abs/1811.11203}{{\tt 1811.11203}}.

\bibitem{Benmachiche:2006df}
I.~Benmachiche and T.~W. Grimm, ``{Generalized N=1 orientifold
  compactifications and the Hitchin functionals},'' {\em Nucl.Phys.} {\bf B748}
  (2006) 200--252,
\href{http://www.arXiv.org/abs/hep-th/0602241}{{\tt hep-th/0602241}}.

\bibitem{Grana:2006hr}
M.~Grana, J.~Louis, and D.~Waldram, ``{SU(3) x SU(3) compactification and
  mirror duals of magnetic fluxes},'' {\em JHEP} {\bf 04} (2007) 101,
\href{http://www.arXiv.org/abs/hep-th/0612237}{{\tt hep-th/0612237}}.

\bibitem{Shukla:2016xdy}
P.~Shukla, ``{Revisiting the two formulations of Bianchi identities and their
  implications on moduli stabilization},'' {\em JHEP} {\bf 08} (2016) 146,
\href{http://www.arXiv.org/abs/1603.08545}{{\tt 1603.08545}}.

\bibitem{Gao:2018ayp}
X.~Gao, P.~Shukla, and R.~Sun, ``{On Missing Bianchi Identities in Cohomology
  Formulation},'' {\em Eur. Phys. J.} {\bf C79} (2019), no.~9, 781,
\href{http://www.arXiv.org/abs/1805.05748}{{\tt 1805.05748}}.

\bibitem{Betzler:2019kon}
P.~Betzler and E.~Plauschinn, ``{Type IIB flux vacua and tadpole
  cancellation},''
\href{http://www.arXiv.org/abs/1905.08823}{{\tt 1905.08823}}.

\bibitem{DeWolfe:2005uu}
O.~DeWolfe, A.~Giryavets, S.~Kachru, and W.~Taylor, ``{Type IIA moduli
  stabilization},'' {\em JHEP} {\bf 07} (2005) 066,
\href{http://www.arXiv.org/abs/hep-th/0505160}{{\tt hep-th/0505160}}.

\bibitem{Camara:2005dc}
P.~G. Camara, A.~Font, and L.~E. Ibanez, ``{Fluxes, moduli fixing and MSSM-like
  vacua in a simple IIA orientifold},'' {\em JHEP} {\bf 09} (2005) 013,
\href{http://www.arXiv.org/abs/hep-th/0506066}{{\tt hep-th/0506066}}.

\bibitem{Palti:2008mg}
E.~Palti, G.~Tasinato, and J.~Ward, ``{WEAKLY-coupled IIA Flux
  Compactifications},'' {\em JHEP} {\bf 06} (2008) 084,
\href{http://www.arXiv.org/abs/0804.1248}{{\tt 0804.1248}}.

\bibitem{Dibitetto:2011qs}
G.~Dibitetto, A.~Guarino, and D.~Roest, ``{Vacua Analysis in Extended
  Supersymmetry Compactifications},'' {\em Fortsch. Phys.} {\bf 60} (2012)
  987--990,
\href{http://www.arXiv.org/abs/1112.1306}{{\tt 1112.1306}}.

\bibitem{Blaback:2013fca}
J.~Blåbäck, U.~Danielsson, and G.~Dibitetto, ``{Accelerated Universes from
  type IIA Compactifications},'' {\em JCAP} {\bf 1403} (2014) 003,
\href{http://www.arXiv.org/abs/1310.8300}{{\tt 1310.8300}}.

\bibitem{Escobar:2018rna}
D.~Escobar, F.~Marchesano, and W.~Staessens, ``{Type IIA flux vacua and
  $\alpha'$-corrections},'' {\em JHEP} {\bf 06} (2019) 129,
\href{http://www.arXiv.org/abs/1812.08735}{{\tt 1812.08735}}.

\bibitem{Marchesano:2019hfb}
F.~Marchesano and J.~Quirant, ``{A Landscape of AdS Flux Vacua},''
\href{http://www.arXiv.org/abs/1908.11386}{{\tt 1908.11386}}.

\bibitem{Maldacena:2000mw}
J.~M. Maldacena and C.~Nunez, ``{Supergravity description of field theories on
  curved manifolds and a no go theorem},'' {\em Int. J. Mod. Phys.} {\bf A16}
  (2001) 822--855, \href{http://www.arXiv.org/abs/hep-th/0007018}{{\tt
  hep-th/0007018}}.
[,182(2000)].

\bibitem{Hertzberg:2007wc}
M.~P. Hertzberg, S.~Kachru, W.~Taylor, and M.~Tegmark, ``{Inflationary
  Constraints on Type IIA String Theory},'' {\em JHEP} {\bf 12} (2007) 095,
\href{http://www.arXiv.org/abs/0711.2512}{{\tt 0711.2512}}.

\bibitem{Hertzberg:2007ke}
M.~P. Hertzberg, M.~Tegmark, S.~Kachru, J.~Shelton, and O.~Ozcan, ``{Searching
  for Inflation in Simple String Theory Models: An Astrophysical
  Perspective},'' {\em Phys. Rev.} {\bf D76} (2007) 103521,
\href{http://www.arXiv.org/abs/0709.0002}{{\tt 0709.0002}}.

\bibitem{Haque:2008jz}
S.~S. Haque, G.~Shiu, B.~Underwood, and T.~Van~Riet, ``{Minimal simple de
  Sitter solutions},'' {\em Phys. Rev.} {\bf D79} (2009) 086005,
\href{http://www.arXiv.org/abs/0810.5328}{{\tt 0810.5328}}.

\bibitem{Flauger:2008ad}
R.~Flauger, S.~Paban, D.~Robbins, and T.~Wrase, ``{Searching for slow-roll
  moduli inflation in massive type IIA supergravity with metric fluxes},'' {\em
  Phys. Rev.} {\bf D79} (2009) 086011,
\href{http://www.arXiv.org/abs/0812.3886}{{\tt 0812.3886}}.

\bibitem{Caviezel:2008tf}
C.~Caviezel, P.~Koerber, S.~Kors, D.~Lust, T.~Wrase, and M.~Zagermann, ``{On
  the Cosmology of Type IIA Compactifications on SU(3)-structure Manifolds},''
  {\em JHEP} {\bf 04} (2009) 010,
\href{http://www.arXiv.org/abs/0812.3551}{{\tt 0812.3551}}.

\bibitem{Covi:2008ea}
L.~Covi, M.~Gomez-Reino, C.~Gross, J.~Louis, G.~A. Palma, and C.~A. Scrucca,
  ``{de Sitter vacua in no-scale supergravities and Calabi-Yau string
  models},'' {\em JHEP} {\bf 06} (2008) 057,
\href{http://www.arXiv.org/abs/0804.1073}{{\tt 0804.1073}}.

\bibitem{deCarlos:2009fq}
B.~de~Carlos, A.~Guarino, and J.~M. Moreno, ``{Flux moduli stabilisation,
  Supergravity algebras and no-go theorems},'' {\em JHEP} {\bf 01} (2010) 012,
\href{http://www.arXiv.org/abs/0907.5580}{{\tt 0907.5580}}.

\bibitem{Caviezel:2009tu}
C.~Caviezel, T.~Wrase, and M.~Zagermann, ``{Moduli Stabilization and Cosmology
  of Type IIB on SU(2)-Structure Orientifolds},'' {\em JHEP} {\bf 04} (2010)
  011,
\href{http://www.arXiv.org/abs/0912.3287}{{\tt 0912.3287}}.

\bibitem{Danielsson:2009ff}
U.~H. Danielsson, S.~S. Haque, G.~Shiu, and T.~Van~Riet, ``{Towards Classical
  de Sitter Solutions in String Theory},'' {\em JHEP} {\bf 09} (2009) 114,
\href{http://www.arXiv.org/abs/0907.2041}{{\tt 0907.2041}}.

\bibitem{Danielsson:2010bc}
U.~H. Danielsson, P.~Koerber, and T.~Van~Riet, ``{Universal de Sitter solutions
  at tree-level},'' {\em JHEP} {\bf 05} (2010) 090,
\href{http://www.arXiv.org/abs/1003.3590}{{\tt 1003.3590}}.

\bibitem{Wrase:2010ew}
T.~Wrase and M.~Zagermann, ``{On Classical de Sitter Vacua in String Theory},''
  {\em Fortsch. Phys.} {\bf 58} (2010) 906--910,
\href{http://www.arXiv.org/abs/1003.0029}{{\tt 1003.0029}}.

\bibitem{Shiu:2011zt}
G.~Shiu and Y.~Sumitomo, ``{Stability Constraints on Classical de Sitter
  Vacua},'' {\em JHEP} {\bf 09} (2011) 052,
\href{http://www.arXiv.org/abs/1107.2925}{{\tt 1107.2925}}.

\bibitem{McOrist:2012yc}
J.~McOrist and S.~Sethi, ``{M-theory and Type IIA Flux Compactifications},''
  {\em JHEP} {\bf 12} (2012) 122,
\href{http://www.arXiv.org/abs/1208.0261}{{\tt 1208.0261}}.

\bibitem{Dasgupta:2014pma}
K.~Dasgupta, R.~Gwyn, E.~McDonough, M.~Mia, and R.~Tatar, ``{de Sitter Vacua in
  Type IIB String Theory: Classical Solutions and Quantum Corrections},'' {\em
  JHEP} {\bf 07} (2014) 054,
\href{http://www.arXiv.org/abs/1402.5112}{{\tt 1402.5112}}.

\bibitem{Gautason:2015tig}
F.~F. Gautason, M.~Schillo, T.~Van~Riet, and M.~Williams, ``{Remarks on scale
  separation in flux vacua},'' {\em JHEP} {\bf 03} (2016) 061,
\href{http://www.arXiv.org/abs/1512.00457}{{\tt 1512.00457}}.

\bibitem{Junghans:2016uvg}
D.~Junghans, ``{Tachyons in Classical de Sitter Vacua},'' {\em JHEP} {\bf 06}
  (2016) 132,
\href{http://www.arXiv.org/abs/1603.08939}{{\tt 1603.08939}}.

\bibitem{Andriot:2016xvq}
D.~Andriot and J.~Blåbäck, ``{Refining the boundaries of the classical de
  Sitter landscape},'' {\em JHEP} {\bf 03} (2017) 102,
  \href{http://www.arXiv.org/abs/1609.00385}{{\tt 1609.00385}}.
[Erratum: JHEP03,083(2018)].

\bibitem{Andriot:2017jhf}
D.~Andriot, ``{On classical de Sitter and Minkowski solutions with intersecting
  branes},'' {\em JHEP} {\bf 03} (2018) 054,
\href{http://www.arXiv.org/abs/1710.08886}{{\tt 1710.08886}}.

\bibitem{Danielsson:2018ztv}
U.~H. Danielsson and T.~Van~Riet, ``{What if string theory has no de Sitter
  vacua?},'' {\em Int. J. Mod. Phys.} {\bf D27} (2018), no.~12, 1830007,
\href{http://www.arXiv.org/abs/1804.01120}{{\tt 1804.01120}}.

\bibitem{Ooguri:2006in}
H.~Ooguri and C.~Vafa, ``{On the Geometry of the String Landscape and the
  Swampland},'' {\em Nucl. Phys.} {\bf B766} (2007) 21--33,
\href{http://www.arXiv.org/abs/hep-th/0605264}{{\tt hep-th/0605264}}.

\bibitem{Obied:2018sgi}
G.~Obied, H.~Ooguri, L.~Spodyneiko, and C.~Vafa, ``{De Sitter Space and the
  Swampland},'' \href{http://www.arXiv.org/abs/1806.08362}{{\tt 1806.08362}}.

\bibitem{Danielsson:2011au}
U.~H. Danielsson, S.~S. Haque, P.~Koerber, G.~Shiu, T.~Van~Riet, and T.~Wrase,
  ``{De Sitter hunting in a classical landscape},'' {\em Fortsch. Phys.} {\bf
  59} (2011) 897--933,
\href{http://www.arXiv.org/abs/1103.4858}{{\tt 1103.4858}}.

\bibitem{Chen:2011ac}
X.~Chen, G.~Shiu, Y.~Sumitomo, and S.~H.~H. Tye, ``{A Global View on The Search
  for de-Sitter Vacua in (type IIA) String Theory},'' {\em JHEP} {\bf 04}
  (2012) 026,
\href{http://www.arXiv.org/abs/1112.3338}{{\tt 1112.3338}}.

\bibitem{Danielsson:2012et}
U.~H. Danielsson, G.~Shiu, T.~Van~Riet, and T.~Wrase, ``{A note on obstinate
  tachyons in classical dS solutions},'' {\em JHEP} {\bf 03} (2013) 138,
\href{http://www.arXiv.org/abs/1212.5178}{{\tt 1212.5178}}.

\bibitem{Andriot:2018wzk}
D.~Andriot, ``{On the de Sitter swampland criterion},'' {\em Phys. Lett.} {\bf
  B785} (2018) 570--573,
\href{http://www.arXiv.org/abs/1806.10999}{{\tt 1806.10999}}.

\bibitem{Andriot:2018ept}
D.~Andriot, ``{New constraints on classical de Sitter: flirting with the
  swampland},'' {\em Fortsch. Phys.} {\bf 67} (2019), no.~1-2, 1800103,
\href{http://www.arXiv.org/abs/1807.09698}{{\tt 1807.09698}}.

\bibitem{Garg:2018reu}
S.~K. Garg and C.~Krishnan, ``{Bounds on Slow Roll and the de Sitter
  Swampland},''
\href{http://www.arXiv.org/abs/1807.05193}{{\tt 1807.05193}}.

\bibitem{Denef:2018etk}
F.~Denef, A.~Hebecker, and T.~Wrase, ``{de Sitter swampland conjecture and the
  Higgs potential},'' {\em Phys. Rev.} {\bf D98} (2018), no.~8, 086004,
\href{http://www.arXiv.org/abs/1807.06581}{{\tt 1807.06581}}.

\bibitem{Conlon:2018eyr}
J.~P. Conlon, ``{The de Sitter swampland conjecture and supersymmetric AdS
  vacua},'' {\em Int. J. Mod. Phys.} {\bf A33} (2018), no.~29, 1850178,
\href{http://www.arXiv.org/abs/1808.05040}{{\tt 1808.05040}}.

\bibitem{Roupec:2018mbn}
C.~Roupec and T.~Wrase, ``{de Sitter Extrema and the Swampland},'' {\em
  Fortsch. Phys.} {\bf 67} (2019), no.~1-2, 1800082,
\href{http://www.arXiv.org/abs/1807.09538}{{\tt 1807.09538}}.

\bibitem{Murayama:2018lie}
H.~Murayama, M.~Yamazaki, and T.~T. Yanagida, ``{Do We Live in the
  Swampland?},'' {\em JHEP} {\bf 12} (2018) 032,
\href{http://www.arXiv.org/abs/1809.00478}{{\tt 1809.00478}}.

\bibitem{Choi:2018rze}
K.~Choi, D.~Chway, and C.~S. Shin, ``{The dS swampland conjecture with the
  electroweak symmetry and QCD chiral symmetry breaking},'' {\em JHEP} {\bf 11}
  (2018) 142,
\href{http://www.arXiv.org/abs/1809.01475}{{\tt 1809.01475}}.

\bibitem{Hamaguchi:2018vtv}
K.~Hamaguchi, M.~Ibe, and T.~Moroi, ``{The swampland conjecture and the Higgs
  expectation value},'' {\em JHEP} {\bf 12} (2018) 023,
\href{http://www.arXiv.org/abs/1810.02095}{{\tt 1810.02095}}.

\bibitem{Olguin-Tejo:2018pfq}
Y.~Olguin-Trejo, S.~L. Parameswaran, G.~Tasinato, and I.~Zavala, ``{Runaway
  Quintessence, Out of the Swampland},'' {\em JCAP} {\bf 1901} (2019), no.~01,
  031,
\href{http://www.arXiv.org/abs/1810.08634}{{\tt 1810.08634}}.

\bibitem{Blanco-Pillado:2018xyn}
J.~J. Blanco-Pillado, M.~A. Urkiola, and J.~M. Wachter, ``{Racetrack Potentials
  and the de Sitter Swampland Conjectures},'' {\em JHEP} {\bf 01} (2019) 187,
\href{http://www.arXiv.org/abs/1811.05463}{{\tt 1811.05463}}.

\bibitem{Ooguri:2018wrx}
H.~Ooguri, E.~Palti, G.~Shiu, and C.~Vafa, ``{Distance and de Sitter
  Conjectures on the Swampland},'' {\em Phys. Lett.} {\bf B788} (2019)
  180--184,
\href{http://www.arXiv.org/abs/1810.05506}{{\tt 1810.05506}}.

\bibitem{BlancoPillado:2006he}
J.~J. Blanco-Pillado, C.~P. Burgess, J.~M. Cline, C.~Escoda, M.~Gomez-Reino,
  R.~Kallosh, A.~D. Linde, and F.~Quevedo, ``{Inflating in a better
  racetrack},'' {\em JHEP} {\bf 09} (2006) 002,
\href{http://www.arXiv.org/abs/hep-th/0603129}{{\tt hep-th/0603129}}.

\bibitem{Kinney:2018nny}
W.~H. Kinney, S.~Vagnozzi, and L.~Visinelli, ``{The zoo plot meets the
  swampland: mutual (in)consistency of single-field inflation, string
  conjectures, and cosmological data},'' {\em Class. Quant. Grav.} {\bf 36}
  (2019), no.~11, 117001,
\href{http://www.arXiv.org/abs/1808.06424}{{\tt 1808.06424}}.

\bibitem{Achucarro:2018vey}
A.~Achúcarro and G.~A. Palma, ``{The string swampland constraints require
  multi-field inflation},'' {\em JCAP} {\bf 1902} (2019) 041,
\href{http://www.arXiv.org/abs/1807.04390}{{\tt 1807.04390}}.

\bibitem{Kehagias:2018uem}
A.~Kehagias and A.~Riotto, ``{A note on Inflation and the Swampland},'' {\em
  Fortsch. Phys.} {\bf 66} (2018), no.~10, 1800052,
\href{http://www.arXiv.org/abs/1807.05445}{{\tt 1807.05445}}.

\bibitem{Kinney:2018kew}
W.~H. Kinney, ``{Eternal Inflation and the Refined Swampland Conjecture},''
  {\em Phys. Rev. Lett.} {\bf 122} (2019), no.~8, 081302,
\href{http://www.arXiv.org/abs/1811.11698}{{\tt 1811.11698}}.

\bibitem{Lin:2018kjm}
C.-M. Lin, K.-W. Ng, and K.~Cheung, ``{Chaotic inflation on the brane and the
  Swampland Criteria},'' {\em Phys. Rev.} {\bf D100} (2019), no.~2, 023545,
\href{http://www.arXiv.org/abs/1810.01644}{{\tt 1810.01644}}.

\bibitem{Han:2018yrk}
C.~Han, S.~Pi, and M.~Sasaki, ``{Quintessence Saves Higgs Instability},'' {\em
  Phys. Lett.} {\bf B791} (2019) 314--318,
\href{http://www.arXiv.org/abs/1809.05507}{{\tt 1809.05507}}.

\bibitem{Raveri:2018ddi}
M.~Raveri, W.~Hu, and S.~Sethi, ``{Swampland Conjectures and Late-Time
  Cosmology},'' {\em Phys. Rev.} {\bf D99} (2019), no.~8, 083518,
\href{http://www.arXiv.org/abs/1812.10448}{{\tt 1812.10448}}.

\bibitem{Danielsson:2018qpa}
U.~Danielsson, ``{The quantum swampland},'' {\em JHEP} {\bf 04} (2019) 095,
\href{http://www.arXiv.org/abs/1809.04512}{{\tt 1809.04512}}.

\bibitem{Dasgupta:2018rtp}
K.~Dasgupta, M.~Emelin, E.~McDonough, and R.~Tatar, ``{Quantum Corrections and
  the de Sitter Swampland Conjecture},'' {\em JHEP} {\bf 01} (2019) 145,
\href{http://www.arXiv.org/abs/1808.07498}{{\tt 1808.07498}}.

\bibitem{Andriolo:2018yrz}
S.~Andriolo, G.~Shiu, H.~Triendl, T.~Van~Riet, G.~Venken, and G.~Zoccarato,
  ``{Compact G2 holonomy spaces from SU(3) structures},'' {\em JHEP} {\bf 03}
  (2019) 059,
\href{http://www.arXiv.org/abs/1811.00063}{{\tt 1811.00063}}.

\bibitem{Dasgupta:2019gcd}
K.~Dasgupta, M.~Emelin, M.~M. Faruk, and R.~Tatar, ``{de Sitter Vacua in the
  String Landscape},''
\href{http://www.arXiv.org/abs/1908.05288}{{\tt 1908.05288}}.

\bibitem{Andriot:2019wrs}
D.~Andriot, ``{Open problems on classical de Sitter solutions},'' {\em Fortsch.
  Phys.} {\bf 67} (2019), no.~7, 1900026,
\href{http://www.arXiv.org/abs/1902.10093}{{\tt 1902.10093}}.

\bibitem{Palti:2019pca}
E.~Palti, ``{The Swampland: Introduction and Review},'' {\em Fortsch. Phys.}
  {\bf 67} (2019), no.~6, 1900037,
\href{http://www.arXiv.org/abs/1903.06239}{{\tt 1903.06239}}.

\bibitem{Blumenhagen:2017cxt}
R.~Blumenhagen, I.~Valenzuela, and F.~Wolf, ``{The Swampland Conjecture and
  F-term Axion Monodromy Inflation},'' {\em JHEP} {\bf 07} (2017) 145,
\href{http://www.arXiv.org/abs/1703.05776}{{\tt 1703.05776}}.

\bibitem{Blumenhagen:2018nts}
R.~Blumenhagen, D.~Kläwer, L.~Schlechter, and F.~Wolf, ``{The Refined
  Swampland Distance Conjecture in Calabi-Yau Moduli Spaces},'' {\em JHEP} {\bf
  06} (2018) 052,
\href{http://www.arXiv.org/abs/1803.04989}{{\tt 1803.04989}}.

\bibitem{Blumenhagen:2018hsh}
R.~Blumenhagen, ``{Large Field Inflation/Quintessence and the Refined Swampland
  Distance Conjecture},'' {\em PoS} {\bf CORFU2017} (2018) 175,
\href{http://www.arXiv.org/abs/1804.10504}{{\tt 1804.10504}}.

\bibitem{Palti:2017elp}
E.~Palti, ``{The Weak Gravity Conjecture and Scalar Fields},'' {\em JHEP} {\bf
  08} (2017) 034,
\href{http://www.arXiv.org/abs/1705.04328}{{\tt 1705.04328}}.

\bibitem{Conlon:2016aea}
J.~P. Conlon and S.~Krippendorf, ``{Axion decay constants away from the
  lamppost},'' {\em JHEP} {\bf 04} (2016) 085,
\href{http://www.arXiv.org/abs/1601.00647}{{\tt 1601.00647}}.

\bibitem{Hebecker:2017lxm}
A.~Hebecker, P.~Henkenjohann, and L.~T. Witkowski, ``{Flat Monodromies and a
  Moduli Space Size Conjecture},'' {\em JHEP} {\bf 12} (2017) 033,
\href{http://www.arXiv.org/abs/1708.06761}{{\tt 1708.06761}}.

\bibitem{Klaewer:2016kiy}
D.~Klaewer and E.~Palti, ``{Super-Planckian Spatial Field Variations and
  Quantum Gravity},'' {\em JHEP} {\bf 01} (2017) 088,
\href{http://www.arXiv.org/abs/1610.00010}{{\tt 1610.00010}}.

\bibitem{Baume:2016psm}
F.~Baume and E.~Palti, ``{Backreacted Axion Field Ranges in String Theory},''
  {\em JHEP} {\bf 08} (2016) 043,
\href{http://www.arXiv.org/abs/1602.06517}{{\tt 1602.06517}}.

\bibitem{Landete:2018kqf}
A.~Landete and G.~Shiu, ``{Mass Hierarchies and Dynamical Field Range},'' {\em
  Phys. Rev.} {\bf D98} (2018), no.~6, 066012,
\href{http://www.arXiv.org/abs/1806.01874}{{\tt 1806.01874}}.

\bibitem{Cicoli:2018tcq}
M.~Cicoli, D.~Ciupke, C.~Mayrhofer, and P.~Shukla, ``{A Geometrical Upper Bound
  on the Inflaton Range},'' {\em JHEP} {\bf 05} (2018) 001,
\href{http://www.arXiv.org/abs/1801.05434}{{\tt 1801.05434}}.

\bibitem{Font:2019cxq}
A.~Font, A.~Herráez, and L.~E. Ibáñez, ``{The Swampland Distance Conjecture
  and Towers of Tensionless Branes},'' {\em JHEP} {\bf 08} (2019) 044,
\href{http://www.arXiv.org/abs/1904.05379}{{\tt 1904.05379}}.

\bibitem{Grimm:2018cpv}
T.~W. Grimm, C.~Li, and E.~Palti, ``{Infinite Distance Networks in Field Space
  and Charge Orbits},'' {\em JHEP} {\bf 03} (2019) 016,
\href{http://www.arXiv.org/abs/1811.02571}{{\tt 1811.02571}}.

\bibitem{Hebecker:2018fln}
A.~Hebecker, D.~Junghans, and A.~Schachner, ``{Large Field Ranges from Aligned
  and Misaligned Winding},'' {\em JHEP} {\bf 03} (2019) 192,
\href{http://www.arXiv.org/abs/1812.05626}{{\tt 1812.05626}}.

\bibitem{Banlaki:2018ayh}
A.~Banlaki, A.~Chowdhury, C.~Roupec, and T.~Wrase, ``{Scaling limits of dS
  vacua and the swampland},'' {\em JHEP} {\bf 03} (2019) 065,
\href{http://www.arXiv.org/abs/1811.07880}{{\tt 1811.07880}}.

\bibitem{Junghans:2018gdb}
D.~Junghans, ``{Weakly Coupled de Sitter Vacua with Fluxes and the
  Swampland},'' {\em JHEP} {\bf 03} (2019) 150,
\href{http://www.arXiv.org/abs/1811.06990}{{\tt 1811.06990}}.

\bibitem{Burgess:2003ic}
C.~P. Burgess, R.~Kallosh, and F.~Quevedo, ``{De Sitter string vacua from
  supersymmetric D terms},'' {\em JHEP} {\bf 10} (2003) 056,
\href{http://www.arXiv.org/abs/hep-th/0309187}{{\tt hep-th/0309187}}.

\bibitem{Achucarro:2006zf}
A.~Achucarro, B.~de~Carlos, J.~A. Casas, and L.~Doplicher, ``{De Sitter vacua
  from uplifting D-terms in effective supergravities from realistic strings},''
  {\em JHEP} {\bf 06} (2006) 014,
\href{http://www.arXiv.org/abs/hep-th/0601190}{{\tt hep-th/0601190}}.

\bibitem{Westphal:2006tn}
A.~Westphal, ``{de Sitter string vacua from Kahler uplifting},'' {\em JHEP}
  {\bf 03} (2007) 102,
\href{http://www.arXiv.org/abs/hep-th/0611332}{{\tt hep-th/0611332}}.

\bibitem{Silverstein:2007ac}
E.~Silverstein, ``{Simple de Sitter Solutions},'' {\em Phys. Rev.} {\bf D77}
  (2008) 106006,
\href{http://www.arXiv.org/abs/0712.1196}{{\tt 0712.1196}}.

\bibitem{Rummel:2011cd}
M.~Rummel and A.~Westphal, ``{A sufficient condition for de Sitter vacua in
  type IIB string theory},'' {\em JHEP} {\bf 01} (2012) 020,
\href{http://www.arXiv.org/abs/1107.2115}{{\tt 1107.2115}}.

\bibitem{Cicoli:2012fh}
M.~Cicoli, A.~Maharana, F.~Quevedo, and C.~P. Burgess, ``{De Sitter String
  Vacua from Dilaton-dependent Non-perturbative Effects},'' {\em JHEP} {\bf 06}
  (2012) 011,
\href{http://www.arXiv.org/abs/1203.1750}{{\tt 1203.1750}}.

\bibitem{Louis:2012nb}
J.~Louis, M.~Rummel, R.~Valandro, and A.~Westphal, ``{Building an explicit de
  Sitter},'' {\em JHEP} {\bf 10} (2012) 163,
\href{http://www.arXiv.org/abs/1208.3208}{{\tt 1208.3208}}.

\bibitem{Cicoli:2013cha}
M.~Cicoli, D.~Klevers, S.~Krippendorf, C.~Mayrhofer, F.~Quevedo, and
  R.~Valandro, ``{Explicit de Sitter Flux Vacua for Global String Models with
  Chiral Matter},'' {\em JHEP} {\bf 05} (2014) 001,
\href{http://www.arXiv.org/abs/1312.0014}{{\tt 1312.0014}}.

\bibitem{Cicoli:2015ylx}
M.~Cicoli, F.~Quevedo, and R.~Valandro, ``{De Sitter from T-branes},'' {\em
  JHEP} {\bf 03} (2016) 141,
\href{http://www.arXiv.org/abs/1512.04558}{{\tt 1512.04558}}.

\bibitem{Cicoli:2017shd}
M.~Cicoli, I.~Garcìa-Etxebarria, C.~Mayrhofer, F.~Quevedo, P.~Shukla, and
  R.~Valandro, ``{Global Orientifolded Quivers with Inflation},'' {\em JHEP}
  {\bf 11} (2017) 134,
\href{http://www.arXiv.org/abs/1706.06128}{{\tt 1706.06128}}.

\bibitem{Akrami:2018ylq}
Y.~Akrami, R.~Kallosh, A.~Linde, and V.~Vardanyan, ``{The Landscape, the
  Swampland and the Era of Precision Cosmology},'' {\em Fortsch. Phys.} {\bf
  67} (2019), no.~1-2, 1800075,
\href{http://www.arXiv.org/abs/1808.09440}{{\tt 1808.09440}}.

\bibitem{Antoniadis:2018hqy}
I.~Antoniadis, Y.~Chen, and G.~K. Leontaris, ``{Perturbative moduli
  stabilisation in type IIB/F-theory framework},'' {\em Eur. Phys. J.} {\bf
  C78} (2018), no.~9, 766,
\href{http://www.arXiv.org/abs/1803.08941}{{\tt 1803.08941}}.

\bibitem{Heckman:2019dsj}
J.~J. Heckman, C.~Lawrie, L.~Lin, J.~Sakstein, and G.~Zoccarato, ``{Pixelated
  Dark Energy},''
\href{http://www.arXiv.org/abs/1901.10489}{{\tt 1901.10489}}.

\bibitem{Heckman:2018mxl}
J.~J. Heckman, C.~Lawrie, L.~Lin, and G.~Zoccarato, ``{F-theory and Dark
  Energy},''
\href{http://www.arXiv.org/abs/1811.01959}{{\tt 1811.01959}}.

\bibitem{Cicoli:2018kdo}
M.~Cicoli, S.~De~Alwis, A.~Maharana, F.~Muia, and F.~Quevedo, ``{De Sitter vs
  Quintessence in String Theory},'' {\em Fortsch. Phys.} {\bf 67} (2019),
  no.~1-2, 1800079,
\href{http://www.arXiv.org/abs/1808.08967}{{\tt 1808.08967}}.

\bibitem{Damian:2018tlf}
C.~Damian and O.~Loaiza-Brito, ``{Two‐Field Axion Inflation and the Swampland
  Constraint in the Flux‐Scaling Scenario},'' {\em Fortsch. Phys.} {\bf 67}
  (2019), no.~1-2, 1800072,
\href{http://www.arXiv.org/abs/1808.03397}{{\tt 1808.03397}}.

\bibitem{Shukla:2019wfo}
P.~Shukla, ``{Dictionary for the type II nongeometric flux
  compactifications},'' {\em Phys. Rev. D} {\bf 103} (2021), no.~8, 086009,
  \href{http://www.arXiv.org/abs/1909.07391}{{\tt 1909.07391}}.

\bibitem{Shukla:2019dqd}
P.~Shukla, ``{$T$-dualizing de Sitter no-go scenarios},'' {\em Phys. Rev. D}
  {\bf 102} (2020), no.~2, 026014,
  \href{http://www.arXiv.org/abs/1909.08630}{{\tt 1909.08630}}.

\bibitem{Hosono:1994av}
S.~Hosono, A.~Klemm, and S.~Theisen, ``{Lectures on mirror symmetry},''
  \href{http://www.arXiv.org/abs/hep-th/9403096}{{\tt hep-th/9403096}}.
[Lect. Notes Phys.436,235(1994)].

\bibitem{Hori:2000kt}
K.~Hori and C.~Vafa, ``{Mirror symmetry},''
\href{http://www.arXiv.org/abs/hep-th/0002222}{{\tt hep-th/0002222}}.

\bibitem{Sethi:1994ch}
S.~Sethi, ``{Supermanifolds, rigid manifolds and mirror symmetry},'' {\em Nucl.
  Phys.} {\bf B430} (1994) 31--50,
  \href{http://www.arXiv.org/abs/hep-th/9404186}{{\tt hep-th/9404186}}.
[AMS/IP Stud. Adv. Math.1,793(1996)].

\bibitem{Shelton:2006fd}
J.~Shelton, W.~Taylor, and B.~Wecht, ``{Generalized Flux Vacua},'' {\em JHEP}
  {\bf 02} (2007) 095,
\href{http://www.arXiv.org/abs/hep-th/0607015}{{\tt hep-th/0607015}}.

\bibitem{Aldazabal:2013sca}
G.~Aldazabal, D.~Marques, and C.~Nunez, ``{Double Field Theory: A Pedagogical
  Review},'' {\em Class. Quant. Grav.} {\bf 30} (2013) 163001,
\href{http://www.arXiv.org/abs/1305.1907}{{\tt 1305.1907}}.

\bibitem{Andriot:2014uda}
D.~Andriot and A.~Betz, ``{NS-branes, source corrected Bianchi identities, and
  more on backgrounds with non-geometric fluxes},'' {\em JHEP} {\bf 07} (2014)
  059,
\href{http://www.arXiv.org/abs/1402.5972}{{\tt 1402.5972}}.

\bibitem{Dixon:1985jw}
L.~J. Dixon, J.~A. Harvey, C.~Vafa, and E.~Witten, ``{Strings on Orbifolds},''
  {\em Nucl. Phys.} {\bf B261} (1985) 678--686.
[,678(1985)].

\bibitem{Strominger:1985ku}
A.~Strominger, ``{TOPOLOGY OF SUPERSTRING COMPACTIFICATION},'' in {\em
  {Workshop on Unified String Theories Santa Barbara, California, July
  29-August 16, 1985}}.
\newblock
1985.
\newblock

\bibitem{Carta:2016ynn}
F.~Carta, F.~Marchesano, W.~Staessens, and G.~Zoccarato, ``{Open string
  multi-branched and Kähler potentials},'' {\em JHEP} {\bf 09} (2016) 062,
\href{http://www.arXiv.org/abs/1606.00508}{{\tt 1606.00508}}.

\bibitem{Farakos:2017jme}
F.~Farakos, S.~Lanza, L.~Martucci, and D.~Sorokin, ``{Three-forms in
  Supergravity and Flux Compactifications},'' {\em Eur. Phys. J.} {\bf C77}
  (2017), no.~9, 602,
\href{http://www.arXiv.org/abs/1706.09422}{{\tt 1706.09422}}.

\end{thebibliography}\endgroup

\end{document}